\makeatletter{}

\documentclass[review,number,fleqn,letterpaper,12pt,oneside]{elsarticle}

\usepackage{palatino}
\usepackage{graphicx}
\usepackage{amsmath}
\usepackage{bm}        \usepackage{amssymb}   \usepackage{lscape}
\usepackage{fancyhdr}
\usepackage{enumerate}
\usepackage{indentfirst}
\usepackage{xspace}
\usepackage{array}
\usepackage{rotating}
\usepackage{units}
\usepackage{textcomp} \usepackage[bookmarks, bookmarksopen, bookmarksnumbered,pdfpagelabels,colorlinks, linkcolor=blue, citecolor=blue, pdfview={Fit}]{hyperref}
\usepackage{marginnote}

\graphicspath{{figures/}{figures/eps/}{figures/pdf/}{figures/jpg/}}

\newcommand{\minerva}{MINERvA\xspace}
\newcommand{\minos}{MINOS\xspace}

\newcommand{\numi}{NuMI\xspace}

\newcommand{\kg}{\ensuremath{\mbox{kg}}\xspace}

\newcommand{\mAmp}{\ensuremath{\mbox{mA}}\xspace}

\newcommand{\nm}{\ensuremath{\mbox{nm}}\xspace}

\newcommand{\mm}{\ensuremath{\mbox{mm}}\xspace}
\newcommand{\cm}{\ensuremath{\mbox{cm}}\xspace}
\newcommand{\m}{\ensuremath{\mbox{m}}\xspace}

\newcommand{\ns}{\ensuremath{\mbox{ns}}\xspace}
\newcommand{\micros}{\ensuremath{\mu \mbox{s}}\xspace}

\newcommand{\Rutgers}{Rutgers, The State University of New Jersey, Piscataway, New Jersey 08854, USA}
\newcommand{\Hampton}{Hampton University, Dept. of Physics, Hampton, Virginia 23668, USA}

\newcommand{\Otterbein}{Department of Physics, Otterbein University, 1 South Grove Street, Westerville, Ohio, 43081, USA}
\newcommand{\JMU}{James Madison University, Harrisonburg, Virginia 22807, USA}
\newcommand{\Florida}{University of Florida, Department of Physics, Gainesville, Florida 32611, USA}
\newcommand{\UCIrvine}{Department of Physics and Astronomy, University of California, Irvine, Irvine, California 92697, USA}
\newcommand{\CBPF}{Centro Brasileiro de Pesquisas F\'\i sicas, Rua Dr. Xavier Sigaud 150, Urca, Rio de Janeiro, RJ, 22290-180, Brazil}
\newcommand{\PUCP}{Secci\'on F\'{\i}sica, Departamento de Ciencias, Pontificia Universidad Cat\'olica del Per\'u, Apartado 1761, Lima, Per\'u}
\newcommand{\INRM}{Institute for Nuclear Research of the Russian Academy of Sciences, 117312 Moscow, Russia}

\newcommand{\Pittsburgh}{Department of Physics and Astronomy, University of Pittsburgh, Pittsburgh, Pennsylvania 15260, USA}
\newcommand{\Guanajuato}{Lascuraín de Retana No. 5, Col. Centro. Guanajuato 36000, Guanajuato. Mexico}
\newcommand{\Athens}{Department of Physics, University of Athens, GR-15771 Athens, Greece}
\newcommand{\Tufts}{Physics Department, Tufts University, Medford, Massachusetts 02155, USA}
\newcommand{\WM}{Department of Physics, College of William \& Mary, Williamsburg, Virginia 23187, USA}
\newcommand{\FNAL}{Fermi National Accelerator Laboratory, Batavia, Illinois 60510, USA}

\newcommand{\MCLA}{Massachusetts College of Liberal Arts, 375 Church Street, North Adams, Massachusetts 01247}
\newcommand{\UMD}{Department of Physics, University of Minnesota -- Duluth, Duluth, Minnesota 55812, USA}
\newcommand{\Northwestern}{Northwestern University, Evanston, Illinois 60208}
\newcommand{\UNI}{12 Facultad de Ciencias, Universidad Nacional de Ingenier'a, Apartado 31139, Lima, Per\'u}
\newcommand{\Rochester}{University of Rochester, Rochester, New York 14610 USA}
\newcommand{\Austin}{Department of Physics, University of Texas, 1 University Station, Austin, Texas 78712, USA}
\newcommand{\USM}{Departamento de F\'isica, Universidad T\'ecnica Federico Santa Mar\'ia, Avda. Espa\~na 1680 Casilla 110-V Valpara\'iso, Chile}

\newcommand{\Chicago}{Enrico Fermi Institute, University of Chicago, Chicago, Illinois 60637, USA}

\newcommand{\deceased}{Deceased.}
\newcommand{\keppelThanks}{now at the Thomas Jefferson National Accelerator Facility, Newport News, VA 23606 USA}
\newcommand{\giulianoThanks}{now at Vrije Universiteit Brussel, Pleinlaan 2, B-1050 Brussels, Belgium}
\newcommand{\LazaThanks}{also at Department of Physics, University of Antananarivo, Madagascar}

\newcommand{\bradfordOverride}{Argonne National Laboratory, 9700 S. Cass Avenue, Lemont, IL 60439, USA}

\usepackage{float}

\journal{Nucl. Instrum. Methods A}

\begin{document}

\begin{frontmatter}

\title{Design, Calibration, and Performance of the \minerva\ Detector}

\makeatletter{}

\author[WM]{L.~Aliaga}  
\author[FNAL]{L.~Bagby}
\author[FNAL]{B.~Baldin}                   
\author[FNAL]{A.~Baumbaugh}
\author[Rochester]{A.~Bodek}   
\author[Rochester]{R.~Bradford\fnref{fnrefbb}}
\fntext[fnrefbb]{Present Address: \bradfordOverride}
\author[USM]{W.K.~Brooks}   
\author[FNAL]{D.~Boehnlein}  
\author[Pittsburgh]{S.~Boyd}    
\author[Rochester]{H.~Budd}                 
\author[INRM]{A.~Butkevich}       
\author[CBPF]{D.A.~Martinez Caicedo}     
\author[CBPF]{C.M.~Castromonte} 
\author[Hampton]{M.E.~Christy}       
\author[Rochester]{J.~Chvojka}                       
\author[CBPF]{H.~da~Motta}                    
\author[WM]{D.S.~Damiani}
\author[Pittsburgh]{I.~Danko}
\author[Hampton]{M.~Datta}                   
\author[FNAL]{R.~DeMaat\fnref{fnrefdec}}
\fntext[fnrefdec]{\deceased}                       
\author[WM]{J.~Devan} 
\author[UMD]{E.~Draeger}                        
\author[Pittsburgh]{S.A.~Dytman}                      
\author[PUCP]{G.A.~D\'{i}az}                 
\author[Pittsburgh]{B.~Eberly} 
\author[WM]{D.A.~Edmondson} 
\author[Guanajuato]{J.~Felix}                        
\author[Northwestern]{L.~Fields}              
\author[CBPF]{G.A.~Fiorentini}                
\author[UNI]{A.M.~Gago}                      
\author[Tufts]{H.~Gallagher}  
\author[Pittsburgh]{C.A.~George}   
\author[FNAL]{C.~Gingu}                 
\author[Northwestern]{B.~Gobbi\fnref{fnrefdec}}
\author[UMD]{R.~Gran}  
\author[FNAL]{N.~Grossman}                        
\author[FNAL]{D.~A.~Harris\corref{cor1}}
\cortext[cor1]{Corresponding Author}
\ead{dharris@fnal.gov}
\author[UMD]{J.~Heaton}                  
\author[Guanajuato]{A.~Higuera}                      
\author[Northwestern]{J.A.~Hobbs}
\author[WM]{I.J.~Howley}  
\author[CBPF,UNI]{K.~Hurtado} 
\author[Austin]{M.~Jerkins}                  
\author[Tufts]{T.~Kafka} 
\author[WM]{M.O.~Kantner}
\author[Hampton]{C.~Keppel\fnref{fnrefck}}
\fntext[fnrefck]{Present Address: \keppelThanks}
\author[FNAL]{J.~Kilmer}                
\author[WM]{A.H.~Krajeski}
\author[Rochester]{H.~Lee}
\author[WM]{A.G.~Leister} 
\author[Rutgers]{G. ~Locke}
\author[USM]{G.~Maggi\fnref{fnrefgm}}
\fntext[fnrefgm]{Present Address: \giulianoThanks}
\author[MCLA]{E.~Maher}                        
\author[Rochester]{S.~Manly}                        
\author[Tufts]{W.A.~Mann}                        
\author[Rochester]{C.M.~Marshall}  
\author[Rochester]{K.S.~McFarland}                  
\author[Pittsburgh]{C.L.~McGivern}                   
\author[Rochester]{A.~M.~McGowan}                   
\author[FNAL]{J.G.~Morf\'{i}n}
\author[Florida]{J.~Mousseau}                      
\author[Pittsburgh]{D.~Naples}                       
\author[WM]{J.K.~Nelson}                     
\author[JMU]{G.~Niculescu}                     
\author[JMU]{I.~Niculescu}                     
\author[WM]{C.D.~O'Connor} 
\author[PUCP]{N.~Ochoa}
\author[FNAL]{J.~Olsen}   
\author[Florida]{B.~Osmanov}
\author[FNAL]{J.~Osta}                      
\author[CBPF]{J.L.~Palomino}                   
\author[Pittsburgh]{V.~Paolone}                       
\author[Rochester]{J.~Park}
\author[Rochester]{G.~N.~Perdue}                     
\author[USM]{C.~Pe\~{n}a}                          
\author[FNAL]{A.~Pla-Dalmau}                  
\author[FNAL]{L.~Rakotondravohitra\fnref{fnreflr}}
\fntext[fnreflr]{Present Address: \LazaThanks}
\author[Rutgers]{R.~D.~Ransome}                    
\author[Florida]{H.~Ray}                          
\author[Pittsburgh]{L.~Ren}     
\author[FNAL]{P.~Rubinov}   
\author[UMD]{C.~Rude}                   
\author[WM]{K.E.~Sassin}
\author[Northwestern]{H.~Schellman}             
\author[Chicago,FNAL]{D.W.~Schmitz}     
\author[WM]{R.M.~Schneider} 
\author[UCIrvine]{C.~Simon}
\author[FNAL]{F.D.~Snider}                     
\author[WM]{M.C.~Snyder} 
\author[UNI]{C.J.~Solano~Salinas}            
\author[Otterbein]{N.~Tagg}                         
\author[Rutgers]{B.~G.~Tice} 
\author[Northwestern]{R.N.~Tilden}  
\author[Athens]{G.~Tzanakos\fnref{fnrefdec}}
\author[PUCP]{J.P.~Vel\'{a}squez}            
\author[Hampton]{T.~Walton}  
\author[UMD]{A.~Westerberg}                     
\author[Rochester]{J.~Wolcott}                    
\author[WM]{B.A.~Wolthuis} 
\author[UMD]{N. Woodward}
\author[Northwestern]{T.~Wytock}     
\author[Guanajuato]{G.~Zavala}                      
\author[WM]{D.~Zhang}  
\author[Hampton]{L.Y.~Zhu}                       
\author[UCIrvine]{B.P.~Ziemer}  
                       
  \address[WM]{\WM}
  \address[FNAL]{\FNAL}
  \address[Rochester]{\Rochester}
  \address[USM]{\USM}
  \address[Pittsburgh]{\Pittsburgh}
  \address[INRM]{\INRM}
  \address[CBPF]{\CBPF}
  \address[Hampton]{\Hampton}
  \address[UMD]{\UMD}
  \address[PUCP]{\PUCP}
  \address[Guanajuato]{\Guanajuato}
  \address[Northwestern]{\Northwestern}
  \address[UNI]{\UNI}
  \address[Tufts]{\Tufts}
  \address[Austin]{\Austin}
  \address[MCLA]{\MCLA}
  \address[Florida]{\Florida}
  \address[JMU]{\JMU}
  \address[Rutgers]{\Rutgers}
  \address[Chicago]{\Chicago}
  \address[UCIrvine]{\UCIrvine}
  \address[Otterbein]{\Otterbein}
  \address[Athens]{\Athens}


\begin{abstract}
The MINERvA\footnote{{\bf M}ain {\bf IN}jector {\bf E}xpe{\bf R}iment {\bf $\nu$-A}} experiment is designed to perform precision studies of neutrino-nucleus scattering using $\nu_\mu$ and ${\bar\nu}_\mu$ neutrinos incident
at 1-20 GeV in the NuMI beam at Fermilab.  This article presents a detailed description
of the \minerva detector and describes the {\em ex situ} and {\em in situ} techniques
employed to characterize the detector and monitor its performance.  
The detector is comprised of a finely-segmented scintillator-based inner tracking region surrounded 
by electromagnetic and hadronic sampling calorimetry.  The upstream portion of the detector includes 
planes of graphite, iron and lead interleaved between tracking planes to facilitate the study of 
nuclear effects in neutrino interactions.  Observations concerning the detector response over sustained
periods of running are reported.  The detector design and methods of operation have relevance to 
future neutrino experiments in which segmented scintillator tracking is utilized.
   
\end{abstract}


\end{frontmatter}

\tableofcontents

\makeatletter{}
\section{Experiment Overview}
\label{sec:ExperimentOverview}

\makeatletter{}
                             
There are many uncertainties in 
the knowledge of the fundamental interactions of neutrinos 
with the nucleon and the more complex
interactions with nuclei.  The properties of the neutrino, such as
its purely weak interaction and its unique flavor sensitivity, make it an ideal
probe, but its small cross sections, together with the difficulty of producing intense
neutrino beams, have impeded detailed investigations of the interaction.  
The intense neutrino beam produced
in the Neutrinos at the Main Injector (NuMI) beamline \cite{refnumi} at Fermilab has opened up new 
possibilities for the detailed study of neutrino interactions
with nuclei.  The \minerva\ experiment takes advantage of
this new opportunity.  \minerva\ is studying neutrino interactions on a variety
of nuclei, including helium, carbon, oxygen, lead, and iron, which complement
the wealth of electron scattering data in helping to understand the weak interaction
in the nuclear environment. 

The field of neutrino oscillations is rapidly advancing towards our understanding 
the complete picture of neutrino masses and mixings~\cite{T2K,NOvA,Hyper-K,LBNE}.
In particular, the next generation of accelerator-based long-baseline experiments aims to 
determine the mass hierarchy of neutrinos and test for CP violation by comparing the 
oscillation probabilities for neutrinos and antineutrinos.  Reaching the level of 
precision needed in these challenging measurements  requires a detailed understanding 
of neutrino- and antineutrino-nucleus scattering processes in the relevant energy range 
of a few hundred MeV to a few GeV.
Currently, however, the specific processes that provide signal and background channels 
for these experiments are, in many cases, either poorly measured or suffer from discrepant 
measurements across various experiments.  
In addition, an important feature of oscillation experiments is the need to measure 
oscillation probabilities as a function of the neutrino energy.  This requires 
experiments to have 
robust models of the relationship between the initial neutrino energy 
 and the visible energy deposited in the detectors.  Such models must incorporate 
the impact of the complex nuclear environment in which the interactions are occurring; 
studies have shown that neglecting these effects can lead to biases in the neutrino energy 
determination.  
Both the impact of the initial state an the interactions of final-state 
particles traversign the parent nucleus must be understood. Through 
measurements of specific interaction cross sections and comparisons among different 
nuclear targets, \minerva\ can
provide data to considerably improve the models of neutrino-nucleus scattering and
 thus to reduce systematic uncertainties in the results from oscillation experiments.

\subsection{\minerva\  Physics Goals and Detector Design}
\label{sec:detectordesign}

The key design features of the \minerva detector have been determined by the physics goals of the experiment.    
An overview of the detector design and its component systems is presented in this section.   
Detailed technical descriptions 
of the main components are provided in subsequent sections. \footnote{The doctoral dissertation of Jesse Chvojka \cite{Chvojka} contains additional detail on aspects of the detector which are beyond the scope of this paper.}

The physics goals of \minerva\ require a detector that can resolve multi-particle final
states, identify the produced particles, track low energy charged particles (for energies
greater than about 100 MeV),
contain electromagnetic showers,
contain high-energy (up to at least 10~GeV) final states, and resolve multiple interactions
in a single beam spill. The detector must include targets with a wide range of nucleon number $A$ to enable studies
of the nuclear dependence of neutrino interactions.   In order to track and resolve
multi-particle final states with low thresholds, the core of the detector must be fully active
with good spatial resolution and of relatively low mass. Full containment of events requires that the inner region 
be surrounded
with electromagnetic and hadronic calorimetry.  Ideally, charge identification would be included
by adding a magnetic field.  This proved impractical for the main detector, as did containment
of high energy muons.  However, by placing \minerva\ immediately upstream of the Main Injector
Oscillation Search (MINOS) near
detector \cite{refminos} (henceforth referred to as MINOS), a neutrino detector composed of
magnetized iron plates interleaved with scintillator planes, 
charge and energy measurements of forward-going muons can be made.

\begin{figure}[t]
\begin{center}
\includegraphics[trim=20mm -27mm 0mm 0mm, width=0.28\textwidth]{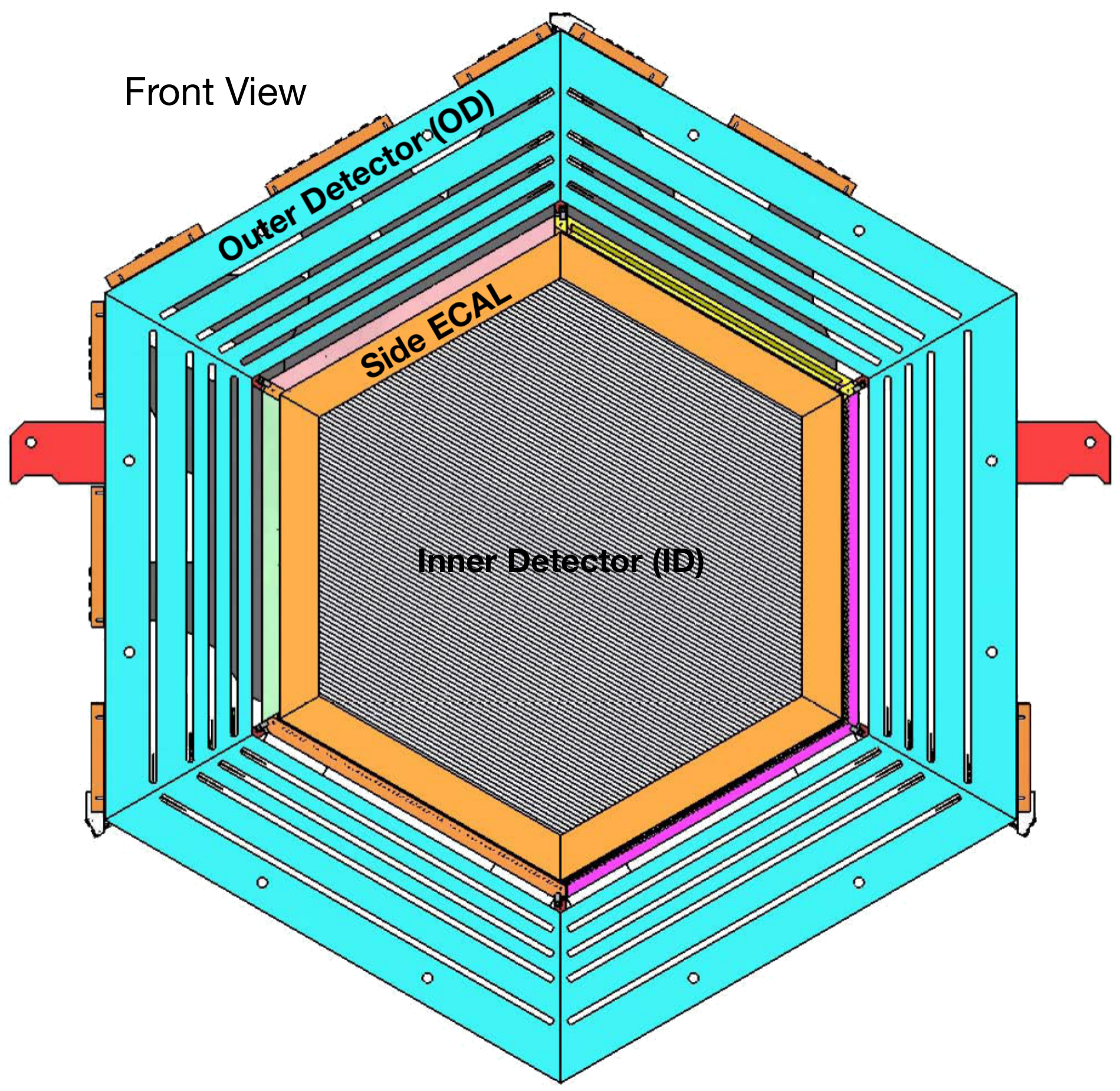}
\hspace{0.02\textwidth}
\includegraphics[width=0.68\textwidth]{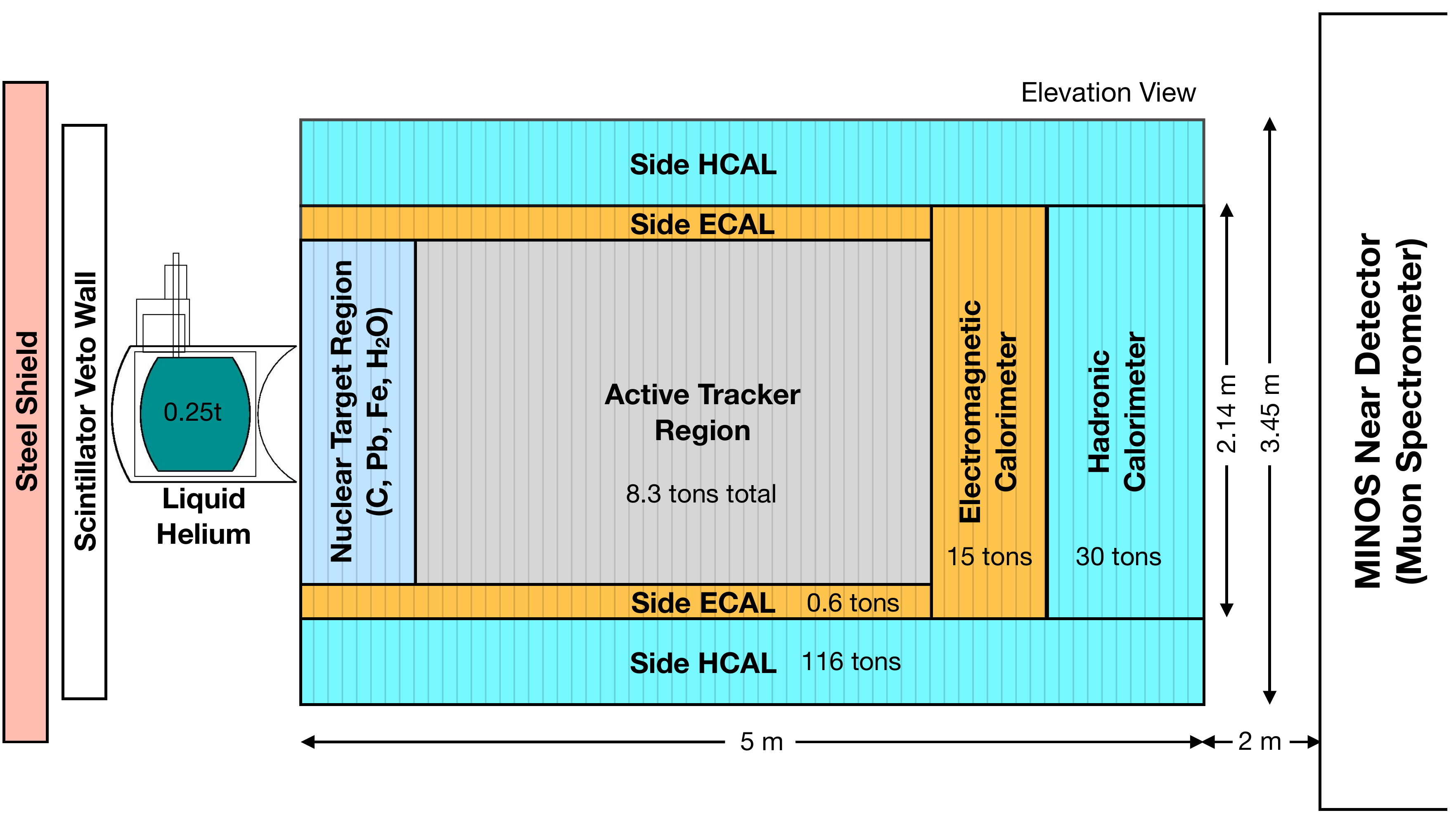}
\caption{Schematic views of the \minerva\ detector.
Left:  front view of a single  detector module.
Right:  side view of the complete detector showing 
the nuclear target, the fully-active tracking region and the
surrounding  
calorimeter regions.}
\label{fig:minerva-schematic}
\end{center}
\end{figure}

A schematic view of the detector is shown in Fig.~\ref{fig:minerva-schematic}. 
Neutrino reactions in the rock upstream of the detector hall can produce hadrons
and muons.  A ``veto wall'' upstream of the main detector shields against lower
energy hadrons from the rock and tags the muons (referred to below as ``rock muons''), which
can traverse all or part of the downstream detector. 
The veto wall consists of a 5~cm thick steel plate, a 1.9~cm thick plane of scintillator, a 2.5~cm thick steel plate,
and a second plane of 1.9~cm thick scintillator.  A cubic meter cryogenic vessel filled with liquid helium, described below, is
placed between the veto wall and the main detector.  

The main \minerva\ detector is segmented transversely into: the
inner detector (ID), with planes of solid scintillator strips mixed with the nuclear
targets; a region of pure scintillator; downstream electromagnetic calorimetry (ECAL)
and hadronic calorimetry (HCAL); and an outer detector (OD)
composed of a frame of steel
with imbedded scintillator,
which also serves as the supporting structure. 
Both the ID and OD are in the shape of  a regular hexagon. For construction and
convenience of handling, a single unit of \minerva\  incorporates both the scintillator 
and outer frame. Up to two planes of scintillator are mounted in one frame, called
a ``module''. Figure~\ref{fig:minerva-schematic} (left) shows a view of a tracking module.
There are three orientations of strips in the tracking planes, 
offset by 60$^\circ$ from each other, which enable a three-dimensional
reconstruction of tracks. The 60$^\circ$ offset fits naturally with the hexagonal transverse
cross section of the detector.  

The \minerva coordinate system is defined such that the
$z$ axis is horizontal and points downstream along the central axis
of the detector, the $y$ axis points upward, and the $x$ axis
is horizontal pointing to beam left, with the $x-y$ origin at the center of the ID.  The
$z$ axis is defined to place the front face of MINOS at $z =1,200$ cm.  In this system the
neutrino beam central axis is in the $y-z$ plane and points downward at $3.34^\circ$.

The core of the \minerva design is the active tracking region, composed purely of scintillator, 
which serves as the primary fiducial
volume where precise tracking, low density of material, and fine sampling ensure that some of the
most difficult measurements can be performed.  These measurements include 
particle identification using energy loss per unit length ($dE/dx$),
and reconstruction of the neutrino interaction vertex 
in the presence of several final state charged particles. The upstream part of the detector
contains solid targets of carbon, iron, and lead interleaved with the scintillator planes.
Because these targets are relatively thin, the ability to precisely reconstruct the
location of the interaction vertex is
crucial for studies of the $A$ dependence.

Electromagnetic calorimetry is accomplished using a 0.2~\cm thick by 15 cm wide lead ``collar'' 
(orange ring in Fig.~\ref{fig:minerva-schematic}) between 
each scintillator plane in the central tracking region.  Additionally there are lead plates, each
0.2~\cm thick and covering the full transverse span of the inner detector, which
are deployed between each scintillator plane within the 10 modules immediately downstream of the tracking region.
For hadronic calorimetry the outer frames of all modules are instrumented with strips of scintillator 
interleaved into the steel.  
Further, in the most downstream 20 modules of the detector, the inner detector scintillator planes 
alternate with 2.54~\cm thick steel plates.  Thus the combination of the outer frame detectors 
and the downstream calorimeter section provides containment of hadrons initiated by interactions in the tracking region.

The \minerva experiment uses several different simulation codes to model the detector and its performance.  
The neutrino interactions are modeled by the GENIE v2.6.2 event generator \cite{genie}.  
The final-state particle interactions in the detector itself are modeled by a GEANT4 
version 9.4.p02~\cite{geant} simulation.  The different detector components such as the 
electronics and scintillator and absorbers are also modeled using GEANT4.  
Both the GENIE and GEANT4 parts of the detector simulation include a detailed model of the 
nuclear makeup of the detector, described in Sec.~\ref{sec:tracking_modules}
 and Sec.~\ref{sec:nuclear_targets}.  The simulation also takes into account the variations 
over time and position of the detector components, and includes accidental detector activity.

\makeatletter{}
\subsection{The \minos Near Detector}
\label{sec:minos}

The \minerva\  detector is situated 2.1~meters upstream of the \minos\  near detector
in the NuMI beamline. The \minos\ magnetic spectrometer is used to momentum analyze muons which exit the \minerva\ 
detector volume in the forward direction. The detector technology and readout are described 
in detail in Ref.~\cite{refminos}.  The segmentation and layout of the MINOS near
detector are described in this section.  
Performance aspects relevant to its use for \minerva
are discussed  in Sec. \ref{sec:tracking}.

\begin{figure*}[t]
\includegraphics[width=0.39\textwidth]{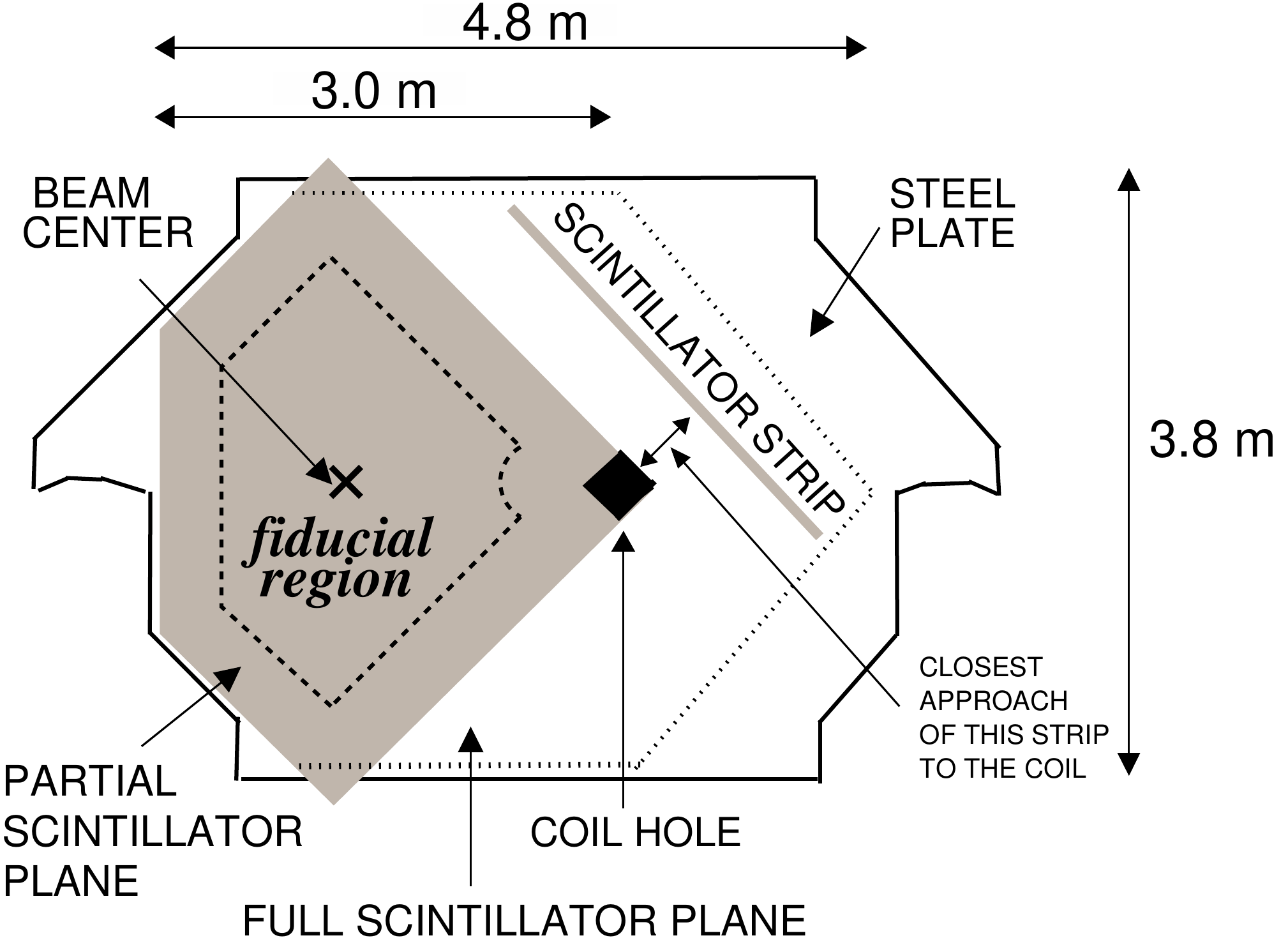}
\hspace{0.01\textwidth}
\includegraphics[trim=0mm -10mm 0mm 0mm, width=0.59\textwidth]{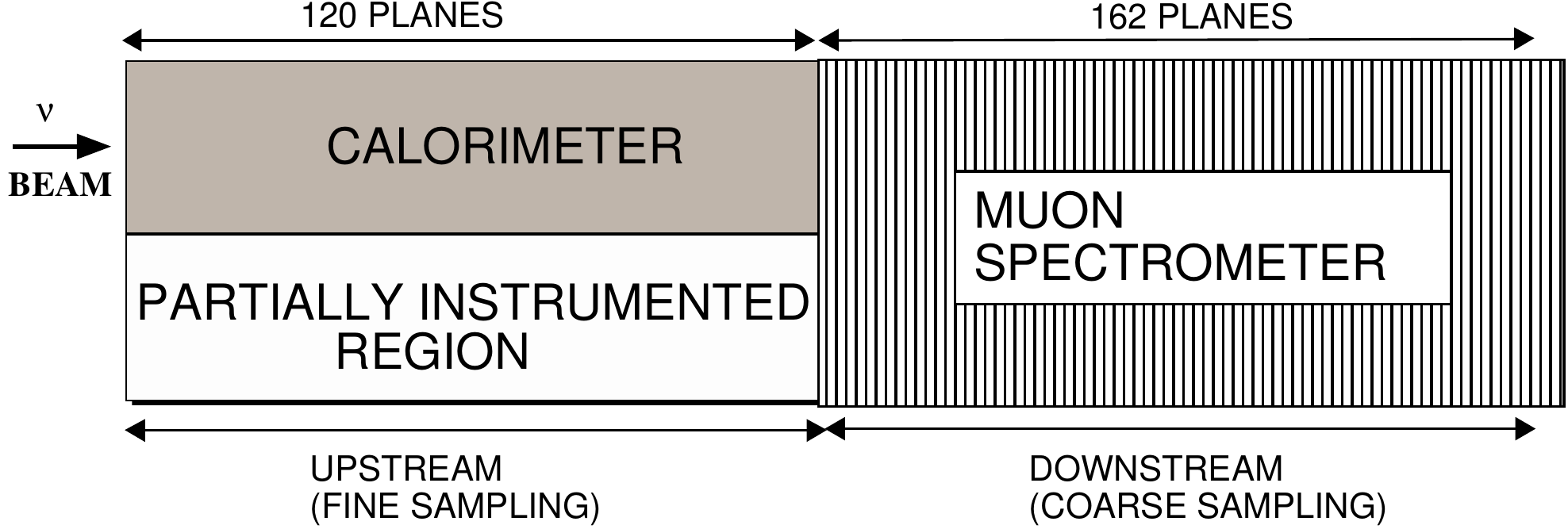}\caption{ \label{long_view}\label{fig:detector} Schematic view of the \minos\ Near Detector.  
Left: transverse view of a Near Detector plane.   The shaded area shows a 
partially instrumented active scintillator plane and the dashed line within 
shows the boundary of the fiducial region. The dotted line shows the outline 
of a fully instrumented scintillator plane.
Right: top view showing the calorimeter and muon spectrometer. 
The drawing is not to scale.  
 Figure taken from Ref.~\cite{minosview}}
\end{figure*}

MINOS, shown in Fig.~\ref{long_view}, is a tracking calorimeter composed of planes
of magnetized iron and plastic scintillator with a total mass of 1~kTon. It has a toroidal magnetic field 
of average strength of 1.3~T, which is produced by a current-carrying coil passing through the entire
length of the detector. The direction of curvature of a charged
particle in the field allows the sign of the 
sign of the track charge to be determined. In the normal operational mode, the field is set to focus the same charge as that selected in the NuMI secondary beam focusing system.  

MINOS consists of 282 steel
plates, each 2.54~cm thick, of which 152 are instrumented with 1~cm thick
scintillator planes. The scintillator planes are made of 4.1~cm wide
strips oriented $\pm$45$^\circ$ with respect to the vertical and
alternating $\pm$90$^\circ$ in successive planes. 
In the upstream region (``calorimeter region''), comprising the first 120 planes, 
a partially-instrumented (or fully-instrumented every fifth plane) 
scintillator plane follows each steel plane. 
In the downstream 162 planes (``muon spectrometer''), there are no partially instrumented planes and
every fifth plane has full scintillator coverage.  In the spectrometer region, the signals from four different pixels in a spectrometer plane are electrically summed prior to digitization.  

\makeatletter{}
\subsection{The \numi Neutrino Beam at Fermilab}
\label{sec:numi}
\minerva\ uses the Fermilab NuMI beamline to produce a high intensity 
beam of muon neutrinos and anti-neutrinos \cite{refnumi}.  
A beam of 120 GeV protons from the Main Injector strikes a 
graphite target over an 8.1 or 9.72~$\mu$s spill every 2.2 seconds.  The two different spill durations correspond to different modes of operation for the accelerator complex.  
The secondary pions and kaons produced by the incident protons are 
then focused by a system of two magnetic horns which direct the 
mesons into a 675~m long decay pipe where most of them decay.  A total of 240~m of rock downstream of the 
decay pipe range out the tertiary muons that are created in the beamline 
concurrently with the neutrinos.  

The focusing horns can be pulsed in either polarity.  When the horns are 
focusing positively charged mesons (forward horn current or FHC) the resulting 
beam is primarily neutrinos, and when the horn is in the reverse horn current 
(RHC) the peak of the beam is primarily anti-neutrinos.  \minerva\ collected 
data with $4.0 \times 10^{20}$
 ($1.7 \times 10^{20}$) protons on target in FHC (RHC) 
over the period of time described in this article (March 2010 through April 2012).  

The NuMI beamline has considerable flexibility and can run with the target in different 
positions relative to the focusing horns. For most of the run the graphite target was located as close to the horns as 
possible to create a beam whose energy of the peak in neutrino flux was approximately 3~GeV,  
a configuration known as the low energy (LE) beam. Of the total exposure roughly 10\% of the protons on target were collected in configurations with the target moved farther upstream of the first horn, which results in a higher 
peak-neutrino-energy flux.  These special runs were taken to better understand the neutrino flux, but the detector calibration and reconstruction procedures remained the same as those described in the remainder of this article.  Future runs will use a horn separation that gives a higher peak neutrino energy.  

As discussed in Sec. \ref{sec:meu}, an important calibration source produced by this beam comes 
from muons resulting from neutrino charged-current interactions that occur in the rock located upstream of the \minerva\ detector.  On average once every two beam pulses a rock muon traverses the entire central tracking region of the \minerva\ detector.

\makeatletter{}
\section{Module Assemblies and Nuclear Targets} 
\label{sec:ModuleAssemblies}

The \minerva detector is comprised of 120 modules suspended vertically and 
stacked along the beam direction. There are four basic types of modules: tracking modules, 
electromagnetic calorimeter modules, hadronic calorimeter modules, and passive nuclear targets.  
The four types of modules are distinguished by their makeup in the ID
region;  the outer module frames are identical except for their thickness. 
The OD frames provide structural support, alignment, and outer hadronic calorimetry.  
Each hexagonal steel frame is 56 cm wide and instrumented with eight strips of 
scintillator of varying lengths to span each wedge in the frame.  The OD frames
are 3.49~cm thick, except in the HCAL, where they are 3.81~cm thick due to the
steel plates being thicker than scintillator planes.  Successive modules are connected with bolts that ensure a 2.5~mm air gap between pairs of scintillator planes.  The air gap between the steel frames is 8.7~mm.  

The subsections below provide detailed descriptions of the construction and material 
composition of each module type.  The material composition of the tracking planes, 
in particular, has relevance for the 
interpretation of cross sections measured by the \minerva detector.  

\makeatletter{}
\subsection{Tracking Modules}
\label{sec:tracking_modules}

Tracking modules consist of two scintillator planes each composed of
triangular scintillator strips, described in Section~\ref{sec:scintillator}.  Each plane
consists of 127 strips glued together with 3M-DP190 translucent epoxy.
Sheets of Lexan cover the planes and are attached with 3M-DP190 gray 
epoxy to make them light tight and to add rigidity.
Black PVC electrical tape is used to seal joints in the Lexan and patch any light leaks. Optical epoxy (Epon Resin 815C and Epicure 3234) provides the coupling between the scintillator and WLS fibers. 

A plane can have one of three different orientations, referred to as X-planes, 
U-planes or V-planes according to the coordinate in the \minerva system in 
which each plane measures particle hit positions.   
X-planes have scintillator strips aligned vertically, hence hits in this view give position 
information in the horizontal or $x$-direction.  The U- and V-planes are rotated 60 degrees 
clockwise and counterclockwise from the X-planes in the $x$-$y$ plane,
 respectively. 
Three different views are 
used in order to avoid ambiguities with reconstructed hit associations that can occur when
multiple tracks traverse two orthogonal planes.  
Each tracking and electromagnetic calorimeter module has one X-plane, and either a U- or V-plane,
with modules alternating between a UX or VX structure with the X-planes always located
downstream of the U- or V-planes.  
The nuclear target region contains 22 tracking modules, and the central tracking region contains 
62 tracking modules.

The tracking modules are designed to perform electromagnetic 
calorimetry using a 0.2 cm thick lead collar 
that starts at roughly 90~cm from the module center and extends to the outer frame.
The collar forms a hexagonal ring whose purpose is to reduce the leakage of electromagnetic 
showers that originate in the central detector.  

The chemical composition and areal density 
(mass per unit surface area) of the planes is determined by
combining measured densities (pure scintillator and coated strips), 
assayed compositions (coated strips and epoxies), and data sheet
values (tape and Lexan).  The estimated areal densities of the epoxy
and tape are based on their usage in plane construction.
The densities and composition of the components are listed in
Table~\ref{tab:tgt-comp}.  The elemental compositions of the strips
and assembled tracker scintillator plane are given in Table~\ref{tab:plane-comp}.
The chemical composition of the components is well known.  There is
some uncertainty in the composition of the coated strips due to the uncertainty
in the coating thickness, which is estimated to have a relative uncertainty
of about 10\%.  This affects most strongly the fraction of the strips which is
scintillator.  The estimated areal density for the scintillator 
plane is 1.65~$\pm$~0.03~g/cm$^2$. 
The estimated areal density of an assembled plane is 2.02~$\pm$~0.03~g/cm$^2$, as described in Sec.~\ref{sec:scintillator}.

\begin{table}
\begin{centering}
  \begin{tabular}{|l|c|c|c|c|c|c|c|c|c|}
    \hline 
    Material & Density (g/cm$^3$) & H & C & N & O & Al & Si & Cl & Ti  \\ \hline
     Scintillator & 1.043 $\pm$ 0.002 & 7.6\% & 92.2\% & 0.06\% & 0.07\% & -  & -  & - & - \\ \hline
     Coating & 1.52  & 6.5\% & 78.5\% & - & 6.0\% & -  & -  & - & 9.0\% \\ \hline
     Lexan & 1.2 & 6.7\% & 66.7\% & -  & 26.7\%  & - & -  & -  & - \\ \hline
     PVC tape & 1.2 & 4.8\% & 38.7\% & -  & -  & - & - & 56.5\% & -  \\ \hline
     DP190 transl. & 1.32  & 10.0\% & 69.0\% & 2.6\% & 17.0\% & -  & -  & 0.5\% & -  \\ \hline
     DP190 gray & 1.70 & 5.0\% & 47.0\% & 1.7\% & 27.0\% & 6.0\% & 6.0\% & 0.05\% & -  \\ \hline
\end{tabular} 
  \caption{Density and elemental composition by mass percentage for the
various materials in the scintillator planes.}
  \label{tab:tgt-comp}
\end{centering}
\end{table}

 \begin{table}
\begin{centering}
  \begin{tabular}{|l|c|c|c|c|c|c|c|}
    \hline 
     Component & H & C & O & Al & Si & Cl & Ti   \\
    \hline
    Strip &  7.59\% & 91.9\% & 0.51\%& - & - & - & 0.77\%  \\ \hline
    Plane &  7.42\% & 87.6\% & 3.18\% & 0.26\% & 0.27\% & 0.55\% & 0.69\%  \\ \hline

 \end{tabular} 
  \caption{Elemental composition of scintillator strips and constructed
planes, by mass percentage.}
  \label{tab:plane-comp}
\end{centering}
\end{table}

\makeatletter{}
\subsection{Electromagnetic and Hadronic Calorimeter Modules}
\label{sec:calorimeter_modules}

An ECAL module is very similar to a
central tracking module. It differs in that it has a 0.2 cm thick sheet of 
lead covering that the entire scintillator plane instead of
a 0.2~cm thick lead collar covering only the outer edge of the scintillator region. 
A transition module is placed between the last central tracking module and first ECAL
module. This module contains a 0.2 cm thick lead sheet on the downstream end of
the last plane in the module so that each plane of the ECAL has a lead absorber
upstream of it. The fine granularity of the ECAL ensures excellent photon and 
electron energy resolution and provides directional  measurement for these
particles.  
There are 10 modules in the ECAL region of the detector 
(Fig. \ref{fig:minerva-schematic}).

The lead sheets used for the side electromagnetic calorimetry were measured using an 
ultrasonic device to determine the variation in thickness along the length of the sheet.  The 
thickness along each piece vary at the 5\% level, and the average thickness of the different 
pieces vary at the 3.5\% level.  The thicknesses of the lead sheets used in the 
downstream electromagnetic calorimetry also vary at this level.    

The HCAL consists of 20 modules that are similar 
to the tracking  modules;  however, instead of two planes of scintillator in each module, 
there is only one plane of scintillator and one 2.54~\cm thick 
hexagonal steel plane in the inner detector region.   The scintillator planes 
located in the HCAL have a repeating pattern of XVXU.

\makeatletter{}
\subsection{Solid Nuclear Target Modules}
\label{sec:nuclear_targets}

\begin{figure}[ht]
\begin{center}
\includegraphics[width=0.95\textwidth]{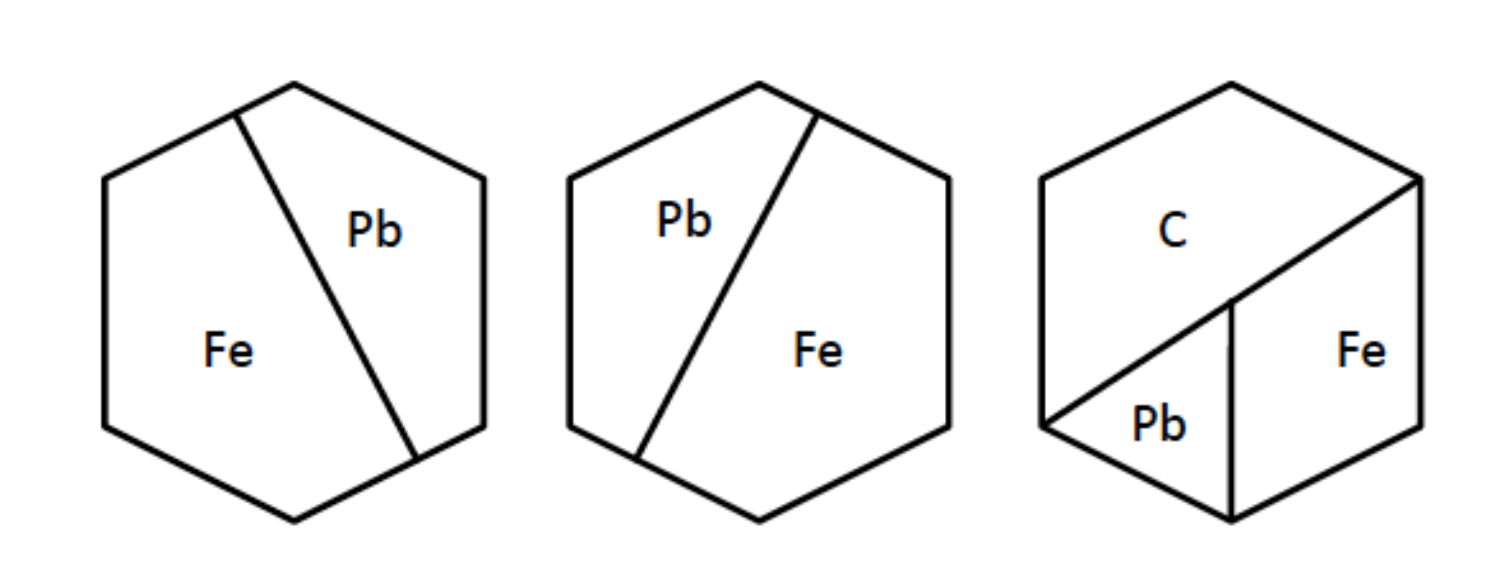}
\caption{Orientation of the nuclear targets looking downstream.  Targets 1 and 5 have
the leftmost orientation, target 2 the middle orientation, and target 3 the rightmost orientation.}
\label{fig:tgt-orient}
\end{center}
\end{figure}

\begin{figure}[ht]
\begin{center}
\includegraphics[width=0.95\textwidth]{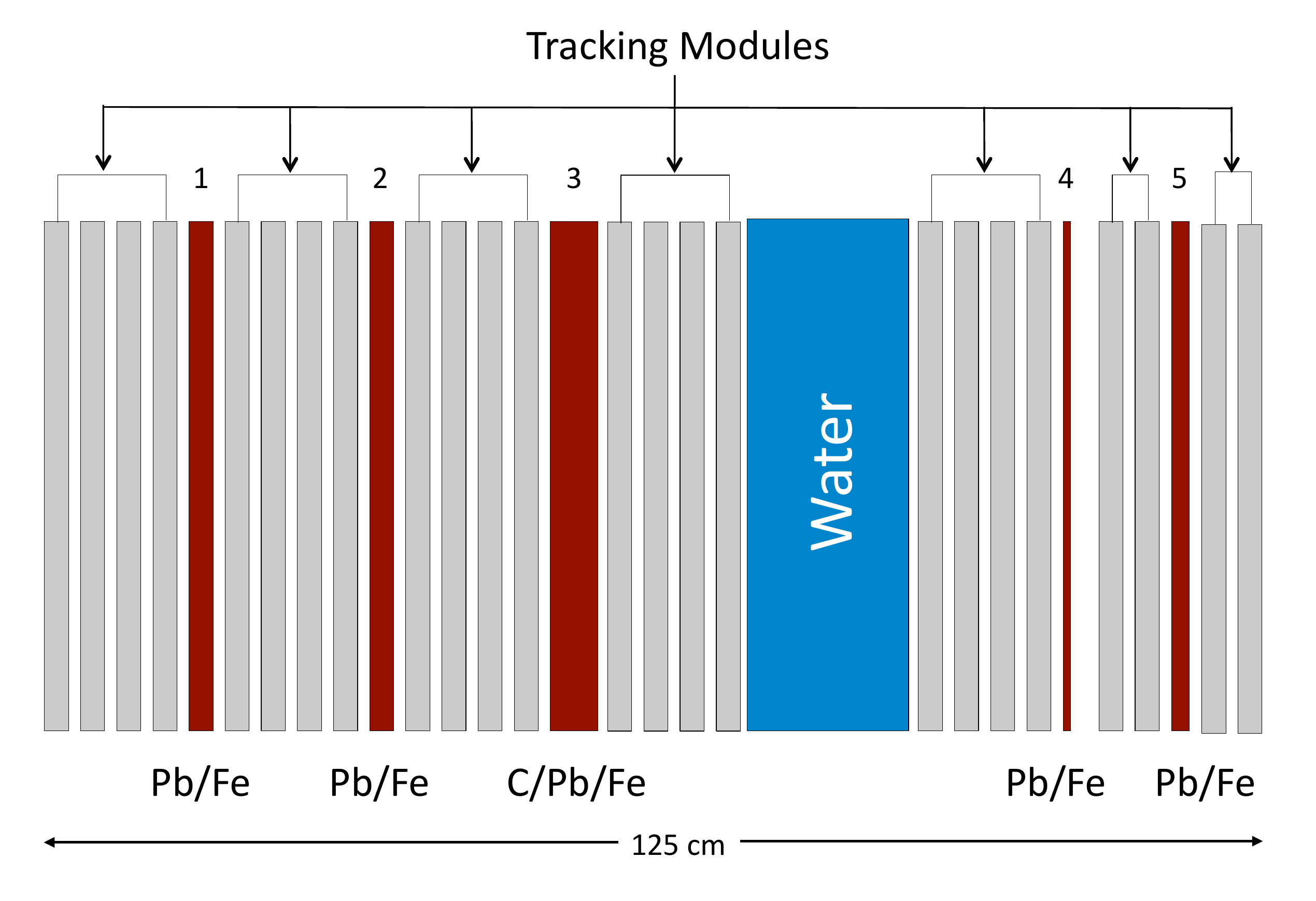}
\caption{Orientation of the nuclear target region along the beamline axis. 
 The thinner targets are located downstream and the thicker targets are located upstream.}
\label{fig:nuke_region}
\end{center}
\end{figure}

 In order to study neutrino interactions on different nuclei,
the most upstream part of the detector includes five layers of passive 
targets, the ``nuclear targets",
 separated by four tracking modules each.
The four modules (eight planes of scintillator) between one target and the next ensure good vertex 
position resolution for events originating in the nuclear targets.  
Each solid nuclear target is mounted in the same instrumented hexagonal steel frame
as the scintillator 
planes for ease of detector construction and for event containment.  
The five targets are configured 
such that the thicker targets are most upstream, and the thinner targets are downstream.  This 
optimizes reconstruction of events occurring upstream.  The thinner targets are included in order 
to study specific reactions that contain low-momentum final-state particles.  Except for the fourth of the five targets,
all contain mixed materials with different orientations in order to minimize the effect of 
acceptance differences for different regions of the detector.  Target 4 is pure lead and 
aids in upstream electromagnetic calorimetry and serves as the thinnest lead target.
Targets 1, 2, and 5 are mixed steel and lead.  
The steel plate section is larger than the lead plate section,
with the dividing line 20.5 cm from the plane center.  Target 3 is composed of graphite, iron and steel.
The graphite covers half the area of the hexagon, the steel one-third, and the lead one-sixth.
The orientation of the planes, as viewed looking downstream, is shown in Fig.~\ref{fig:tgt-orient}.  
The orientation of the planes along the axis of the beamline is shown in Fig.~\ref{fig:nuke_region}.  
The  composition by element of the targets is given in Table~\ref{tab:nucl-tgt-comp}.

\begin{table}
\begin{center}	
  \begin{tabular}{|l|c|c|c|c|c|c|c|}
    \hline 
    Material & Density (g/cm$^3$) & 
	       C & Si & Mn & Fe & Cu & Pb \\ \hline
     Steel & 7.83 $\pm$ 0.03 & 0.13\% & 0.2\% & 1.0\% & 98.7\% &-  & - \\ \hline
     Lead & 11.29 $\pm$ 0.03 & -   & -  &  -  & -  & 0.05\% & 99.95\% \\ \hline
     Graphite & 1.74 $\pm$ 0.01 & $>$99.5\% & -  & -  & -  & -  & - \\ \hline
\end{tabular} 
  \caption{Density and element composition by mass percentage for the
nuclear targets.}
  \label{tab:nucl-tgt-comp}
  \end{center}	
\end{table}

The residual radioactivity of the lead was measured by  
the Fermilab Radionuclide Analysis Facility to be below the maximum 
allowable radioactivity of 0.15 $\gamma$/sec/kg with energy above 0.5 MeV.  
The fiducial area for the mixed targets is bounded by a hexagon
with an 85~\cm apothem, and a 2.5~\cm cut on each side of the boundary
between materials.  The $z$-location of the center of each target 
and the fiducial mass of each material for each target is
given in Table~\ref{tab:fid-mass-nucl}.  The estimated uncertainty on the
fiducial masses due to density and thickness variations is less than 1\%.

\begin{table}
\begin{center}
  \begin{tabular}{|l|c|c|c|c|c|c|c|}
    \hline 
    Target & $z$-location  & Thickness & Fiducial Area & Fiducial Mass  & Total Mass   \\
      & (cm) & (cm) & (cm$^2$) &  (kg) &  (kg)  \\
    \hline
    1-Fe & 452.5 & 2.567 $\pm$ 0.006 & 15999 & 322  & 492 \\ \hline
    1-Pb & 452.5 & 2.578 $\pm$  0.012 & 9029  & 263 & 437 \\ \hline
    2-Fe & 470.2 & 2.563 $\pm$ 0.006 & 15999 & 321  & 492 \\ \hline
    2-Pb & 470.2 & 2.581 $\pm$  0.016 & 9029  & 263 & 437  \\ \hline
    3-Fe & 492.3 & 2.573 $\pm$ 0.004 & 7858  & 158 & 238 \\ \hline
    3-Pb & 492.3 & 2.563 $\pm$  0.004 & 3694  & 107 & 170 \\ \hline
    3-C & 492.3 & 7.620 $\pm$ 0.005 & 12027 &  160 & 258 \\ \hline
   Water & 528.4 & 17-24 &25028 &  452  &  627 \\ \hline 
    4-Pb & 564.5 & 0.795 $\pm$  0.005 & 25028  & 225 & 340 \\ \hline
    5-Fe & 577.8 & 1.289 $\pm$ 0.006 & 15999 & 162  & 227 \\ \hline
    5-Pb & 577.8 & 1.317 $\pm$  0.007 & 9029  & 134 & 204 \\ \hline
 \end{tabular} 
  \caption{Nuclear target locations, thickness and fiducial mass.  The total mass is
for the entire plate of target material. The location is in the MINERvA coordinate system, which is defined in the text.}
  \label{tab:fid-mass-nucl}
  \end{center}
\end{table}

\makeatletter{}
\subsection{Water Target}
\label{sec:water_target}

\begin{figure}[ht]
\begin{center}
\includegraphics[width=0.7\textwidth]{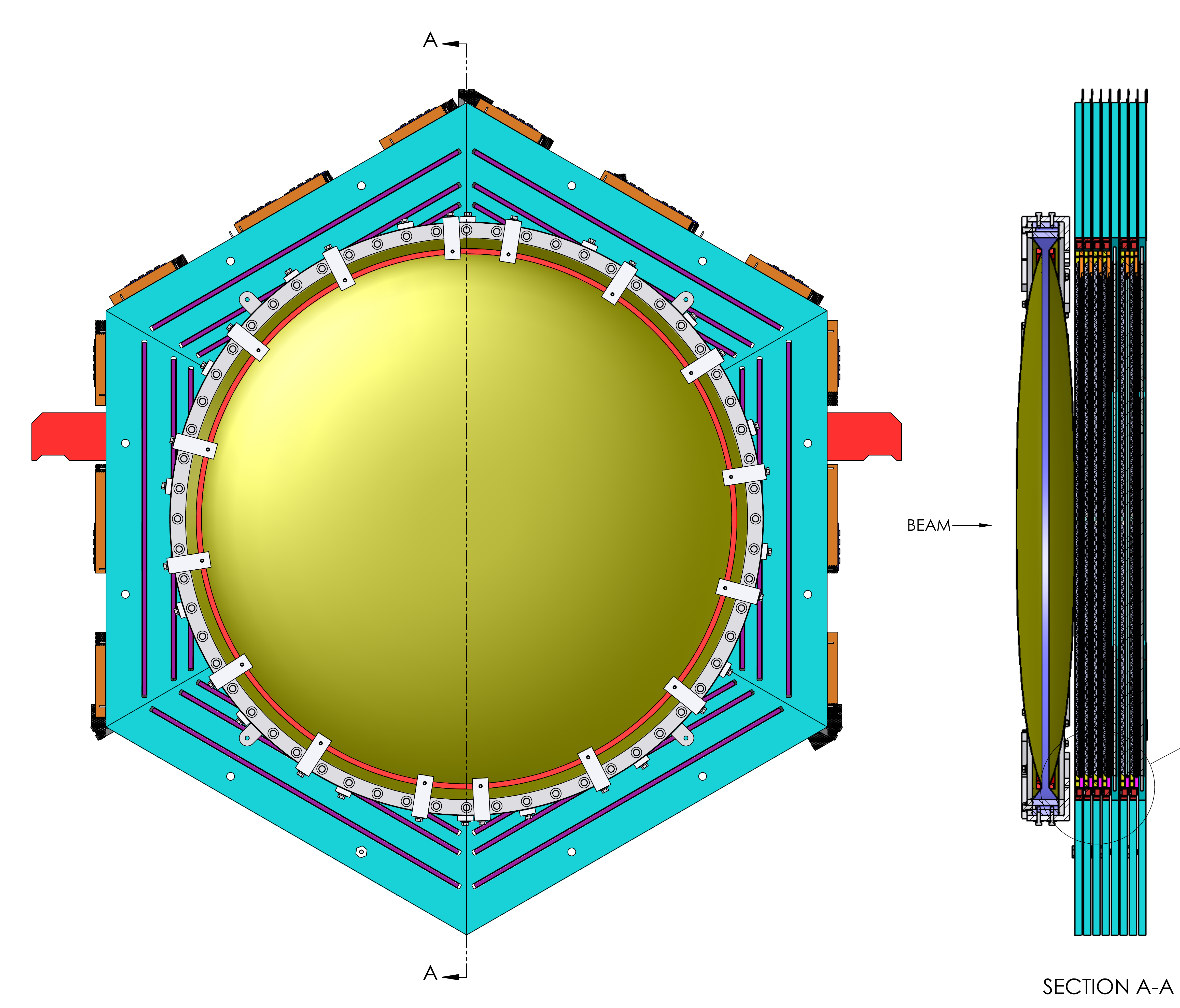}
\caption{Schematic drawing of the \minerva water target, showing a front view and
side view of the center section.}
\label{fig:water}
\end{center}
\end{figure}

A water target is positioned between solid targets 3 and 4, with a mean position of 530.8~\cm.
It consists of a circular steel frame with a diameter slightly larger than the \minerva inner 
detector size, and Kevlar\textregistered\  (polymerized  C$_{14}$H$_{10}$N$_2$O$_2$)
sheets stretched across the frame as shown in Fig. \ref{fig:water}. 
The shape of the water target is not as well known as that of the solid
targets.  When the target
is filled the lower part expands more than the upper part, and it is not possible
to access the entire target in order to make precise measurements.  The
shape of the target is estimated via a finite element analysis and
compared with the actual volume, which is determined by measuring the volume
of water when the target is emptied.  The thickness varies from about 17~\cm
at the edge of the fiducial region to 24~\cm at the thickest part.  The estimated
mass within an 85~\cm apothem hexagon is 452~\kg with an uncertainty
of about 3\%.  The Kevlar walls add  0.1 g/cm$^2$, for a total mass of 2.5 kg.
  The water
target chemical composition by mass is 88.5\% O and 11.1\% H with negligible amounts of C and N.

\makeatletter{}
\subsection{Liquid Helium Target}
\label{sec:helium_tgt}

The \minerva\  cryogenic helium target is located immediately upstream of the active detector 
and was filled with liquid helium during the latter parts of the run.  Its design reflects the following considerations:  the largest 
volume possible for increased statistics; a minimum of material to be
traversed in reaching the tracking detectors by particles originating from interactions in
the helium; the largest acceptance possible for neutrino events scattering off helium; and an
acceptable level of backgrounds produced from the containment vessels and support structures.
The cryogenic target consists of an aluminum cryostat capable of holding approximately 2300~l 
of cryogen.  The cryostat consists of an inner 
vessel containing the cryogen, which is thermally isolated from an outer vacuum vessel, and which hangs from a 
set of four Kevlar ropes.  The vacuum region contains layers of thin aluminum baffles for minimizing the 
radiative heat transfer from the inside of the outer vessel to the inner vessel.  The inner vessel itself 
consists of a cylinder with an inner diameter of 152 cm, length of 100 cm, and a wall thickness of 0.635 cm.  
The ends of the inner vessel are capped with 0.635 cm thick 
American Society of Mechanical Engineers (ASME) flanged and dished heads, which are welded to 
the cylinder.  The outer vessel cylinder has an inner diameter of 183 cm and a wall thickness of 0.952 cm.  In 
order to minimize energy loss and rescattering of final state particles entering \minerva, the amount of 
aluminum on the downstream end of the cryostat was minimized in the design.  Consequently, the 
flanged and dished head on the upstream end has a thickness of 0.635 cm, while the hemispherical head on 
the downstream end has a nominal thickness of 0.160 cm and a radius of curvature of 107 cm with the center of 
curvature downstream.

The target is instrumented with the following: a set of four load cells (one on each support leg) for 
measuring the total weight of the cryostat;  a cryogen depth gauge in the inner vessel fill tube for 
determining the volume of cryogen; a cryogen temperature sensor; and  a pressure sensor for measuring 
the pressure of the vapor in thermodynamic equilibrium with the cryogen.  Since the equation of state for 
helium is known, the temperature and pressure sensors allow an accurate 
determination of cryogen density.  Combining the density and depth gauge measurements 
provides a determination of the cryogen total mass, which is complementary to the 
measurement from the load cells.  In addition, the temperature is regulated to within 25~mK via a feedback 
loop which controls a heater.  The regulation of the helium density is thus estimated to better than 0.5\%.

\makeatletter{}
\section{The Optical System}
\label{sec:OpticalSystem}

Light signals from the over 32,000 scintillator strips in \minerva must be 
converted to electrical pulses which carry accurate timing information 
and have amplitudes proportional to the energies deposited.  This section 
describes the \minerva optical system, which begins with the light generated 
when charged particles pass through the individual scintillator strips.   
Scintillation light is collected in wavelength shifting fibers at the 
center of each bar and transmitted through clear optical cables to 64-anode 
photo-multiplier tubes (PMTs) mounted above the detector.  Each PMT is housed in a 
metal cylindrical ``box'' which contains an additional set of fibers,
called an optical decoder unit (ODU), to guide the 
light to the front face of the  PMT.  Each box also houses the associated 
PMT base.  The front-end electronics board that services the PMT is mounted to the
exterior of the box on one endplate.

\makeatletter{}
\subsection{Scintillator}
\label{sec:scintillator}

\minerva\ uses extruded plastic scintillator technology for the tracking
detectors of the ID and embedded scintillators in the OD.
The blue-emitting extruded plastic scintillator strips are read out with a green
wavelength shifting (WLS) fiber placed in the center of the strips.
The extruded scintillator strips are made from polystyrene pellets (Dow Styron 
663 W) doped with 1\% (by weight) 2,5-diphenyloxazole (PPO) and 0.03\% (by weight) 
1,4-bis(5-phenyloxazol-2-yl) benzene (POPOP).
This scintillator composition was previously utilized in the MINOS scintillator strips \cite{refminos}.  
PPO and POPOP are used for their spectroscopic properties.
The strips are co-extruded with a white reflective coating based on 15\% TiO$_2$ (by weight) in polystyrene.

\begin{figure}[htbp]
  \begin{center}
  \includegraphics[width=1.0\textwidth]{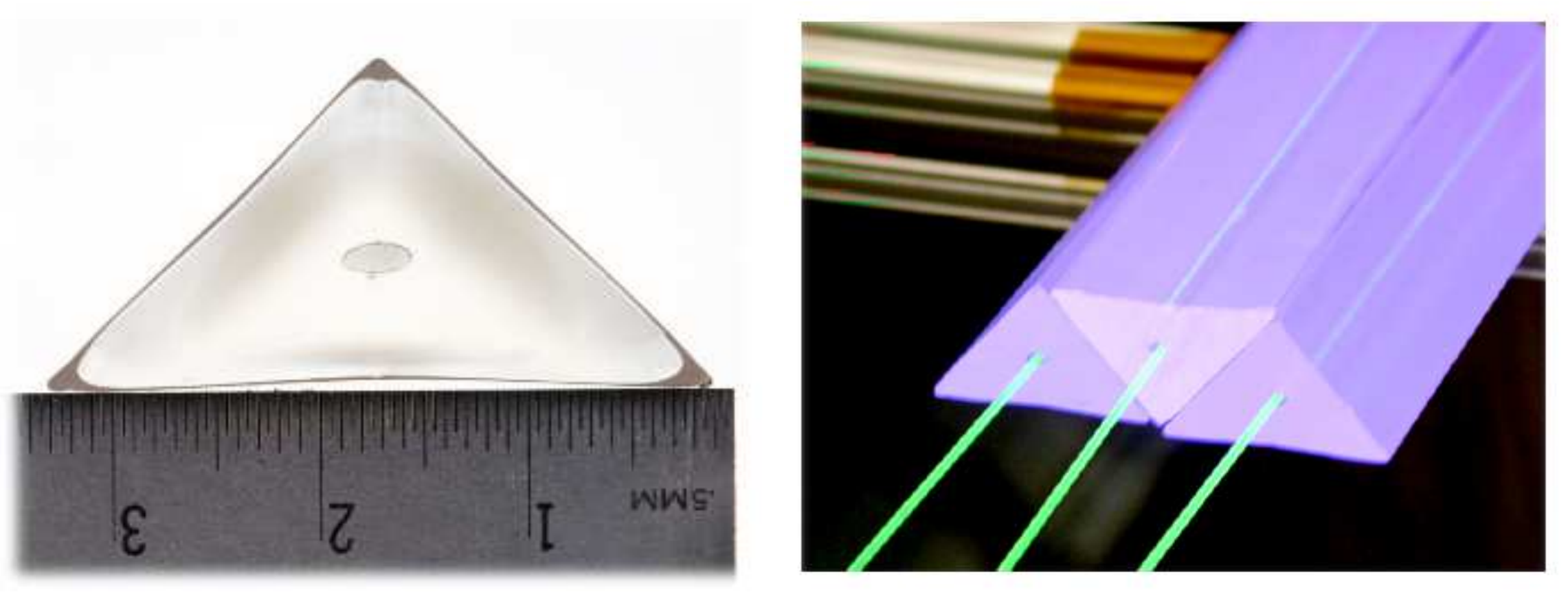}
  \caption{The \minerva\ scintillator strips are triangular in cross section (left) and 
range from 122~\cm to 245~\cm in length.  Planes are built by stacking the triangular 
strips as shown in the right figure.  This configuration ensures that any charged 
particle traversing the plane creates a scintillation signal in a minimum of two strips.}
  \label{fig:strips}
  \end{center} 
\end{figure}
 
The ID scintillator strips are triangular in cross-section with a height of  
17~$\pm$~0.5~\mm and width of 33~$\pm$~0.5~\mm (Fig. \ref{fig:strips}).  Each ID strip has a 2.6~$\pm$~0.2~\mm diameter hole centered at 
8.5~$\pm$~0.25~\mm above the widest part of the triangle.  Both ends of the 
scintillator strips are painted with white EJ-510 TiO$_2$ Eljen paint \cite{eljen}.
The OD scintillator strips have two different rectangular cross sections.
For 90\% of the detector the OD scintillator strips have a base of 19~$\pm$~0.5~\mm and a height of 16.6~$\pm$~0.5~\mm.
For the hadron calorimeter region the OD steel is thicker, 
hence the OD scintillator strips are also thicker to improve hermeticity.
The OD scintillator strips have a 3.5~$\pm~$0.2~\mm diameter hole.
Both ID and OD scintillator strips were extruded in the Extrusion Line Facility
at Fermilab supported by both Fermilab and the Northern Illinois Center for Accelerator and Detector Development.
The extrusion equipment allows for a continuous process from
the polystyrene pellets received in boxes to the final product, with little manual handling.
Only the dopants and the TiO$_2$ pellet mixture have to be periodically added 
to the gravimetric feeder and co-extruder hopper, respectively, which dispenses the extrusions.
The extrusion line operates under a nitrogen purge from the dryer which uses high pressure nitrogen to dry the
polystyrene pellets in the die.

\makeatletter{}\subsection{Wavelength Shifting Fibers} 
\label{sec:wls_fibers}

\label{sect:wlsprep}

The scintillator strips are read out by 1.2~mm diameter, 
175~ppm Y-11 doped, S-35, multiclad wavelength shifting (WLS)
fibers produced by the Kuraray corporation. 
\minerva\ reads out only one end of its WLS fibers.
To maximize light collection, the unread end of each fiber is mirrored.
This mirroring procedure consists of 3 steps: polishing the end
to be mirrored using a technique called ``ice-polishing'' \cite{icepolish}, 
then depositing a 2500 \AA\  thick reflective coating of 99.999\% pure aluminum by sputtering, 
and finally applying a protective layer of Red Spot UV Epoxy to the mirrors.

\begin{figure}[htbp]
\begin{center}
\includegraphics[width=0.95\textwidth]{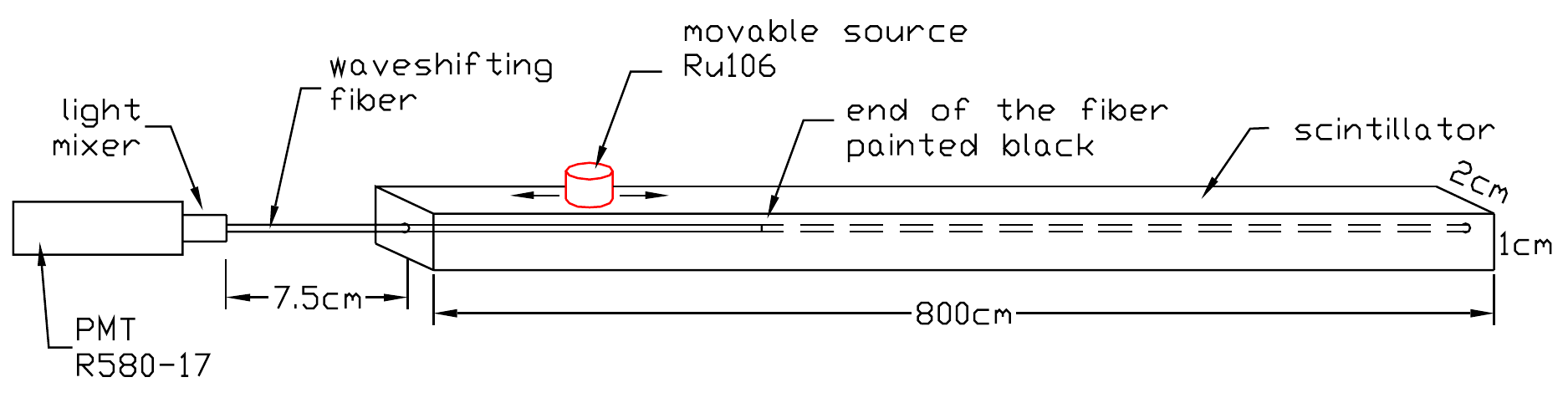}
\caption[MINOS Scanner]
{A schematic diagram of the scanner used to measure the attenuation length of WLS fibers.}
\label{fig:minos_scanner}
\end{center}
\end{figure}
 
\begin{figure}[htbp]
\begin{center}
\includegraphics[width=0.45\textwidth]{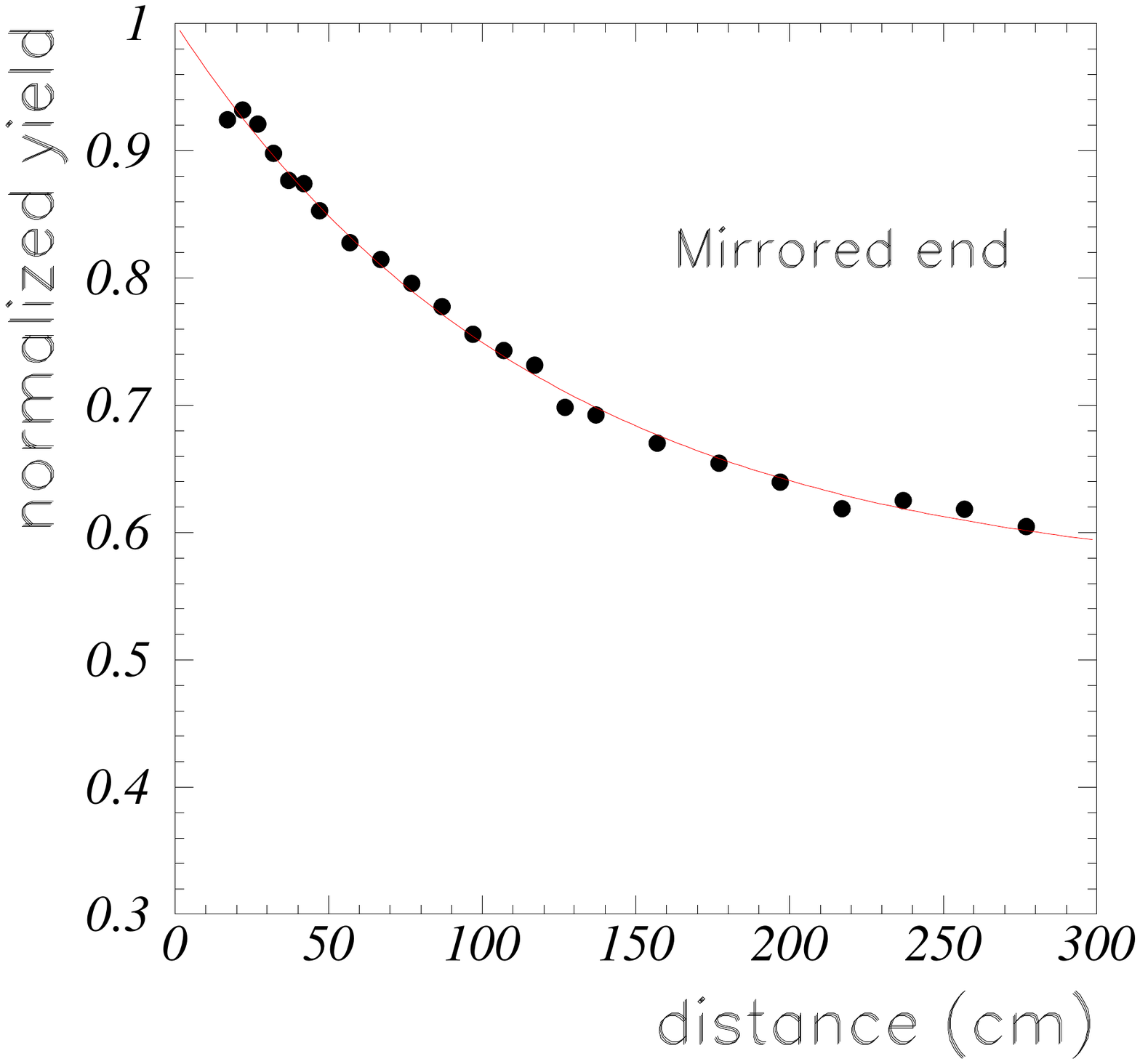}
\includegraphics[width=0.45\textwidth]{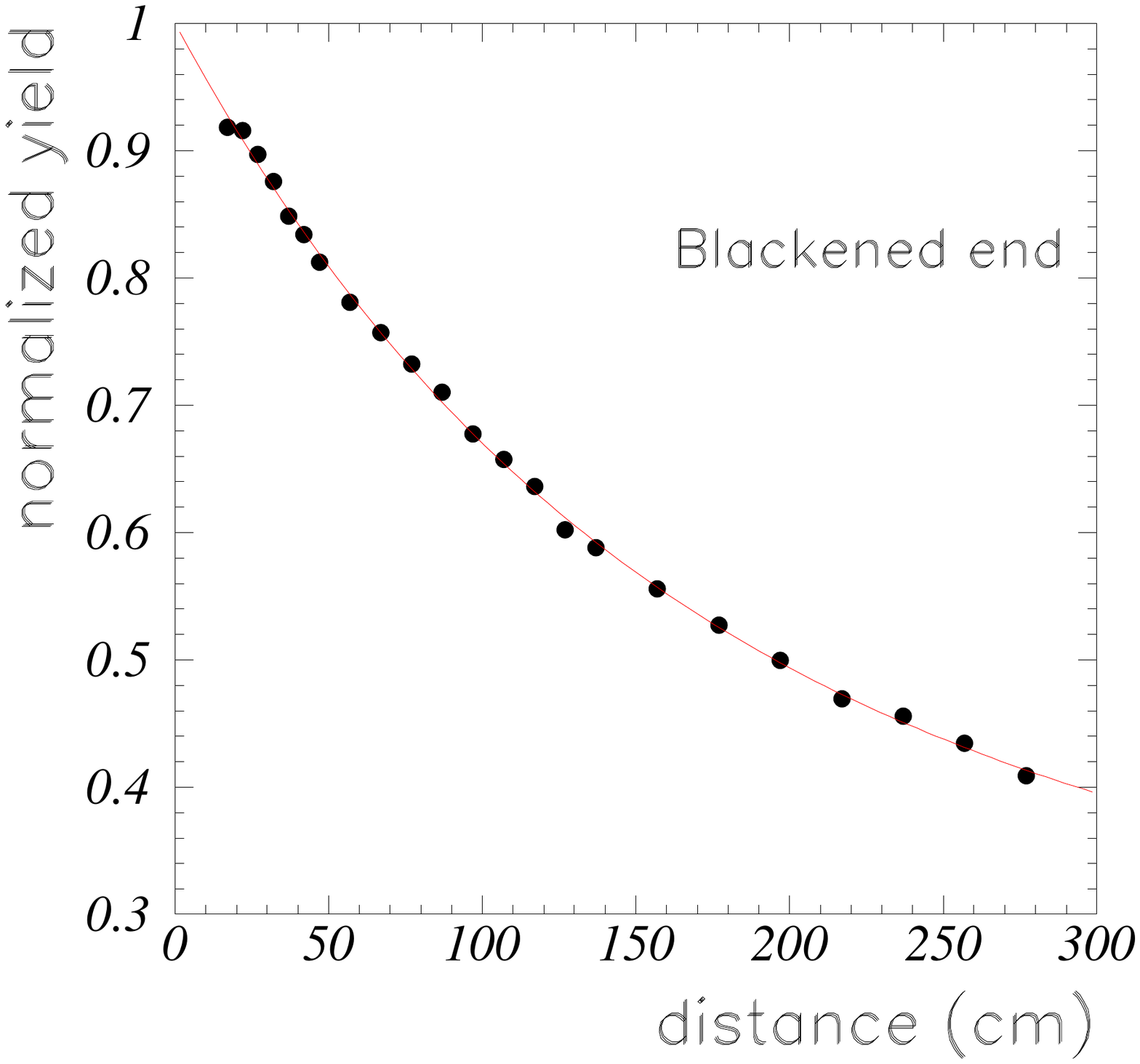}
\caption[Measurement of a fiber using the MINOS Scanner]
{Two measurements of the same fiber using the WLS scanner. The 
fiber is first measured with the non-readout end mirrored (left). Next, the mirror is 
cut off, the end painted black and the fiber is remeasured (right). The
fit is a double exponential function, $p(1)e^{-x/p(2)}+p(3)e^{-x/p(4)}$, 
where $x$ is the distance to the phototube and $p(i)$ are the fit parameters.
}
\label{fig:fiber_fit}
\end{center}
\end{figure}

After this process, a destructive measurement of
the mirror reflectivity is made with five 2.1 m mirrored fibers from each sputtering session.
 The unmirrored ends are inserted 
into a 10 cm by 10 cm scintillator tile and illuminated with   
a Cs-137 source collimated by a lead cone which sits on top 
of the tile.
The mirrored end is then cut off at 45$^\circ$ and
painted with black paint, Model Master Flatblack 4768, and the fiber is remeasured.
The reflectivity is then defined using the measured intensities in the two cases as 
$R = 1- I_{\mathrm{black}}/I_{\mathrm{mirror}}$. 
The average mirror reflectivity is 83\% with a standard deviation of 7\%.

The fibers are manufactured in batches, called 
preforms. Five 320~\cm fibers from each preform 
are tested using a fiber scanner, shown in 
Fig.~\ref{fig:minos_scanner}, to determine if the attenuation
is acceptable.  To measure the attenuation, a fiber is inserted into the long scintillator bar and 
read out at one end using a R580-17 Hamamatsu PMT attached to a picoammeter 
for different positions of a $^{106}$Ru source along its length. 
Figure~\ref{fig:fiber_fit} shows an example of the same fiber being scanned when the terminated end has been mirrored versus when it has been blackened. 
The data are fit to a double exponential and extrapolated to 320~\cm.
The quality control is based on the measured attenuation length and the amount of 
light at 320~cm from the readout end since
the longest WLS fibers in \minerva\ are 320~cm.
The measured attenuation varied over different batches from 
0.31 to 0.37, with standard deviations varying from 0.015 to 0.033.

\makeatletter{}
\subsection{Optical Connectors and Clear Fiber Cables} 
\label{sec:clear_fibers}

The WLS fibers are terminated with optical connectors from 
Fujikura-DDK \cite{ddk_con},
referred to as DDK connectors.  Each connector groups eight fibers. The DDK connectors are used to connect to cables containing eight clear optical fibers which transmit light from the WLS fibers to the PMT boxes above the detector.  These connectors were originally developed by DDK for the plug upgrade for the CDF experiment at 
Fermilab \cite{Albrow}, in consultation with Tsukuba University.

\begin{figure}[htbp]
\begin{center}
\includegraphics[height=0.25\textheight]{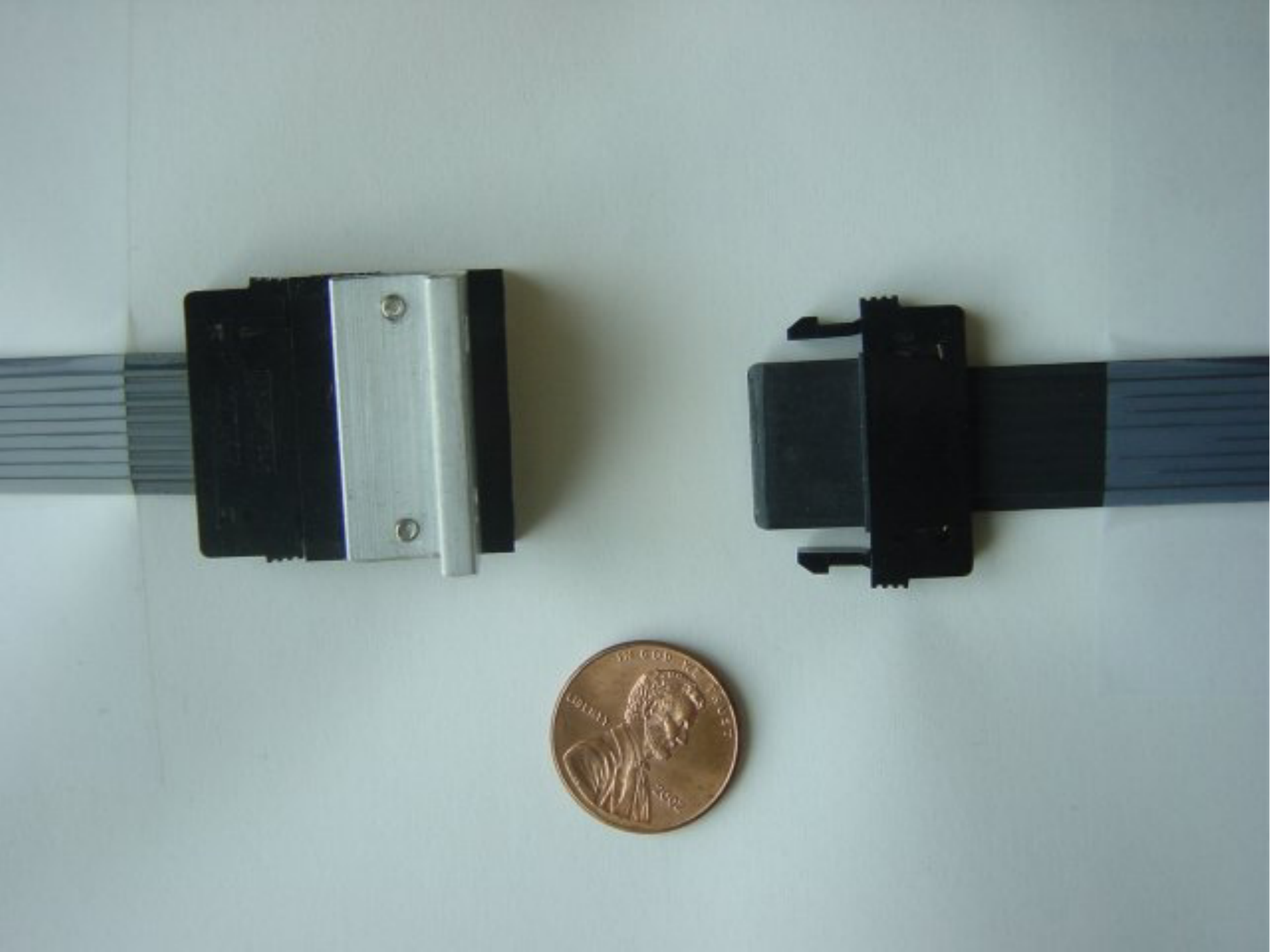}
\hspace{0.02\textwidth}
\includegraphics[height=0.25\textheight]{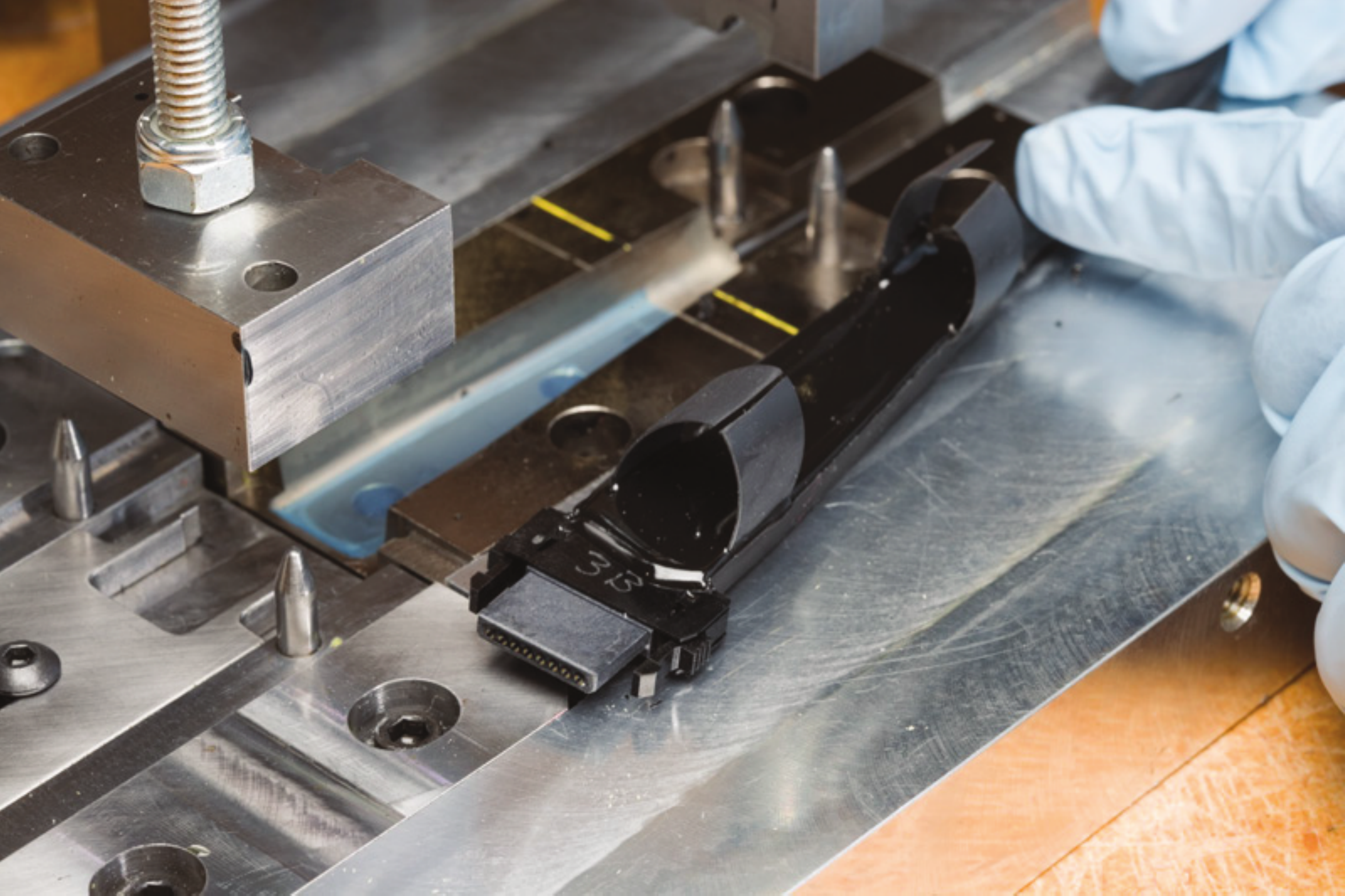}
\caption[DDK optical connector parts]
{The DDK connector  used to connect 8 WLS fibers to 8 clear optical fibers which carry signals to the PMTs mounted above the detector (left).  The aluminum mold used to attach the light-tight boot to the optical cable shown in the foreground (right). 
}
\label{fig:DDK_connectors}
\end{center}
\end{figure}

The DDK connectors consist of a ferrule, clip,
and box (Fig.~\ref{fig:DDK_connectors}). They snap
together without screws or pins.  The clear optical cables are made of 1.2~\mm, 
S-35 multiclad clear fiber from Kuraray to match the WLS fiber.  
Clear fibers are also used in the PMT boxes to transmit light from
the cable to the PMT via the ODU, described in
Section \ref{sec:pmts}.
There are four different lengths of clear cables used in the detector depending on 
where the WLS bundles exit the detector: 1.08~m, 1.38~m, 3.13~m, and 6~m. 

To make a clear fiber cable, the fibers are first cut to the correct length.  
The fibers are then inserted into a ferrule
and the tops are
taped against a horizontal piece of metal.  BC600 epoxy is then
placed in the pocket of the ferrule with a syringe. After the
epoxy cures (the next day), two clips are placed on the fibers
and a light-tight tube is placed over the entire length of the 
fibers except for approximately 5 cm near the ends where the 
fibers enter the ferrules. 
The free end of the 
fibers are placed in a 
second ferrule and epoxied in place. After curing, the fibers on both
ends of the cable are trimmed to about 0.3 cm at the connector in
anticipation of  polishing. After the ferrules and fiber ends
are polished, the clips are pushed up onto the two ferrules.

To ensure light-tightness the fibers are surrounded by a 0.6~\cm thick opaque sheath of INSUL \#4900/3. 
Figure~\ref{fig:DDK_connectors} (right) shows the mold used to surround the 
region near the connectors with a light-tight boot. The mold is
filled with custom epoxy ``MP5405 bk with extra carbon black''
from Heigl Technologies \cite{heiglweb}.
The boots are made fire-resistant by putting on a single coat of
Performix HCF from Plasti Dip, a flexible acrylic water-based coating.

\begin{figure}[htbp]
\begin{center}
\includegraphics[width=0.8\textwidth]{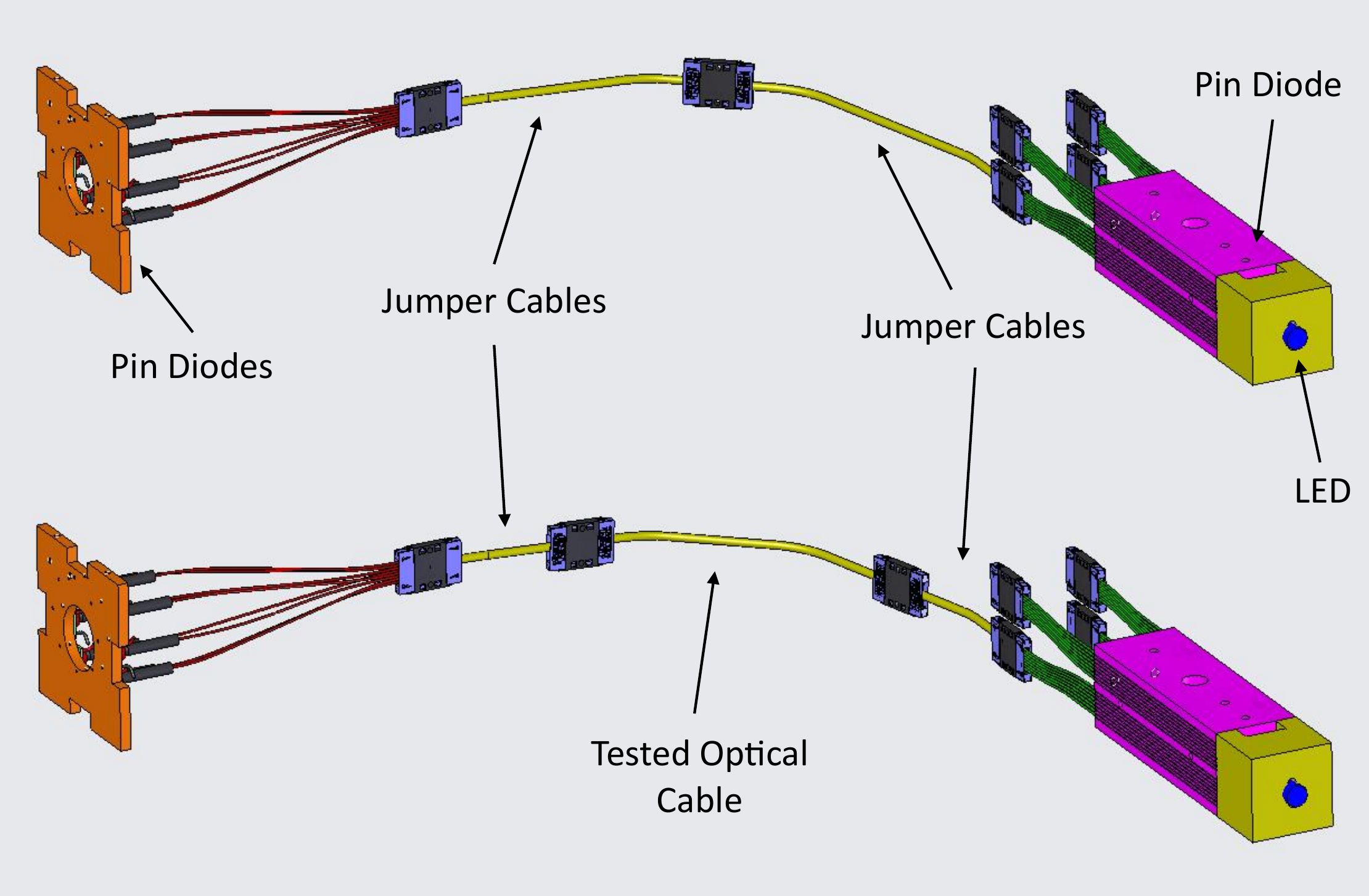}
\caption[Schematic of optical QC apparatus used to QC the optical cables ]
{Schematic of the setup used to measure the transmission coefficients of the clear optical cables.}
\label{fig:cab_test_schematic}
\end{center}
\end{figure}

Before use in the experiment, the light transmission of each fiber in each 
connector is measured on a dedicated test stand.
Figure~\ref{fig:cab_test_schematic} shows a schematic diagram of
the apparatus used for this measurement.
A light-emitting diode (LED) shines on the fibers in a repeatable way with normalization provided by
 a calibration pin diode. Each cable is
connected from the LED source box to a readout box by optical jumper
cables.  The jumper cables help to preserve
the fiber optical surfaces of both boxes by reducing the number
of connections made to them. 
The transmission is determined by measuring the light transmission
through the apparatus without the cable (the Direct Measurement),
then with the cable inserted, (the Cable Measurement).  
The transmission is defined to be Cable Measurement/Direct Measurement. 
The attenuation length is determined by fitting the measurements 
from 0.5~m, 1.08~m, 1.38~m, and 3.13~m length cables to an exponential. 
The fit gives an attenuation length of 7.83~\m for the clear fiber. 
The attenuation in the clear fiber can then be subtracted to determine the transmission across the connector.

Measurements of the production ODUs yield transmissions with an average
value of 87.5\% with a standard deviation of 3.5\%.  The connector transmissions
averaged about 87.0\% with a standard deviation of 4.8\%.  
Fibers are visually checked
for breaks or cracks, and are tested for light tightness and for correct fiber ordering
 at the end of the assembly.

\makeatletter{}
\subsection{Photomultiplier Tubes and Optical Boxes} 
\label{sec:pmts}

The light output for a minimum ionizing
particle (MIP) traversing a scintillator strip in the detector is sufficient to enable the use of a low quantum efficiency photosensor.
However, a timing resolution of better than $\sim$5~\ns is important for distinguishing overlapping events within a single spill of the NuMI beamline and for measuring time-of-flight and decay times of charged mesons created in neutrino interactions.   
With these considerations in mind, the multi-anode
photomultiplier tube (PMT) model number H8804MOD-2 manufactured by Hamamatsu Photonics 
was chosen to serve as the experiment's signal readout photosensor.  It is essentially the same
PMT as was used by MINOS \cite{minos-pmt}.

The H8804MOD-2 PMT has an 8~$\times$~8 array of pixels laid out on a 2~\cm~$\times$~2~\cm grid, 
i.e. 64 pixels per PMT with each pixel having an effective size of 2~$\times$~2~mm$^2$. 
The general properties of the H8804MOD-2 PMT are listed in Table  \ref{tab:pmt1} and the 
operating characteristics provided by the manufacturer are given in Table \ref{tab:pmt2}.    
The PMTs for the detector are required to have a minimum quantum efficiency of 12\% at 520 nm and a maximum-to-minimum pixel gain ratio less than three.  A total of 507 PMTs are used in the fully instrumented detector.

\begin{table}[htbp]
\centering
  \begin{tabular}{|l|l|l|} \hline
    Parameter & Description$/$Value \\  \hline
    Spectral Response & 300 -- 650 nm \\  \hline
    Peak Wavelength & 420 nm \\  \hline
    Photocathode Material & Bialkali \\  \hline
    Photocathode Minimum Effective Area & 18$\times$ 18 mm$^2$ \\  \hline
    Window Material & Borosilicate Glass  \\  \hline
    Dynode Structure & Metal Channel Dynodes  \\  \hline
    Number of Stages & 12  \\  \hline
    Weight & 30 g \\  \hline
    Operating Ambient Temperature & -30 -- 50 $^o$C \\  \hline
    Average Anode Current & 0.1 mA \\   \hline
  \end{tabular} 
 \caption{General properties of the Hamamatsu H8804MOD-2 multi-anode photomultiplier tube.}  
  \label{tab:pmt1}
\end{table}

\begin{table}[htbp]
\centering
  \begin{tabular}{| l | c | c | c | c |} \hline
    Parameter & Min. & Typical & Max. & Unit \\  \hline
    Luminous (2856K) Cathode Sensitivity & 60 & 70 & - & $\mu$A/lm \\  \hline
    Quantum Efficiency at 420~nm & - & 20 & - & \% \\ \hline
    Anode Dark Current & - & 2 & 20 & nA \\ \hline
    Anode Pulse Rise Time & - & 1.4 & - & ns \\ \hline
    Electron Transit Time & - & 8.8 & - & ns \\ \hline
    Pulse Linearity $\pm$2\% & - & 30 & - & mA \\ \hline
  \end{tabular}
\caption{Hamamatsu H8804MOD-2 multi-anode photomultiplier tube operating characteristics at 25$^o$ C. }
  \label{tab:pmt2}  
\end{table}

\begin{figure}[htbp]
\begin{center}
  \includegraphics[width=.8\columnwidth]{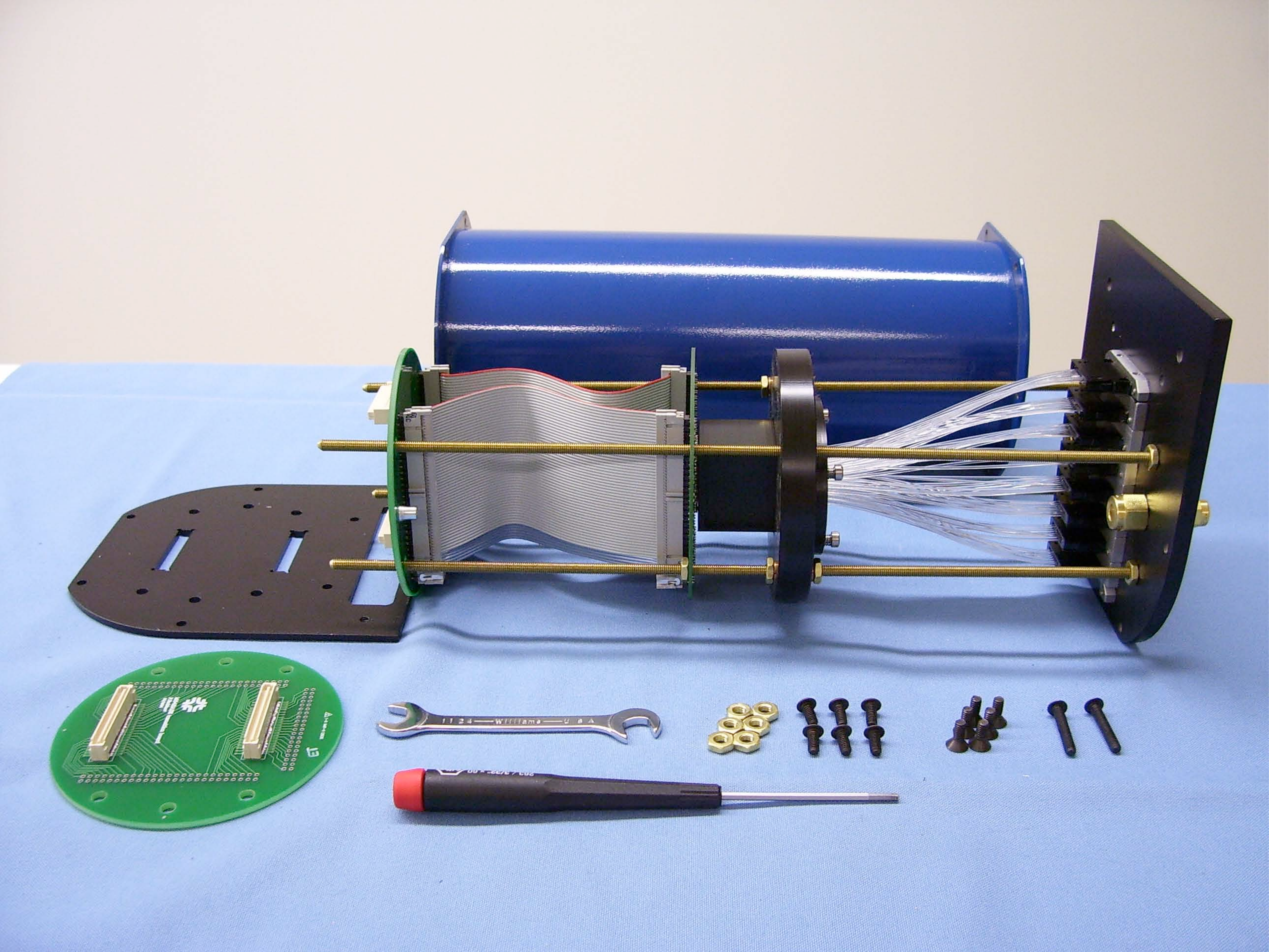}
\caption{Optical box with interior mounting frame, prior to assembly.
The frame holds the fiber cookie and PMT alignment plate
within the shielding enclosure. The Optical Decoder Unit can be seen between the PMT
and face plate on the right.}
\label{fig:Box-parts}
\end{center}
\end{figure} 

Each PMT is housed in an individual light-tight cylindrical enclosure
made of 2.36~mm thick steel.   
The boxes provide mechanical protection to the PMTs, facilitate
the routing of signal and voltage cables, and provide shielding from ambient magnetic fields 
due to the close proximity of the \minos\ detector.  The cylindrical housings are capped
with machined steel endplates.
Each endplate has a gasket and a room temperature vulcanizing (RTV) silicone
seal to ensure that no light can leak in from the outside.
The front-end board (FEB) that services the PMT is mounted directly to the outside of one of the endplates 
to reduce input capacitance and allow easy access for connections, testing, and replacement.
Within the optical box, the PMT is mounted to a base circuit board 
that includes a Cockcroft-Walton high voltage generator and 
provides signal routing to the FEB.  On the other endplate is a set of eight connectors to 
receive clear fiber cables, which connect to the ODU inside the box.
Precise alignment of individual fibers to PMT pixels is assured by routing the fibers onto a mounting ``cookie." 
The fiber mounting cookie together
with the PMT are firmly held within a mounting holder which guarantees their relative alignment.
Each box supports two additional optical fiber ports 
terminated by diffusers which allow all pixels of the PMT
to be ``flashed'' in a controlled way by an external light injection system (Sec. \ref{sec:light_injection}).
 An optical box with its component parts before assembly is shown in 
Fig.~\ref{fig:Box-parts}.  

\begin{figure}[htbp]
  \begin{center}
  \includegraphics[width=.9\columnwidth]{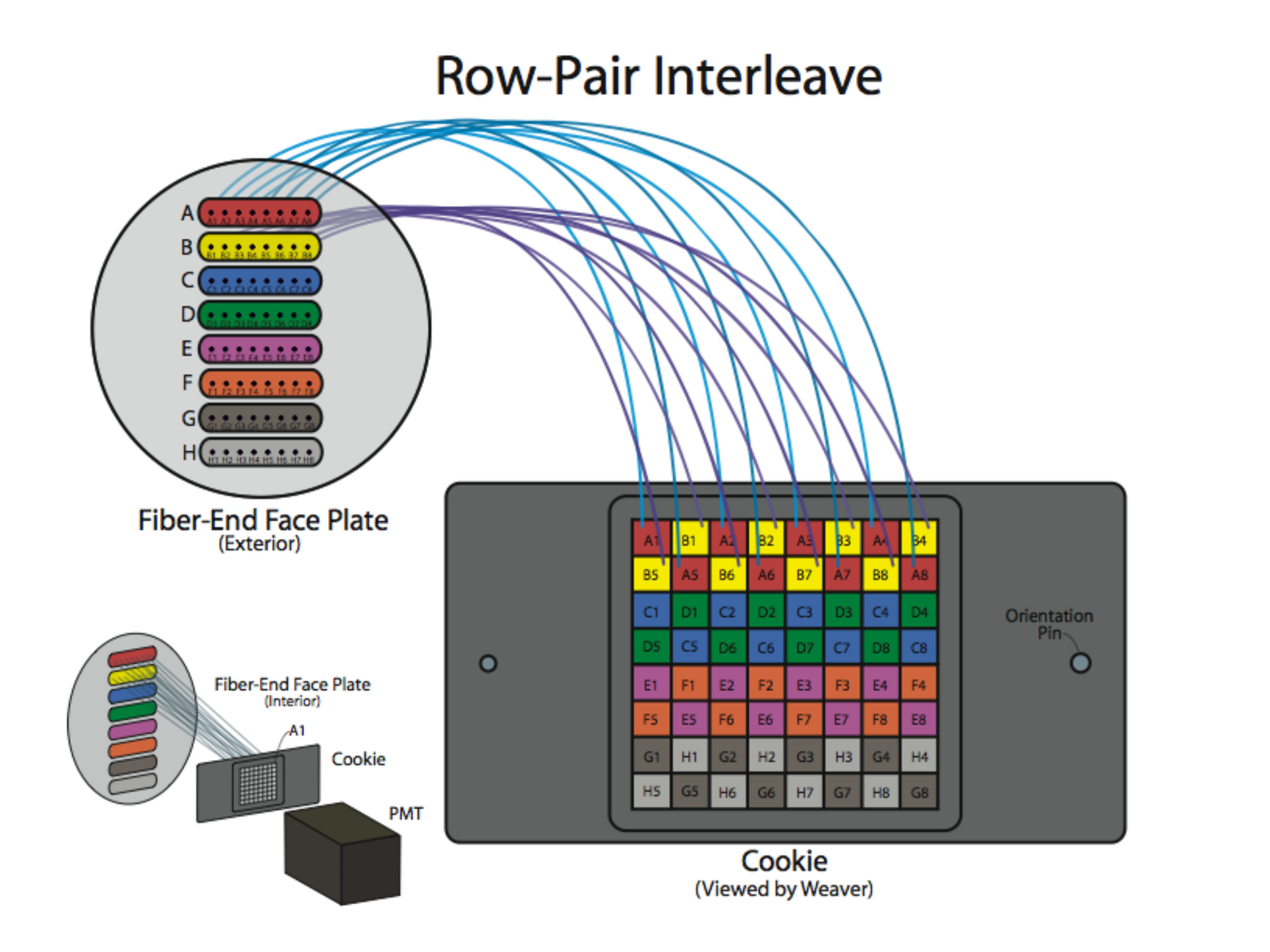}
  \caption{Diagram showing the pattern of the fiber weave within each PMT box.    The weave carries light 
signals arriving on optical fibers going into the box fiber port
  to the pixel grid on the front face of the PMT.    The end of each fiber of the weave is epoxied into a 
machined fiber cookie; the cookie is precisely aligned, via its
  mounting to the PMT holder, to witness marks located on the PMT pixel grid.}
  \label{fig:weave}
  \end{center} 
\end{figure}

One requirement in designing these assemblies is that light signals from physically 
adjacent scintillator strips in the detector do not go into physically adjacent pixels on a phototube.  
This is because nearest neighbor pixels are the most likely to experience signal cross-talk.  
The ODU uses a special weave of the optical fibers 
to ensure this separation, as shown in Fig. \ref{fig:weave}.

\minerva\  optical boxes are deployed in the vicinity of the 
energizing coils and magnetized steel of the MINOS near detector. 
In some regions occupied by the box array, the ambient magnetic fields are 
typically 5--10 gauss and can be has high as 30 gauss. 
Because the efficiency of a PMT begins to 
degrade when subjected to an ambient field exceeding 5 gauss, some shielding must be provided.  
  Several steps have been taken to reduce the field strength at the PMTs:  
the 2.36~\mm steel provides some shielding; the boxes are oriented perpendicular to the 
residual magnetic field;  
a steel ``mirror plane" has been placed between \minos\ and \minerva\ in the 
region of the return coil; and  40 of the \minerva PMT boxes closest 
to \minos\ have been augmented with an interior shielding of high permeability metal cylinders.

\makeatletter{}
\subsection{Readout Electronics and Data Acquisition} 
\label{sec:electronics}

Fast analog signals from the PMTs are fed to the FEBs attached to the optical box. 
The FEBs digitize timing and pulse-height signals, provide high voltage for 
the PMTs and communicate with VME-resident readout controller modules 
over a Low-Voltage Differential Signaling (LVDS) token-ring.  
Pulse-heights and latched times over a 16~\micros readout gate are 
recorded for all channels at the end of each 10~\micros spill of the NuMI beam.  
The additional 6~\micros ensure that any delayed detector activity, such as from Michel 
electrons from muon decays, is recorded.  

The FEB design is based on the D0 TriP-t \cite{tript} Application-Specific Integrated Circuit (ASIC)
 which is a redesign of the readout ASIC of the fiber tracker and   
preshower systems of the D0 experiment at Fermilab.  Each FEB services one PMT (64 channels) which requires 6 
TriP-t chips per board. The TriP-t chips are controlled by a commercial 
Field-Programmable Gate Array (FPGA) using custom firmware. The readout chain is connected at
both ends to a custom VME module called the Chain Read Out Controller (CROC), which serves
up to 4 FEB chains,  with each chain having up to ten FEBs.  The CROCs receive timing and trigger
commands from a custom module, the CROC Interface Module (CRIM).  There is no hit-based
trigger, but rather a timing based integration gate synchronized to the FNAL Main Injector timing signal.
 A full description of the \minerva\ data acquisition (DAQ) system is given in Ref. \cite{daqnim}.
 
\makeatletter{}
\section{Event Formation and Hit Calibration Chain}
\label{sec:EventForming}

\begin{figure}[htbp]
\begin{center}
\includegraphics[width=1.0\textwidth]{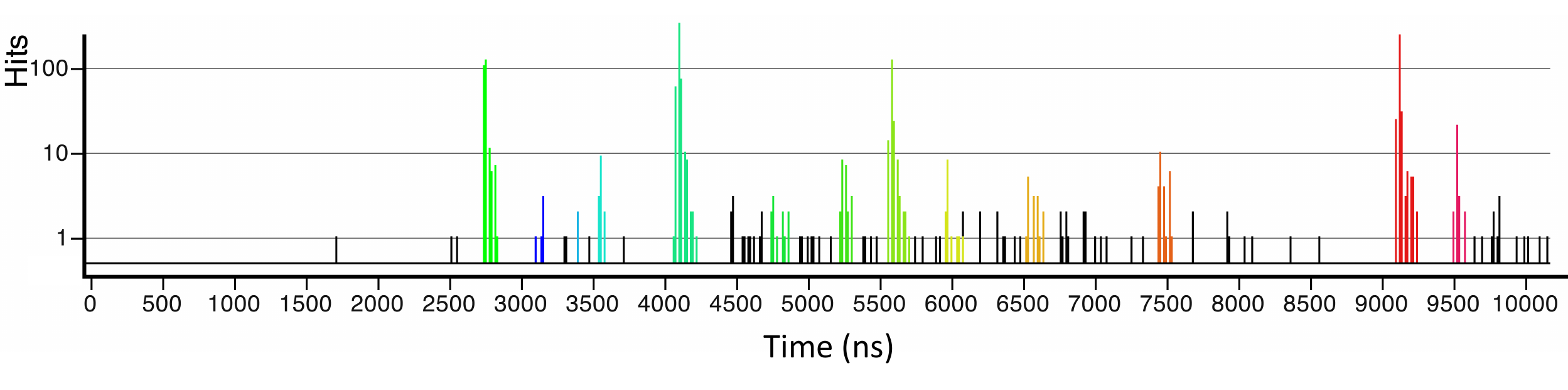}
  \caption{\small Time profile of PMT signals recorded in a typical readout gate 
(corresponding to a single 10~\micros spill of the NuMI beam).  In this particular gate, 
2,743 scintillator strips in the detector produce signals above the TDC discriminator threshold 
of about 0.7 photoelectrons. The colored groupings of hits indicate peaks found by the 
offline event forming algorithm.  The black entries are groupings below the algorithm's threshold 
for forming an event.}
  \label{fig:gatetimes}
  \end{center}
\end{figure}

The neutrino beam is sufficiently intense to produce  multiple neutrino interactions in the \minerva detector within one
10 $\mu$sec beam spill.  The use of a non-triggered 
integration-style readout system requires that individual interactions be separated offline using timing information.
Figure \ref{fig:gatetimes} shows the hit time profile of a typical readout gate.  
Several isolated groupings of activity are clearly visible.

\makeatletter{}

\begin{figure}[htbp]
\begin{center}
\includegraphics[width=0.7\textwidth]{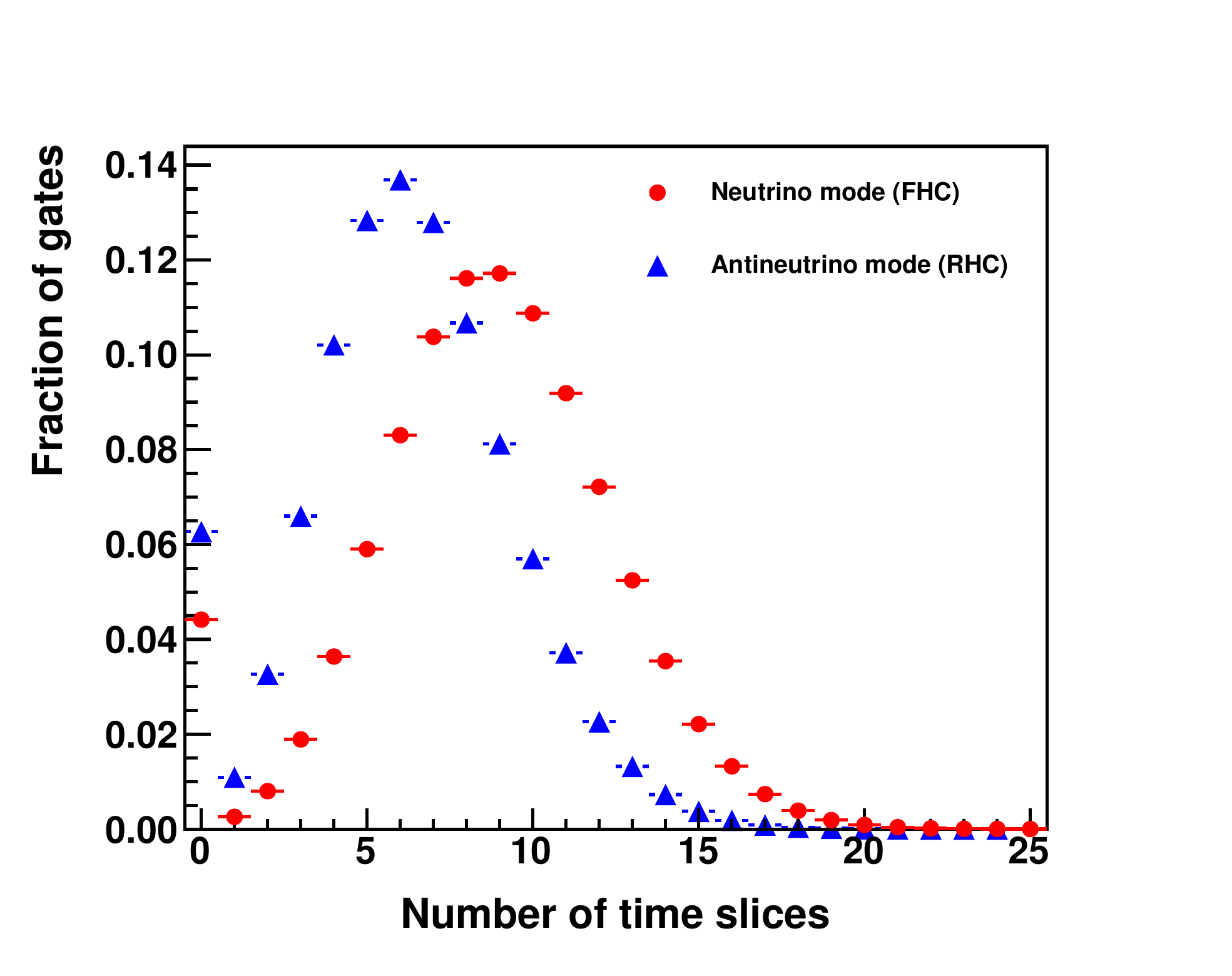}
  \caption{\small The number of time slices per readout gate for a period of forward horn current (neutrino mode) running from 22 March to 12 July, 2010, and reverse horn current (antineutrino mode) running from 5 November, 2010 to 24 February, 2011.  The zero bin is over-populated due to occasional gates where a trigger is recorded but no protons are delivered.  For these data the average protons on the NuMI target is about $344\times 10^{12}$ per spill.}
  \label{fig:nslices}
  \end{center}
\end{figure}

Raw time-to-digital converter (TDC) data are first corrected for propagation delays 
to the center of each scintillator strip. Clusters of activity in time are grouped without 
consideration of any spatial relationships between the scintillator channels by an offline peak-finding algorithm
to create ``time slices''.
Time slices are initiated when hits firing the discriminator within an 80~\ns time window exceed a total charge threshold of 10 photoelectrons (corresponding to 2/3 of the signal over a plane for an normally-incident minimum ionizing particle).  The window then slides forward until the threshold requirement is no longer met.  Hits which do not fire the discriminator are then added if they share a TriP-t with a hit already in the collection.  The colored peaks in Fig. \ref{fig:gatetimes} indicate  groupings of hits made with this algorithm in one readout gate.  Figure \ref{fig:nslices} shows the number of time slices per gate over a period of FHC (``neutrino mode'') and RHC (``antineutrino mode'') running at typical intensities.
With the exception of electrons from the decay of stopped muons
(Michel electrons), activity from a single neutrino interaction is usually contained in a single time slice.  Michel electrons are identified by using spatial information to stitch together two time slices.

\makeatletter{}

\begin{figure}[htbp]
\begin{center}
\includegraphics[trim=15mm 0 0 0, width=0.5\textwidth]{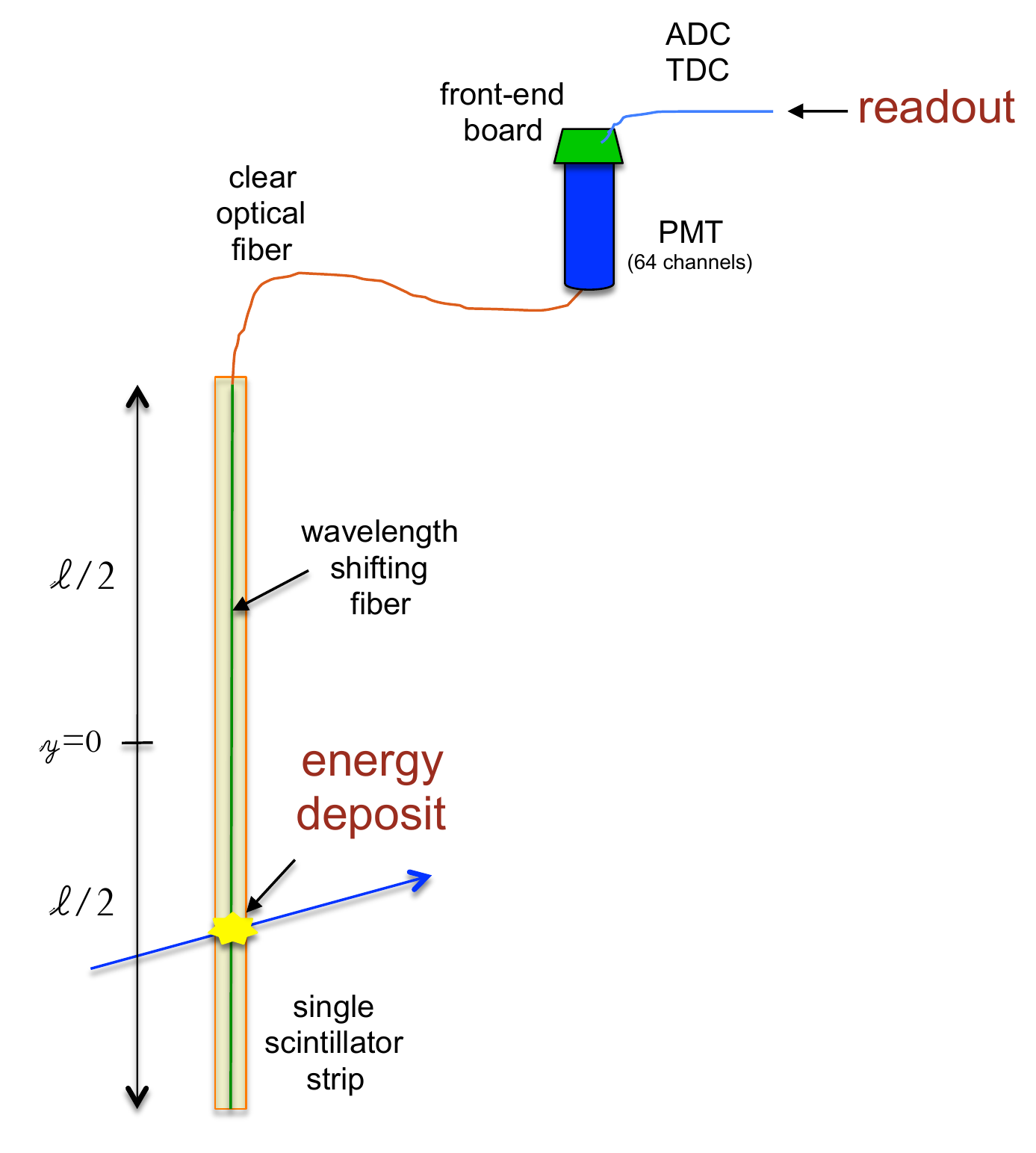}
  \caption{\small Schematic diagram of a single optical readout channel in \minerva.  
Raw ADC counts are converted into energy through a series of calibrations that 
account for the various components in the chain.}
  \label{fig:hitcal}
  \end{center}
\end{figure}

Raw analog-to-digital converter (ADC) data must be calibrated to provide an 
estimate of the energy deposited in each scintillator strip.  
 Differences between channels as well as variations of 
individual channel responses over long periods of time must be taken into account.  
Figure \ref{fig:hitcal} shows a schematic of the components in each readout
 channel as described in Sec. \ref{sec:OpticalSystem}.  
There are four effects which must be accounted for in order to convert the ADC value to 
an estimate of the energy deposited. (1) photons are attenuated as they travel along the wavelength 
shifting fiber to the end of the strip.  This attenuation depends on details of the 
construction process for each channel and has been measured using a radioactive source on  
each strip in the detector prior to installation as described in Sec. \ref{sec:mapper}.  
(2) The light signal is attenuated in the clear optical fibers, which have a characteristic attenuation 
length of 7.83~\m as described in Sec. \ref{sec:clear_fibers}. 
(3) Photons reaching the PMT generate photoelectrons which are amplified in the dynode chain.  
The PMT gains (output charge per photoelectron) are monitored \emph{in situ} for every channel 
as described in Sec. \ref{sec:pmt_gains}. 
(4) The readout charge is digitized on the FEB.  The ADC conversion function was
 measured with a bench-top charge injection setup for every channel on every FEB before installation in 
the detector as described in Sec. \ref{sec:feb_response}.

Additional channel-to-channel response variations can exist due to a variety of small effects 
that are difficult to characterize individually.  These include construction differences, 
the varying quality of connections between components, temperature or humidity dependence, and other effects.  
The relative differences between channels can be precisely monitored in real time using the 
high statistics sample of through-going rock muons produced upstream of the detector,  
as described in Sec.~\ref{sec:s2s}.  The rock muon sample also provides a means for determining the absolute energy scale, 
as discussed in Sec.~\ref{sec:meu}.  
  
Each of the above effects is accounted for in the event reconstruction through  multiplicative correction factors applied to the raw ADC data.  The energy deposited in scintillator strip $i$ is estimated according to
\begin{equation}
\label{eq:chain}
E_i = \left[ C(t) \cdot S_i(t) \cdot \eta^{att}_i \cdot e^{\ell_{i}/\lambda_{clear}} 
\cdot G_i(t) \cdot Q_i(\mathrm{ADC}) \right] \times \mathrm{ADC}_i 
\end{equation}

\noindent where $C(t)$ is the time dependent overall energy scale constant for the entire 
detector and $S_i(t)$ is the relative correction factor for channel $i$. The correction 
factor for attenuation within the scintillator strip, $\eta^{att}_i$, comes from the 
point-by-point source map data described in Section \ref{sec:mapper}.  
The factor $e^{\ell_{i}/\lambda_{clear}}$ 
is a correction for the attenuation in the clear optical fiber of length $\ell_i$ and 
attenuation constant $\lambda_{clear} = $ 7.83~\m.  The function $G_i(t)$ is the measured PMT pixel gain, 
and $Q_i(\mathrm{ADC})$ is the ADC-to-charge conversion factor for the FEB channel used to read out strip $i$.

\makeatletter{}
\section{\emph{Ex situ} Calibrations of Optical System Components}
\label{sec:ExSituCalibrations}

The previous section introduced the factors that need to be applied to translate between 
ADC counts and energy deposited in the scintillator.  Although several of those factors are 
measured throughout the course of the run with the operating detector, there are others which
 are best measured by \emph{ex situ} measurements prior to final assembly of the detector.  
In Sec. \ref{sec:OpticalSystem} the measurements of the WLS fiber attenuation, reflection, 
and clear fiber cable transmission are described.  However, the response of the scintillator 
strips in a full module after construction must also be measured, as well as the functionality 
of the PMTs and their associated electronics.  

This section describes the \emph{ex situ} measurements that are made before the detector is assembled.  
These measurements ensure that the components met the experimental requirements
and provide several of the constants needed to reconstruct energy depositions in the detector.  

\makeatletter{}

\subsection{Module Mapper}
\label{sec:mapper}

In order to translate the light output from each channel in the detector  into energy deposits for further analysis, it is necessary to determine the attenuation of optical pulses as a function of position along the scintillator strip.  
A custom ``module mapper'' makes this measurement on all channels in the detector, one module at a time.  
The mapper also serves as the final quality check for each module before it is deployed in the detector, 
identifying dead channels and those having anomalous response distributions.  

Corrections need to be made for optical attenuation of scintillation light in order to determine the amount of energy deposited in the scintillator by through-going charged particles.  In addition, local anomalies in the optical path (compromised WLS fiber or imperfect coupling between WLS fiber and scintillator, for example) can affect light collection.  These anomalies are identified and characterized allowing appropriate corrections to be applied to physics data.

\begin{figure}[ht!]
\begin{center}
\includegraphics[width=0.8\textwidth]{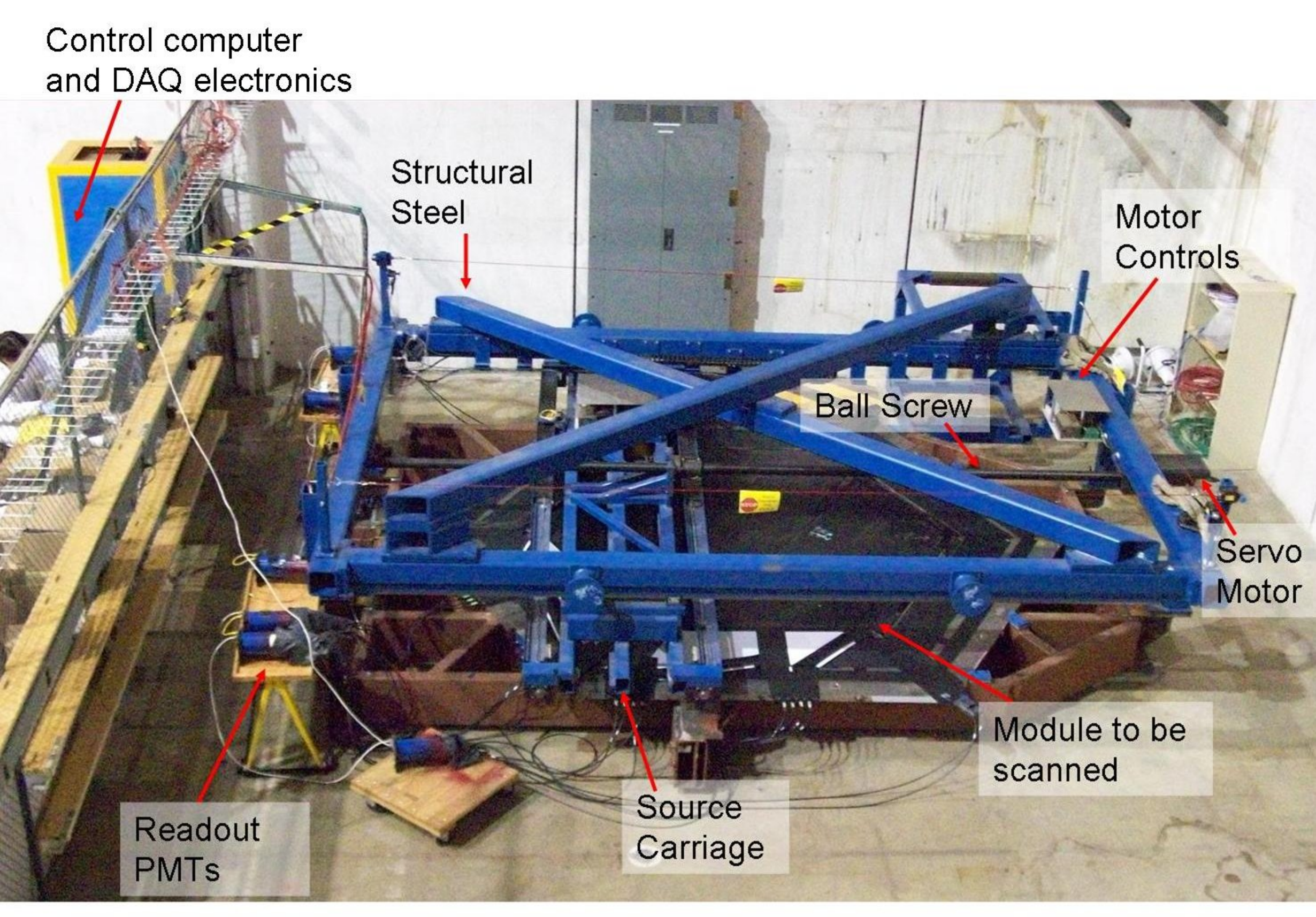}
\end{center}
\caption[Module Mapper]{The module mapper uses two Cs-137 sources to characterize the optics of \minerva detector modules.  Sources were mounted on the source carriage.  Ball screws controlled by servo motors provide independent motion to both sources.  The scintillator response is read out by a DAQ system, and the entire scan is coordinated by the control computer.}
\label{fig:f5}
\end{figure}

The module mapper is shown in Fig.~\ref{fig:f5}.  The scanner provides independent motion to two Cs-137 radioactive sources.  As the sources are moved through a pre-defined scan pattern over a detector module, 
the scintillator response is recorded by a series of Hamamatsu M-64 PMTs interfaced to a computer running custom data-acquisition software. 

A set of three computer-controlled servo motors coupled to ball 
screws move the radioactive sources through a pre-defined scan pattern. A hexagonal source location grid is optimized to minimize the total scan time while providing sufficient transverse and longitudinal resolution for strips in both the upper and lower plane.  A grid with transverse spacing of 19.05~mm in the two transverse dimensions was measured, with a total of 18,012 measurements in the inner tracking region.  The outer detector scintillator was measured over a grid that had 10~mm transverse and 100~mm longitudinal spacings, corresponding to about 1100 measurements per outer detector scintillator.  

The raw data stream records one ADC distribution at each source location for all channels instrumented (254 channels, for a typical two-layer tracking module).  The data stream must therefore be sorted to isolate the response of a single strip for grid positions centered on that strip's coordinate system.

The mapper data analysis consists of three stages: identifying the response of each strip at each individual radioactive source position; reconstructing a transverse scan for many points perpendicular to each strip by ordering the strip response and source positions appropriately; and reconstructing the strip response as a function of longitudinal distance along the fiber.  Offline analysis refinements (lead correction, layer correction) provide the final attenuation constants for the detector data calibration.

Careful tracking of the pedestal value (zero optical signal) for each channel is crucial for 
determining the response to the radioactive source.  Due to the nature of the DAQ electronics (operated in a mode in which the mean of the ADC distribution remains constant) the pedestal value for a given electronics channel shifts to lower ADC counts when the corresponding optical channel is illuminated.  This pedestal shift has been determined to be directly proportional to the current from the PMT pixel.  In order to measure the pedestal shift due to the response to the source illumination of a fiber, two reference pedestals are recorded: one prior to the illumination and one after the illumination.  The mean of these two source-absent pedestal locations is used as the reference point.  Then the ADC distribution for this channel is recorded as the source is brought in proximity to the corresponding scintillator strip.     

After the strip response is measured at a number of transverse distances from the strip at a given longitudinal position, the pedestal shift is plotted as a function of the transverse position.  This distribution is then fit to a Lorentzian profile to locate the center of the fiber position and the maximum response amplitude.  Based on the source mapping program there were 15 dead channels identified (out of over 32,000) prior to installation. 

After fitting many such transverse scans to obtain the maximum response as a function of longitudinal position along the strip, an attenuation response curve is determined.  The response curve is used to correct for the attenuation
once the position of a hit along a strip is determined by  later tracking algorithms.  The response curves made with the radioactive source technique are in agreement with what is measured in the data using minimum ionizing particles, but are measured to a higher precision using the radioactive source. 

\makeatletter{}
\subsection{PMT Testing}
\label{sec:pmt_testing}

\begin{figure}
\begin{center}
  \includegraphics[width=.7\columnwidth]{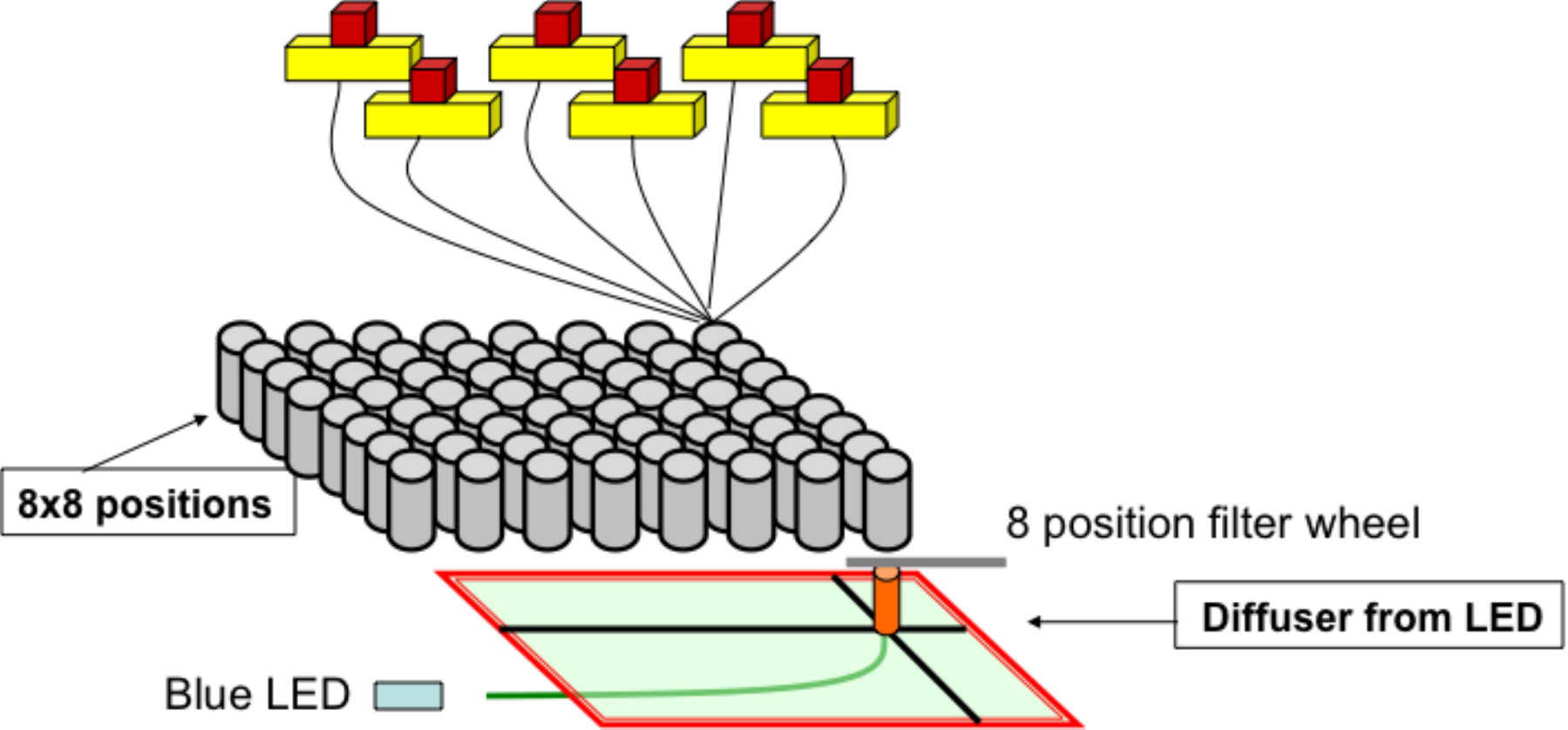}
  \caption{ A schematic diagram of the PMT test stand.  The motion control positions the LED 
    beneath a bundle of six clear optical fibers, which lead to the centers of individual pixels 
    on six PMTs.  Light from the LED is collected by a green WLS fiber, which shines 
    on a bundle of six clear fibers so that one pixel on each of the 6 PMTs is illuminated 
    simultaneously.}
  \label{fig:pmt_test_stand}
\end{center} 
\end{figure}

Before installation in its steel enclosure, each of the 507 deployed PMTs is tested for efficiency, linearity, pixel-to-pixel gain variation, 
dark noise, and cross-talk prior to installation in the steel enclosure of its optical box. Figure~\ref{fig:pmt_test_stand} shows a schematic diagram of the light-tight test stand used 
to collect these data.  Six PMTs are mounted onto cookies on the test stand.
Each cookie connects to one of the endpoints of 64 clear optical fibers, such that 
the center of each pixel is above the center of a fiber.
The other ends of the fibers connected to the same pixel on each PMT are bundled 
together; for example, the six fibers attached to each PMT's pixel two are grouped.
A blue LED beams light into one end of a green wavelength-shifting fiber, while the other 
end of the fiber is attached to a four-axis motion control system so it can be used to 
illuminate one bundle of pixels at a time.  Thus the WLS fiber is used to illuminate 
the same pixel on all PMTs simultaneously. The fourth axis of the motion control 
sets the amount of light filtered from the WLS fiber, and uses a mechanical neutral density filter.  

Once the PMTs are installed in optical boxes, they are subjected to additional tests to better understand the operational properties of the entire unit.  Specifically, any misalignment between the PMT and the cookie that holds the clear fibers from the ODUs could introduce additional cross-talk that would not be measured on the test setup described above.  
The optical box testing uses the same 
light source as is used for testing the optical cables, but without the PIN diode readout.
The test setup additionally uses
special optical cables where only one pixel in a row of pixels is illuminated.  
In this way, four pixels that are far from each other can be tested simultaneously for cross-talk, and PMT boxes which 
exhibit high cross-talk can be repaired before installation on the detector.  
The cross-talk used in the detector simulation is that measured using muons in an {\em in situ} technique (see
Sec.~\ref{sec:cross_talk}).  
  DA

\makeatletter{}\subsection{FEB Response Measurements} 
\label{sec:feb_response} 
 
Each PMT is read out by a 64-channel FEB which contains six (32-channel) TriP-t ASIC chips. The signal from each PMT anode segment 
is divided capacitively in the ratio of $1:4:12$ and routed into separate TriP-t channels to provide a low, a medium, and a high gain response to the same input charge, thereby increasing the dynamic range of the electronic circuits.
 
Theh 507 deployed FEBs were tested prior to installation in the detector. The tests included burn-in, HV control, basic input/output functionality, discriminator, digital control, charge calibration, and cross-talk measurements.

The charge calibration measurements are performed using a custom built test stand  
that injects a series of external charges into four selected input channels. The external charge is provided by 10 pF capacitors charged to 4 V by a 
pulse generator and discharged to the FEB input. 
The pulse from the generator is divided equally using a passive voltage divider and routed to four charge injection capacitors at the same time using remote controlled relay switches.
Injecting charge to four well-separated FEB input channels is a compromise to minimize the impact of cross talk on the measurement and to maximize the number of channels tested one time.

\begin{figure}[t]
\centerline{
\includegraphics[width=0.7\textwidth,angle=0]{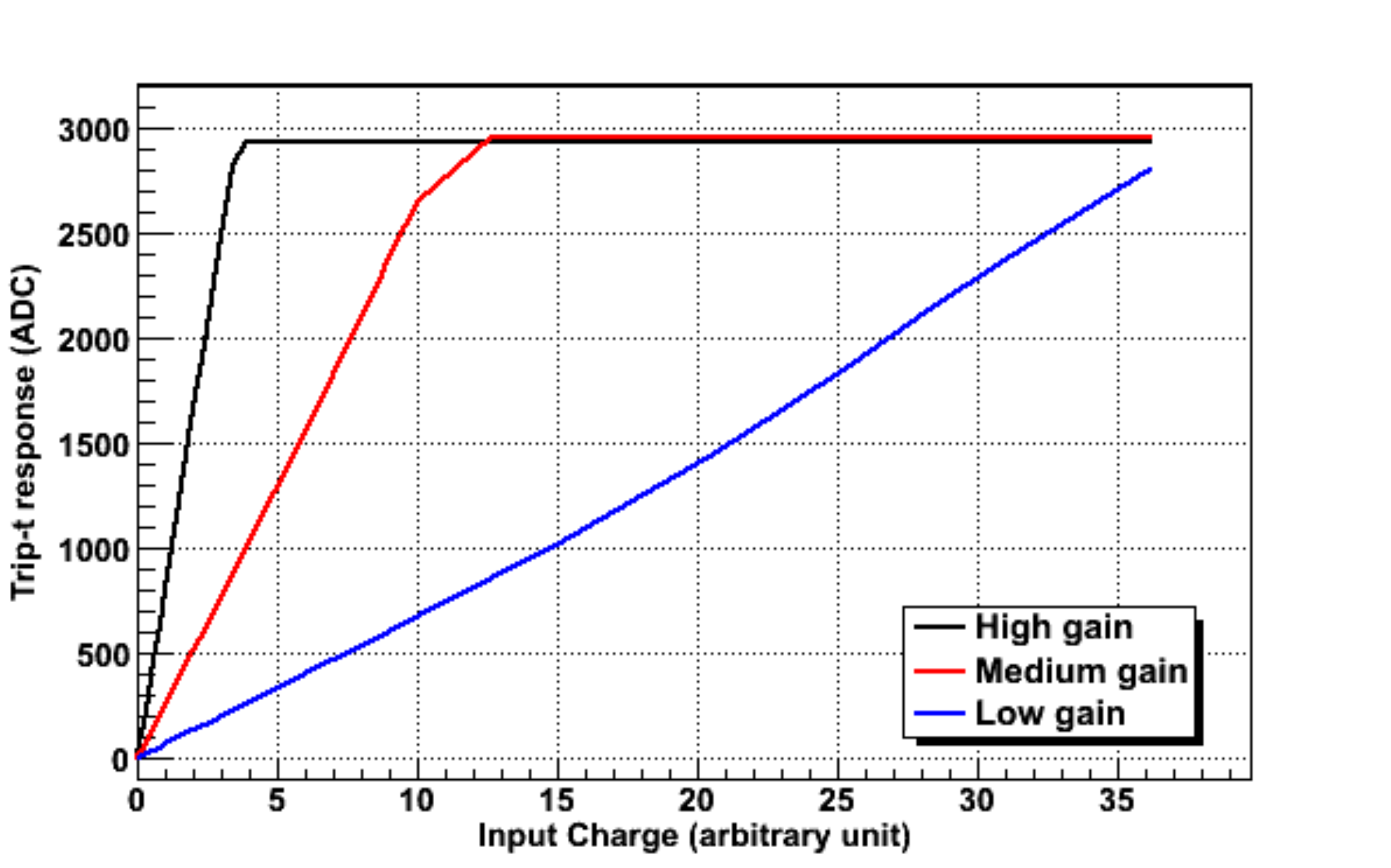}}
     \caption{The digitized high, medium, and low gain response of one electronics 
			channel as a function of input charge for a typical \minerva\  front end circuit board.}
\label{fig:feb_response}
\end{figure}

The DAQ of the charge-calibration test stand is a replica of the \minerva\  detector DAQ, consisting of a VME crate containing a CRIM and a CROC that read out the FEB being tested via Ethernet cable using the LVDS protocol. The data is passed to the DAQ via optical fiber by a PC/VME controller. The CRIM also provides the external trigger to the pulse generator for synchronous charge injection. The relay switches and the pulse generator voltage levels are controlled via a general purpose interface bus (GPIB) 
interface and their operation has been integrated into the DAQ software for automatic testing.  
A typical response of one FEB channel as a function of the input charge is  
shown in Fig.~\ref{fig:feb_response}.

The response of the electronics cannot be characterized by a simple linear function of the input charge. 
In order to characterize the non-linearity, the high, medium, and low gain response of each channel 
is fitted to a tri-linear function that consists of three distinct linear segments. This 
simple parameterization of the response is accurate within approximately $1\%$ over the full dynamic range.
As a result, each electronics channel has three sets of six parameters (the slope and the starting point of the three linear segments) describing its high, medium, and low gain response. These parameters are stored in an offline database and used  to convert the raw ADC response to linearized charge. The charge is calculated from the high, medium, or low gain response (in that order) if the corresponding digitized output is below the saturation of the electronics (approximately 2500 ADC counts).

\makeatletter{}
\section{\textit{In situ} Calibrations and Monitoring}
\label{sec:InSituCalibrations}

Several factors needed to convert ADC counts to an energy deposition
are best measured in the assembled detector.  Because these
factors can vary over the course of a multi-year run, the time dependence must be 
accounted for in the reconstruction and the detector simulation.  The rock muons generated
from neutrino interactions upstream of the detector provide a natural calibration source.  
Through-going muons provide a standard candle to set the overall energy scale, timing calibration, and a measurement of the 
cross-talk between adjacent phototube pixels.  They can also be used to determine the relative alignment of the 
ID modules and the relative light yield of each of the scintillator strips in the detector. 

Some calibration factors, such as the pedestal value of each channel,
are best measured in the absence of beam-induced detector activity.  
Other factors, such as the time-varying gain of each channel in each PMT, cannot be measured to the desired accuracy with rock muons.
In order to track the gains of the PMTs (which vary at a different rate from the light yield of the scintillator), 
\minerva uses a separate calibration system that injects light into the optical boxes in between beam spills.  

The following subsections describe the suite of calibration constants that are measured \emph{in situ}, 
either using rock muons that occur during the beam spill, or using special calibration triggers taken between beam spills.  

\makeatletter{}
\subsection{Pedestal Monitoring}
\label{sec:pedestals}

Regular monitoring of detector noise during beam-absent time periods is necessary to establish the 
reference point, or pedestal, which needs to be subtracted from  beam-on signals.  During standard beam-on operation, 
pedestal levels are measured for all 32,448 channels during a special mixed beam/pedestal subrun 
that occurs once every 32 subruns, about 10.5 hours apart.  This subrun collects approximately 750 gates from 
each channel over the course of roughly 27 minutes.  Each readout gate is open for 
16 $\mu$s \cite{daqnim} and captures the noise from cosmic rays, radioactivity, electronic sources, and the PMTs dark current.

\begin{figure}[ht]
\begin{center}
  \includegraphics[width=.9\columnwidth]{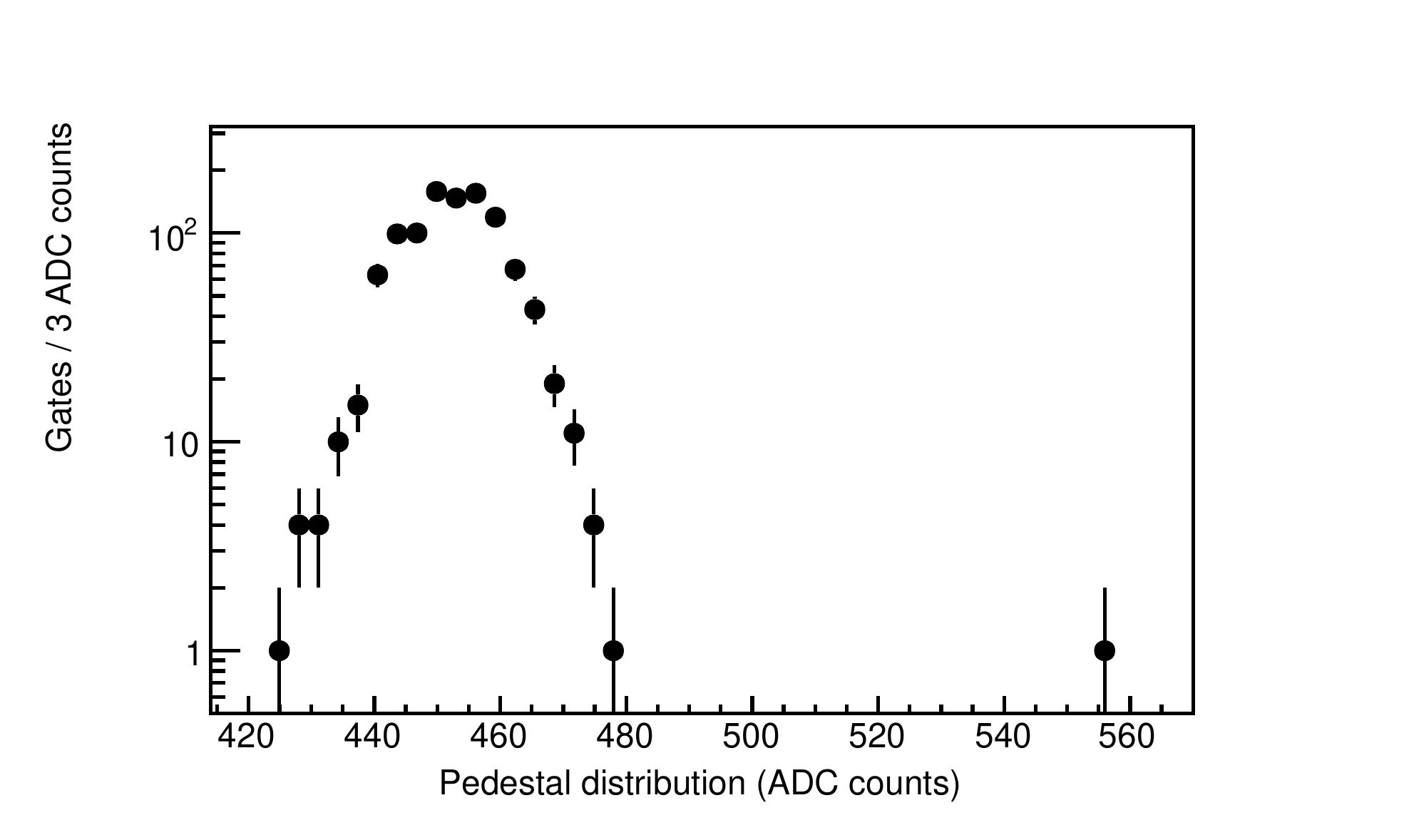}
  \caption{Example of a channel for a single pedestal gate with a measured signal of about 100 ADC counts 
    above the pedestal level.  Such gates are removed from the distribution before calculating pedestal means and widths. }
  \label{fig:outlier}
  \end{center} 
\end{figure}

Background activity event displays have been hand-scanned for a sample of about 40,000 pedestal gates.  A major contributor to this background is
cosmic ray muons (at 18 Hz over the entire detector) penetrating the detector after passing through the 100~m rock overburden.  Each cosmic ray muon 
event produces a single high signal during a pedestal gate for the illuminated channels.  An example of a signal which is well 
above the normal pedestal distribution for a representative channel is shown in Fig. \ref{fig:outlier}.

It is necessary to eliminate the high-side readings that occur from background particle entry 
into the detector and from spurious electronic signals.  An outlier removal method known 
as Peirce's Criterion \cite{peirce} is utilized to identify such pedestal readings in the raw data stream.  
After identification and removal of such outlying pedestal measurements, the mean and RMS pedestal is computed for 
each channel over the course of the subrun.

Pedestal mean values vary by about 7\% across all channels, and each channel's pedestal is observed to be stable to 
within 2\% during a pedestal subrun.  A summary of the 
pedestal variation for the entire detector during a single pedestal subrun is given in Table \ref{tab:ped_det_var}.  
The number of gates per channel varies due 
to the removal of spurious pedestal values taken during a subrun.
\begin{table}
\begin{center}
\begin{tabular}{|l|l|l|l|} \hline
            & Mean (ADC)  & RMS (ADC)     & Gates      \\ \hline
High gain   & 432$\pm$30  & 7.76$\pm$0.41 & 746$\pm$5  \\ \hline
Medium gain & 436$\pm$30  & 6.65$\pm$0.29 & 745$\pm$5  \\ \hline
Low gain    & 440$\pm$29  & 6.36$\pm$0.27 & 746$\pm$5  \\ \hline
\end{tabular}
\caption{Pedestal variation across 32576 detector channels for a representative subrun.}
\label{tab:ped_det_var}
\end{center}  
\end{table}
No coherent drift of pedestal values has been observed over long time scales.  The pedestal mean computed
from the mixed pedestal/beam subrun is used as the reference point against which all signal levels are
calculated for the 10.5 hours until the next pedestal sample is taken.  The RMS of pedestal means over time
is much smaller than the RMS of the pedestal values within one subrun, justifying the use of the latter as an 
estimate of the uncertainty on the true pedestal at the moment a beam signal is recorded.

\makeatletter{}
\subsection{PMT Gain Monitoring}
\label{sec:pmt_gains}

\minerva uses \textit{in situ} calibration data to monitor fluctuations in the single PE PMT gain 
for each channel in the detector.  The calibration source is LED light from 
a light injection (LI) system that is similar to that used by the MINOS experiment \cite{minosLI}.  
The LI system is triggered once after each 
beam spill.  At this rate (0.5~Hz), enough LI data is collected to measure the PMT gains across 
the entire detector once per day.  This measurement rate is sufficient to monitor gain fluctuations, 
as significant gain changes tend to occur over a span of weeks.

\subsubsection{Light Injection System}
\label{sec:light_injection}

Each optical box (Sect.~\ref{sec:pmts}) has two ports which receive clear optical fibers from the LI 
system.  The light is spread out inside the PMT box by transmission through a
polypropylene diffuser. The light reaches each of the PMT pixels through the small
space between each clear fiber and its holder.  Thus, the signal that reaches
the PMT is small (few PEs), but uniform within a range of about 30\%.  
This enables a rapid and accurate gain check for the entire PMT.

The core of the system is a blue AlGaInP (472 \nm) LED which has a current limit of 20 \mAmp.
It is driven with a pulse generated by a custom circuit~\cite{minosLI}.  Because
the pulse has a width of about 30 ns, the current can be much larger ($\sim$200 \mAmp).
The system has 23 LEDs. Each LED provides light to 50 clear fibers through an optical fanout,
and two fibers are routed to each optical box.  The optical
hardware and electronics are in two boxes located 40 feet from the PMTs.  The pulse amplitude,
width, and timing are controlled through an RS232 interface driven by the DAQ computer.

\begin{figure}[t]
\begin{center}
  \includegraphics[width=0.9\columnwidth]{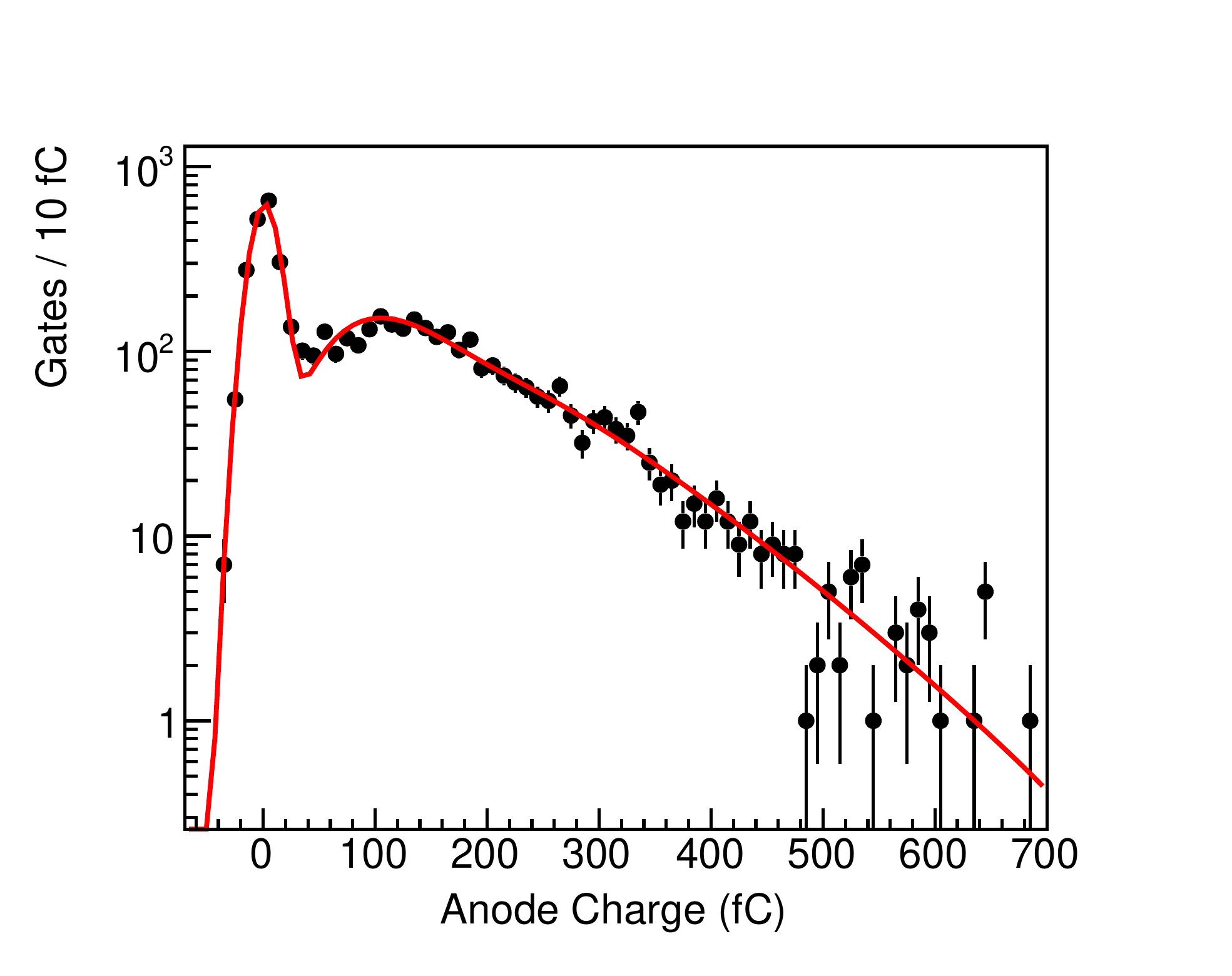}
  \caption{
    A typical 1~PE LI spectrum.  The fit function is a simple noiseless parameterization of the PMT output.  
}
  \label{fig:LI_fit}
    \end{center}

\end{figure} 

Figure \ref{fig:LI_fit} shows a typical LI spectrum in one channel, where the 
light injection system was set to pulse at a 1~PE level.  The large peak near zero is the 
pedestal.  Subsequent peaks are too broad to be 
distinguished from each other.

\subsubsection{Gain Calculation}
 The PMT gain (i.e.electrons/PE count) is determined using 
LI data
together with modeling of the photostatistics in the 
PMT dynode chain.  

The gain \(g\) of a pixel is defined as

\begin{equation} \label{eq:defGain}
g = \frac{\overline{Q}}{\lambda e} 
\end{equation}
where \(\overline{Q}\) is the mean of the pixel's pedestal-subtracted anode charge
distribution, \(\lambda\) is the mean number of PEs arriving at
the first dynode, and \(e\) is the magnitude of the electron charge.  Note that the quantum
efficiency of the photocathode and the collection efficiency of the first dynode
are not included in this definition of the gain.  Channel-to-channel differences in the
latter quantities are accounted
for in the relative strip-to-strip calibration (see Sec. \ref{sec:s2s}).

For a PMT with \(n\) dynodes, the probability distribution \(P_n(q)\) of the
charge \(q\) measured at the PMT anode is equal to
the probability distribution of the number of PEs observed at the anode
\cite{rademacker} convolved with a Gaussian to account for electronic noise. 
To good approximation each dynode amplifies according to a Poisson distribution, and the
amplification is linear with the number of incoming PEs, the variance \(\sigma^2\) 
of \(P_n(q)\) is
\begin{equation}
\sigma^2 = \sigma_p^2 + \lambda g^2 e^2 + \lambda g^2 e^2 w^2.
\label{eq:chargeConvWidth}
\end{equation}
Here \(\sigma_p\) is the electronic noise (pedestal) width and the gain \(g\) is 
identified as the product of the individual dynode gains \(g_i\).  The parameter \(w^2\) is defined as
\begin{equation} \label{eq:w2}
w^2 \equiv \sum_{j=1}^n \left(\prod_{i=1}^j \frac{1}{g_i}\right).
\end{equation}
Typically \(w^2\) has values between 0.2 and 0.3.  Equation
(\ref{eq:chargeConvWidth}) states that the variance of the PMT anode charge distribution is
the sum of the variance of the pedestal, the variance of the incoming PE
distribution, and the variance due to the statistical broadening of the dynode
chain.  

By solving Eqs. (\ref{eq:defGain}) and (\ref{eq:chargeConvWidth}) simultaneously, 
the gain $g$ can be written as a function of
\(w\) and the mean and standard deviation of the anode charge distribution: 
\begin{equation} \label{eq:measGain}
g = \frac{\sigma^2 - \sigma_p^2}{\overline{Q} (1+w^2) e }.
\end{equation}
\noindent Equation (\ref{eq:measGain}) is not sufficient to calculate the pixel gains because the
inclusion of \(w\) 
introduces \(n\) unmeasured parameters \(g_i\).  The parameter \(w\) can be expressed solely as a function of the 
total gain by noting that each \(g_i\) is proportional to a power of
\(V_i\), the potential difference that accelerates each PE arriving at the \(i\)th
dynode
\begin{equation} \label{eq:gnVn}
g_i \propto V_i^{\alpha}.
\end{equation}
The exponent \(\alpha\) typically has values between 0.7 and 0.8 \cite{hamamatsu}.  \minerva uses
\(\alpha = 0.75\), which introduces less than a 1\% systematic uncertainty into
the gain measurement.  

\begin{figure}[t]
\begin{center}
\includegraphics[width=0.9\columnwidth]{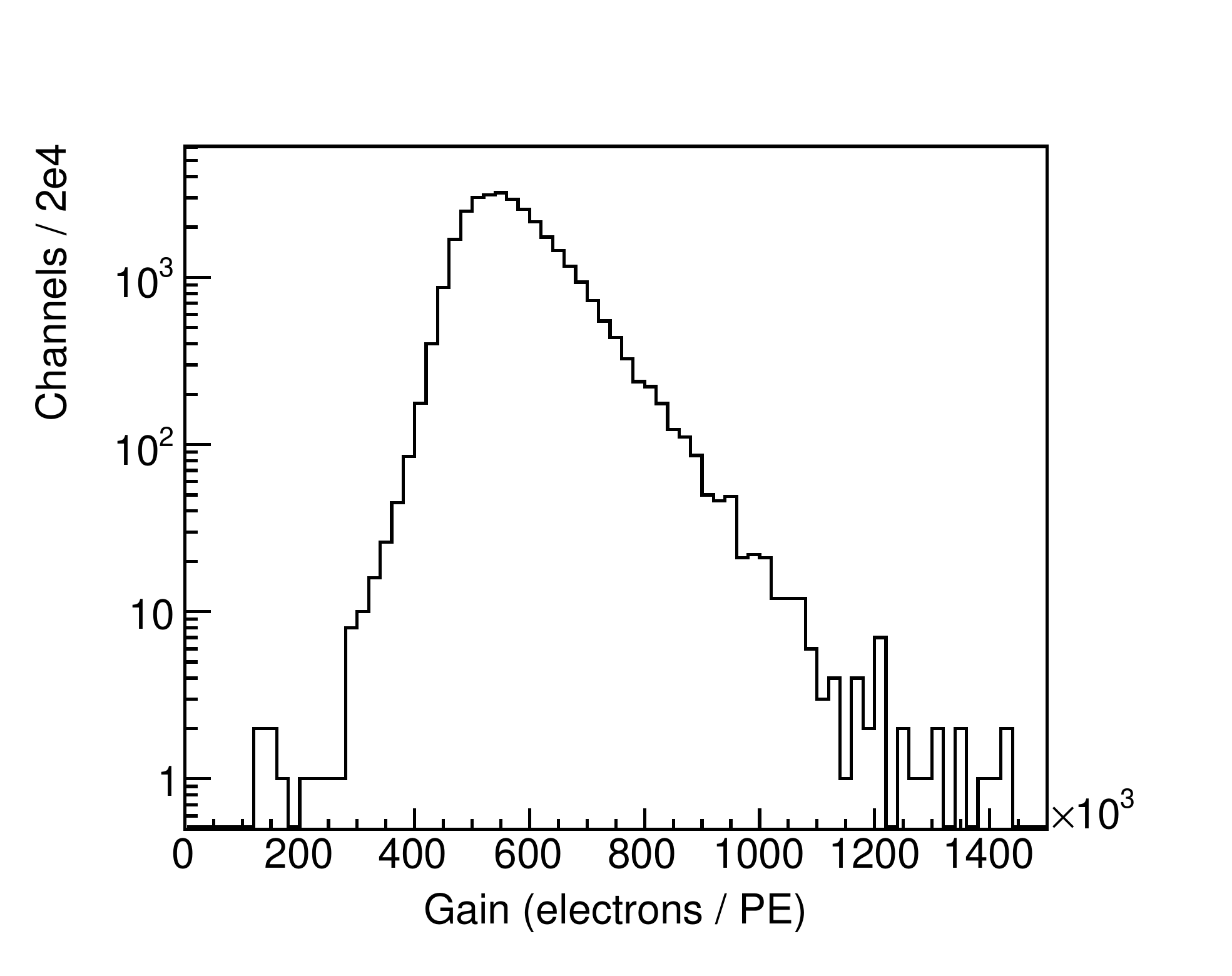}
\caption{ The measured distribution of gains of all PMT pixels in  \minerva on April 1, 2010. }
\label{fig:allGains}
\end{center}
\end{figure}

\begin{figure}
\begin{center}
\includegraphics[width=0.9\columnwidth]{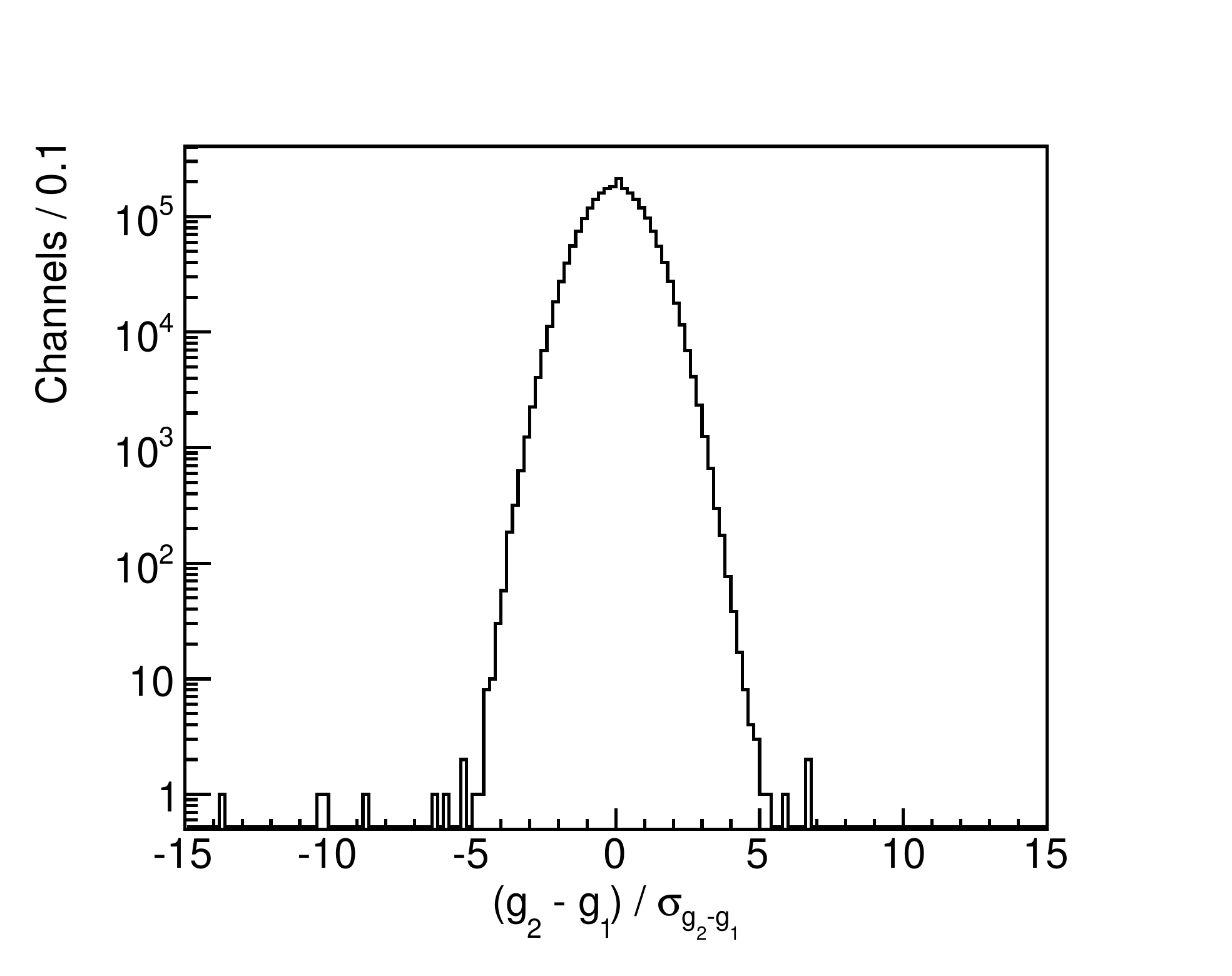}
\caption{ Statistical significance of gain fluctuations between consecutive measurements.  
The mean and width of the distribution are 0.002 and 0.97 respectively, indicating that gain measurement 
fluctuations over one day time periods are 
primarily statistical.  The plot includes 70 gain measurements for each pixel, 
performed between March 1 and April 29, 2012.  A total of 102 dead 
and underperforming channels were removed out of the 32,448 channels in the detector. }
\label{fig:gainChangeSig}
\end{center}
\end{figure}

The measured distribution of PMT gains in \minerva is shown in Fig. \ref{fig:allGains}.  
Figure \ref{fig:gainChangeSig} shows the statistical significance of gain fluctuations
in consecutive measurements, demonstrating that daily fluctuations in the measured gains are mostly
accounted for by the measurement's statistical uncertainty of 3\% to 5\%.
Additionally, this gain measurement procedure is used to set the operating HV
for each PMT box at the start of the data taking.  
The HVs are tuned such that the lowest-gain pixels have uniform 
gains across the detector.  This is done by measuring the gain of the eight
lowest-gain pixels on each tube, then adjusting the HV to give an average
gain of 4.38$\times$10$^5$.

\begin{figure}[t]
\includegraphics[width=0.9\textwidth]{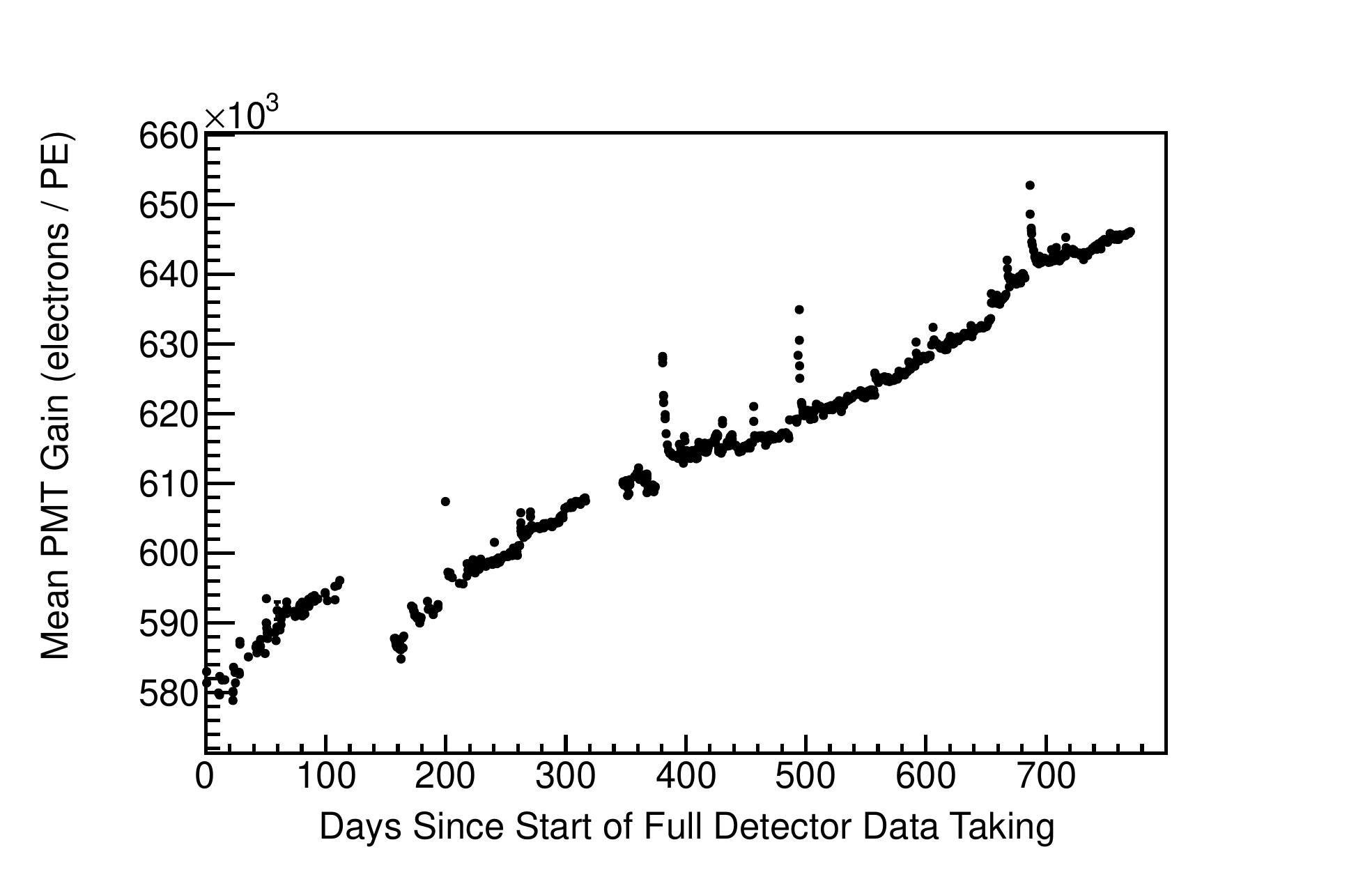}
\caption{ Average PMT gain vs. time.  The nominal HVs were adjusted between days 120 and 160, 
causing a decrease in the mean gain. 
The gradual increase in gain over time is due to aging; the nature of the aging is dependent upon operating conditions and initial conditioning.  
The sharp peaks, e.g. day 380, correspond to times when the HV was turned off and on again.  
The gains typically stabilize within 24 hours after HV is re-applied. }
\label{fig:meanGainVTime}
\end{figure}

The results of the HV tuning procedure can be seen in Fig.~\ref{fig:allGains}, 
as this data was collected shortly after the tuned 
target HVs were applied.  The gains vary over the two years of data collection as shown in Fig.~\ref{fig:meanGainVTime}.  

\makeatletter{}
\subsection{Scintillator Plane Alignment}
\label{sec:alignment}

The active \minerva\ detector is built of 120 mechanically-independent modules, 
hung in series upon a rack. Small variations in the relative module positions are inevitable. 
 The tracking and electromagnetic calorimeter modules have two active scintillator planes while the hadronic calorimeter modules have only one.
Co-modular 
planes, although sharing support from one steel frame, can be perturbed independently 
by stresses and strain within the frame and manufacturing tolerances. Typical offsets are comparable to or less than the strip widths of 33 mm.
These offsets are determined by an alignment procedure and corrected for in event reconstruction.

The plane-based alignment procedure treats the 127 individual strips in a plane
as a rigid unit. The first-order effect is an offset in  the coordinate that corresponds
to the direction of measurement in that plane.
 This can be 
expressed as a translation in the direction of the measurement. In addition to the translational parameter, a parameter for rotation
about the $z$-axis is also introduced. Rotations are found to be on the order of a few milliradians. The translation and 
rotation parameters characterize the 
plane alignment model completely. This alignment is not sensitive to the component 
perpendicular to the measurement direction or to other 
degrees of freedom of the plane, such as rotation about the $y$-axis.
 The shift is measured by comparing the energy deposited by tracks of through-going 
rock muons up to their points of intersection with the 33 mm base of the strip. 

\begin{figure}[htbp]
\begin{center}
\includegraphics[width=0.9\textwidth]{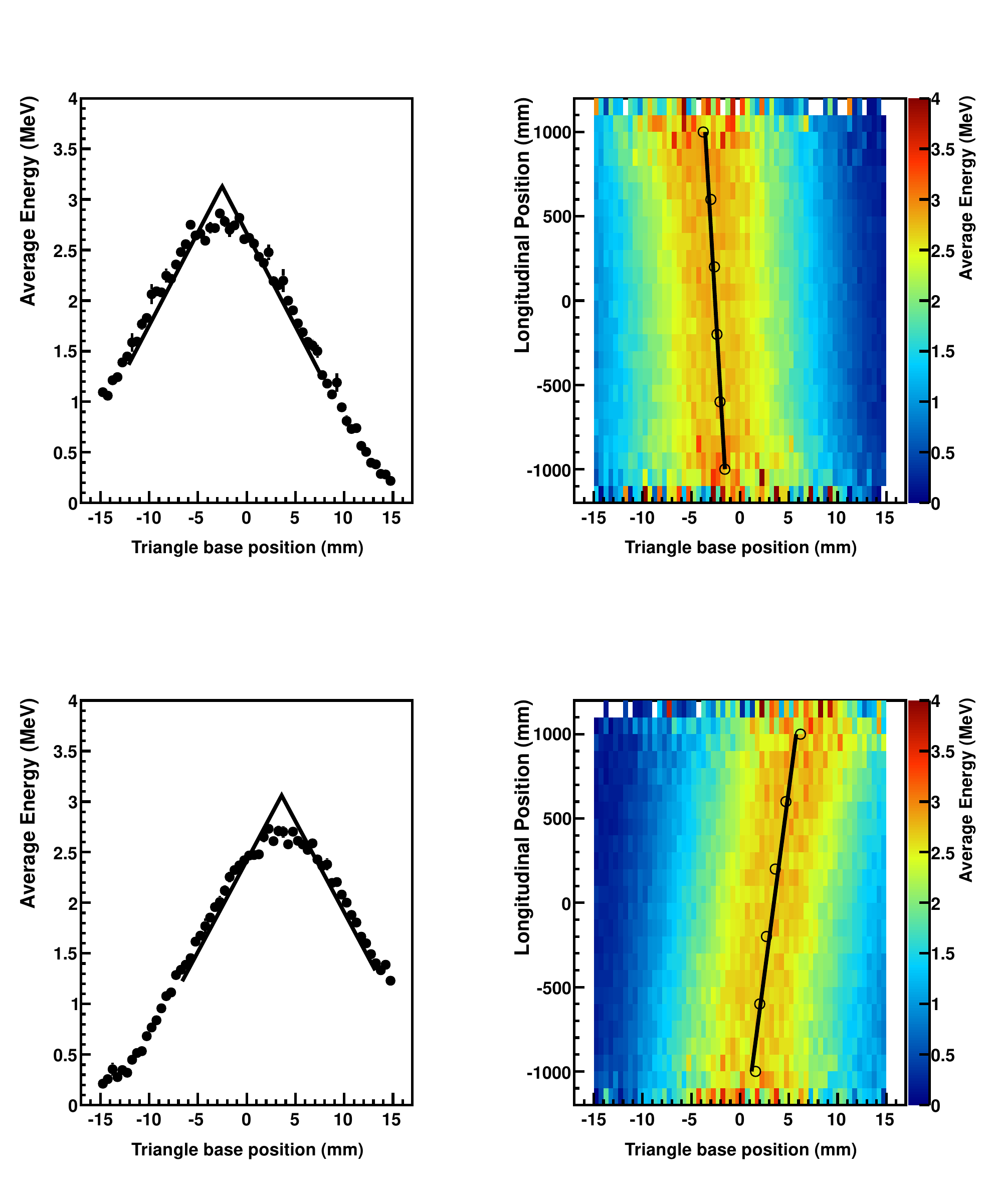}
  \caption{\small The alignment fits for module 50, plane 2 (top) and module 61, plane 1 (bottom).
At left, the average energy as a function of triangle base position is fitted to yield the 
shift parameter. The rounded tip is due to the fiber hole in the scintillator. At right, the rotation 
parameter is determined from the dependence of the shift on longitudinal position.}
  \label{fig:alignplot}
  \end{center}
\end{figure}

The energy is corrected for 
normal incidence such that the maximum occurs at the strip center where the 
muon path length is greatest.  The average energy in bins of triangle base position 
is fitted to the shape of the strip as shown in Fig.~\ref{fig:alignplot}; the 
shift parameter is the peak of the fit.  This procedure is performed in 6 bins 
along the strip.  The shift is then plotted as a function of longitudinal position 
and the rotation parameter is extracted from the slope.
One iteration of this procedure is sufficient to align all the planes. 
The residual alignment parameter uncertainties are $< 1$ mm and $< 1$ mrad. No further reduction in residuals is obtained
if a second iteration of alignment is performed.

\makeatletter{}
\subsection{Relative Strip-to-Strip Response Variations}
\label{sec:s2s}

Variations in light level between ID strips can be caused, 
for example, by differences in the composition of batches of scintillator, 
by air bubbles in the epoxy used to fill the fiber hole, or by couplings between the 
optical fibers and photomultiplier tube.  These variations are corrected by 
applying a multiplicative constant to strip energy deposits so that the response is uniform throughout the detector.
The constants are determined from the path length-normalized peak energy deposited by through-going rock muons.  The calibration consists of three stages.  Initially, an iteration of strip-by-strip corrections is generated using the truncated mean energy \cite{auchincloss}.  Secondly, dead channels are identified and a second iteration of strip-by-strip corrections is produced.  Finally, the peak energy is equalized plane by plane.  The constant for each strip is the product of the output of each iteration, normalized so that the average constant is 1.0 over the entire inner detector.

The calibration is performed in time intervals such that every strip is intersected by at least several hundred rock muon tracks in a given period.  This imposes a requirement of hundreds of thousands of rock muons for adequate statistics.  The boundaries of the intervals are defined by hardware changes in the detector, such as when a front end board or PMT is replaced.
The energy spectrum of the incident rock muons is not measured.  
However, the peak energy loss per unit path length for a minimum ionizing particle is known
 to be a slowly varying function of the energy of the particle.  
Fitting for the peak requires higher statistics than can be obtained
 for each strip in a single interval, so this is done only on a plane-by-plane basis.  
For the strip-to-strip correction, the truncated mean is used as a proxy for the peak.  

Differences in the relationship between the truncated mean and the peak could arise due to 
differences in the shape of the energy distributions between channels.  Scintillator aging and absorber 
effects are expected to be among the leading causes of such shape differences.  
Scintillator within a plane was manufactured and installed at the same time, 
so aging effects might exist between planes but are uniform across strips within a plane.  
Absorber effects are uniform across a plane in the downstream calorimeters.  

Non-uniform absorbers exist in the upstream nuclear target region and in the side ECAL.  
This could introduce differences in the shape of the response within a plane.  
The size of this difference is estimated by comparing the size of the fitted peak correction of Eq. (\ref{eq:pk}) between the central tracking region and downstream ECAL region.  This correction is essentially the amount by which the truncated mean overestimates the peak; it is 1.6\% higher on average in the 
ECAL than in the tracking region.  A similar difference might exist in the side ECAL but is not taken into account.
The truncated mean is computed iteratively.  The mean for the first iteration is the full mean of events from 0 to 20 MeV per centimeter path length.  In each successive iteration, the mean is calculated using events that fall between 50\% and 150\% of the mean from the previous iteration.  Eight iterations are used; however the procedure typically converges within four iterations.

Events for which there is reconstructed path length but with null energy are counted separately.  A channel is considered dead when zero-energy tracks represent more than 30\% of all events.  In addition to removing dead channels in which the zero-energy percentage is near 100\%, this cut also eliminates extremely low-light channels where the expected number of photoelectrons from a minimum-ionizing hit has a high Poisson probability of fluctuating to zero.  This effectively places an upper bound on the correction
constant at about 5.0. Any strip that would require a correction larger than this is treated as a dead channel and ignored by the reconstruction in both data and simulation.  It is also required that the RMS of the energy in a strip not exceed four times the truncated mean.  This cut addresses rare channels where the energy response changes dramatically over a calibration interval.  

The constant $C_{i}$ for strip $i$ is 

\begin{equation}
\label{eq:tm}
C_i = \frac{\frac{1}{x_i}}{\frac{1}{N} \sum\limits_{j} \frac{1}{x_j}}
\end{equation}

\noindent where $x_{i}$ is the truncated mean energy in strip $i$, and $N$ is the number of good channels in the inner detector. The sum in the denominator is over all good channels, indexed by $j$.  This definition guarantees that the average constant is 1.0, so that application of the constants will correct for strip-to-strip energy differences without affecting the overall energy scale.

The fitted peak correction is performed only after data is processed with the first iteration of strip-to-strip constants based on the truncated mean.  This ensures that the energy distribution in a plane is not due to large variations between constituent strips.  The correction factor $C^j$ for plane $j$ is

\begin{equation}
\label{eq:pk}
C^j = \frac{\frac{E^j}{p^j}}{\frac{1}{n} \sum\limits_{k} \frac{E^k}{p^k}}
\end{equation}

\noindent where $E^{j}$ is the truncated mean energy averaged over a plane, $p^{j}$ is the fitted peak energy for a plane, $n$ is the number of planes, and the sum in the denominator is over planes.  The applied strip-to-strip constant is the product of the constant from Eq. (\ref{eq:tm}) for two iterations and the plane-to-plane constant from Eq. (\ref{eq:pk}).

\begin{figure}[t]
\begin{center}
\includegraphics[width=0.9\textwidth]{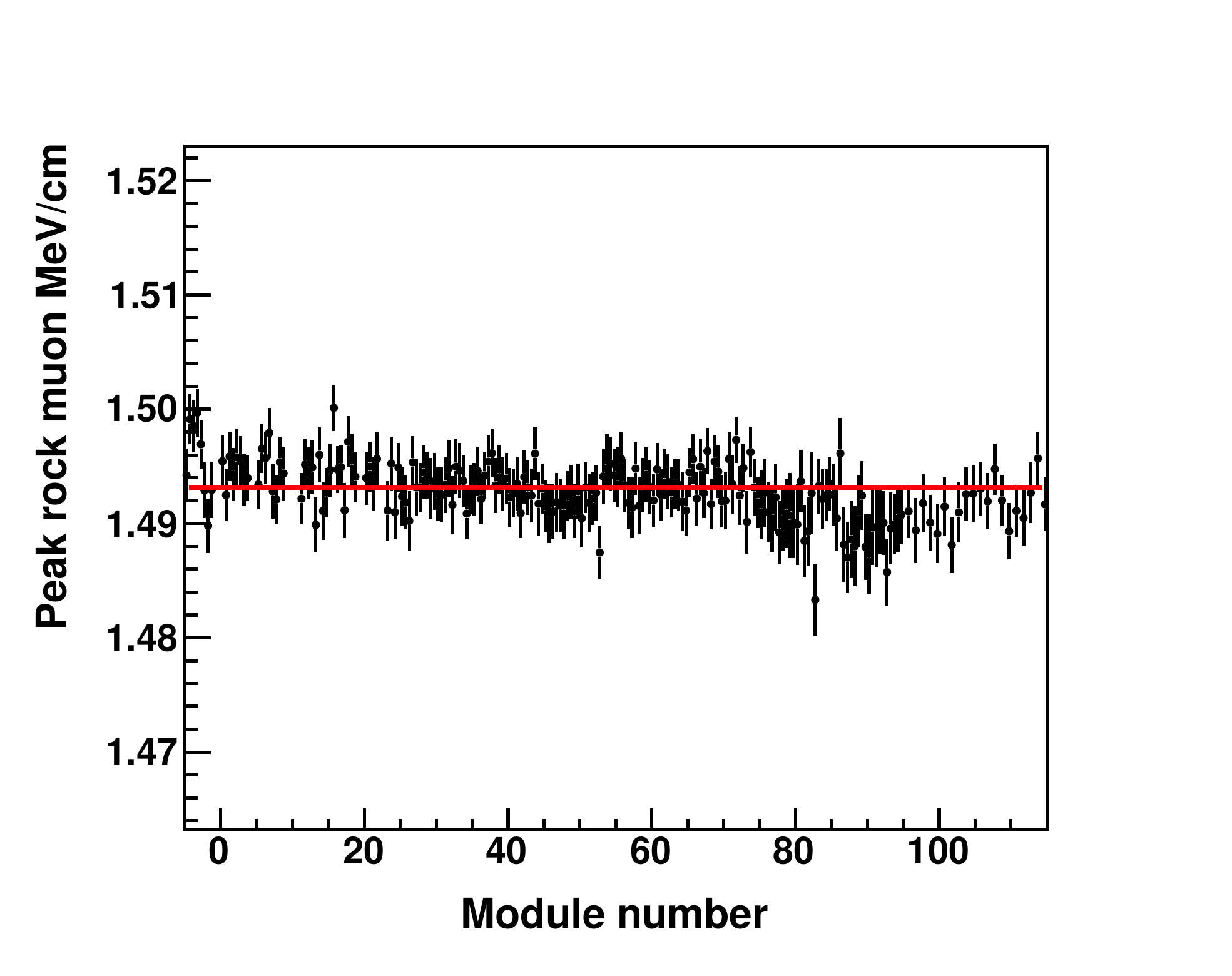}
  \caption{\small The peak energy per unit path length is fitted for each plane.  The resulting peaks, with uncertainties from the fit, are fitted to a linear distribution with zero slope since the plane-to-plane energy is expected to be flat.}
  \label{fig:p2p}
  \end{center}
\end{figure}

After applying the calibrations, the plane-to-plane peak energy is roughly uniform over the modules, as shown in Fig. \ref{fig:p2p}.  
Non-statistical fluctuations, such as in the ECAL region from module 85 to 95, could be reduced by further iterations of fitting but are already approximately the size of the fit uncertainty after one iteration.

\makeatletter{}
\subsection{Absolute Energy Scale Determination Using Minimum Ionizing Particles}
\label{sec:meu}

The absolute energy scale of \minerva is calibrated using the well understood muon energy loss
in the active scintillator.  The absolute energy scale 
is described in terms of a ``muon equivalent unit'' (MEU factor) of energy deposited in a plane of scintillator.  
This factor converts attenuation and strip corrected number of photoelectrons to an energy.  The MEU factor is tuned 
using clusters of energy depositions (cluster energies) from rock muon tracks in data and the simulation.  
Details of cluster energy formation are presented in Sec. \ref{sec:cluster_formation}.

The data sample consists of rock muons whose track in \minerva has a matching track in MINOS.  The muon momentum at its entrance point to \minerva is estimated by correcting the MINOS momentum reconstructed by range or curvature for energy loss in \minerva. The reconstructed position and momentum vector is used as input to a data-driven simulation sample.

\begin{figure}[htb]
\center
\includegraphics[width=0.48\textwidth]{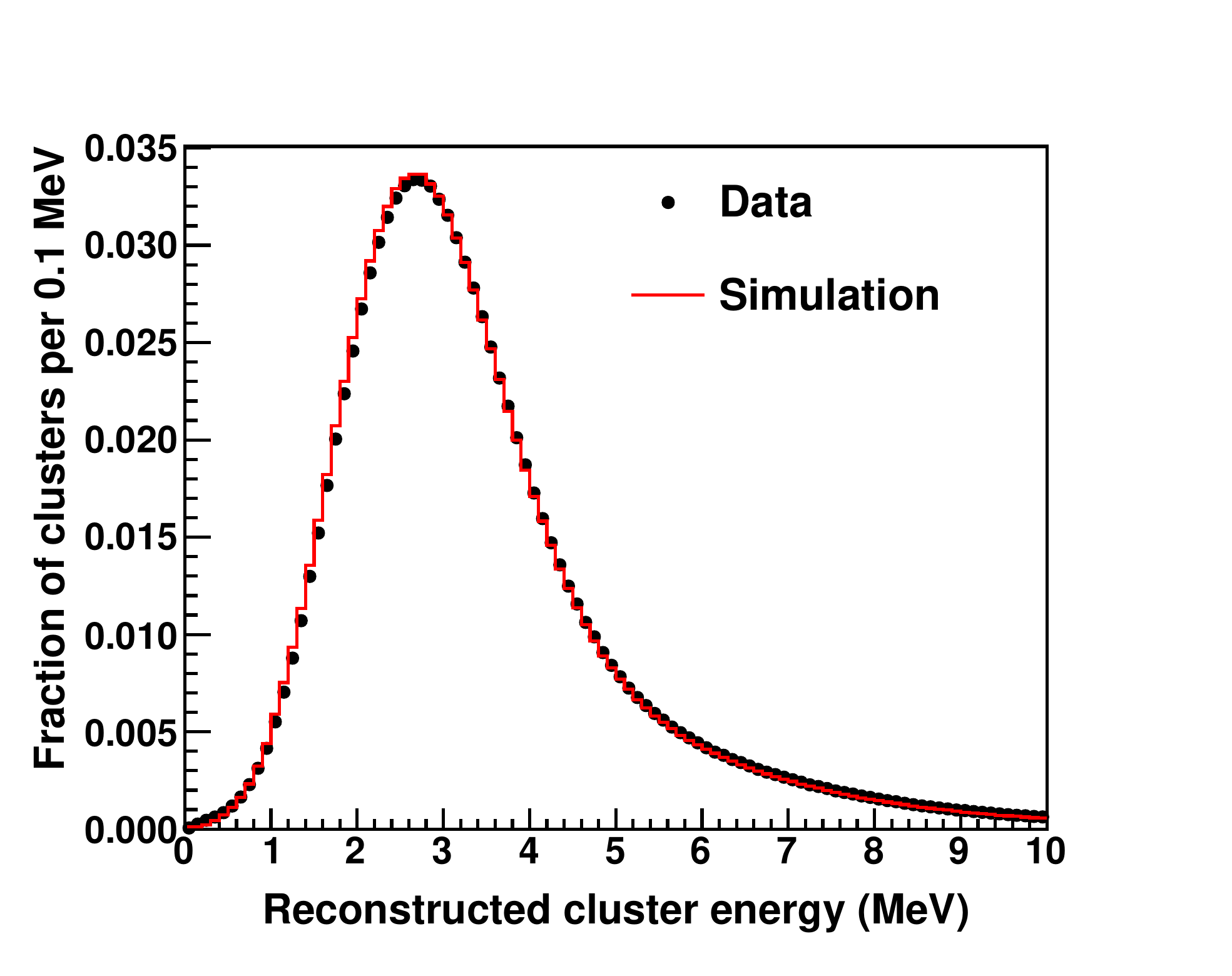}
\includegraphics[width=0.48\textwidth]{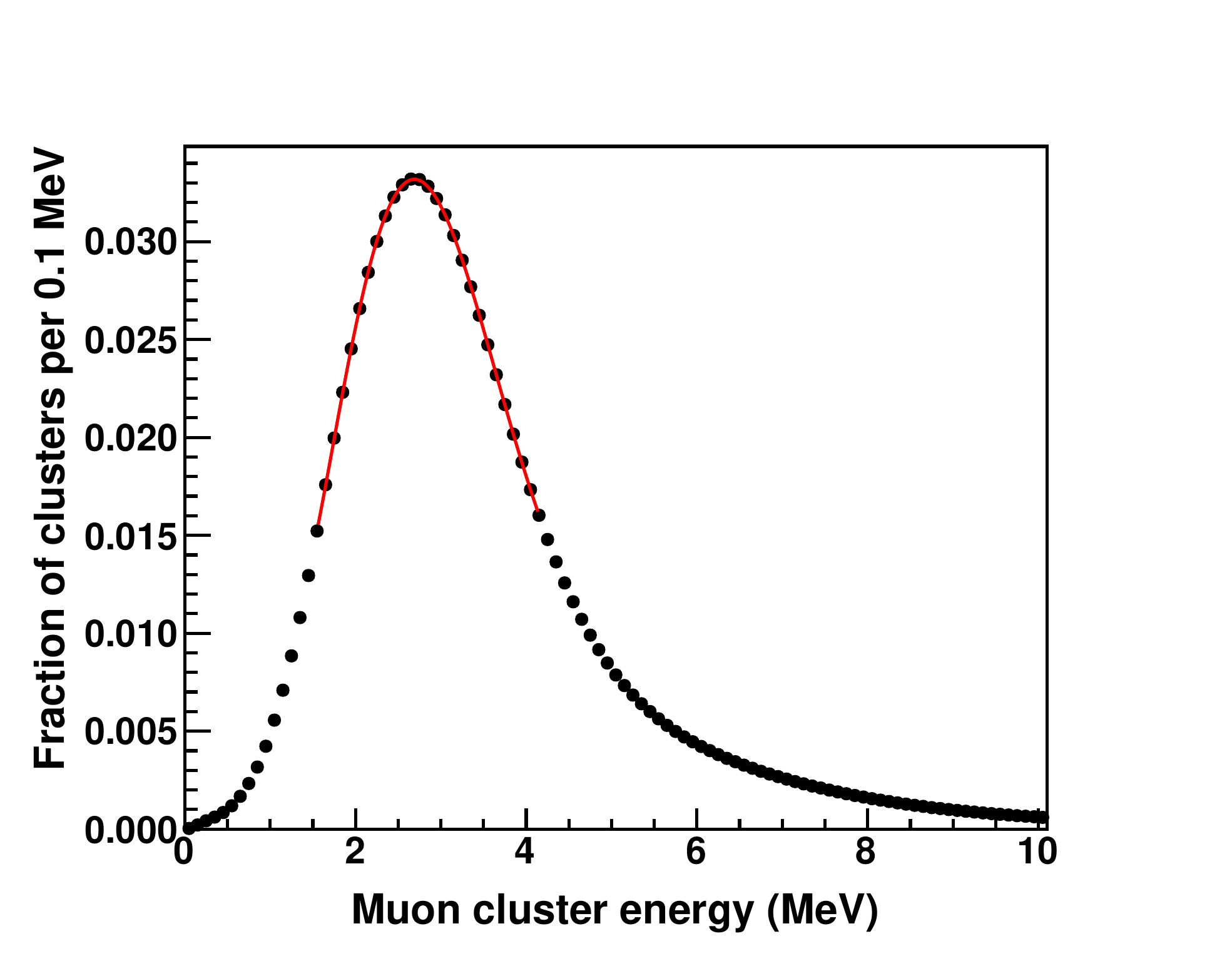}
\caption{The energy distribution of clusters along a muon track in data and the simulation (left) and the resulting fit to the peak of the data distribution (right).}
\label{fig:data_reco_cluster_energy_tuning}
\end{figure}

\begin{figure}[htb]
\center
\includegraphics[width=0.9\textwidth]{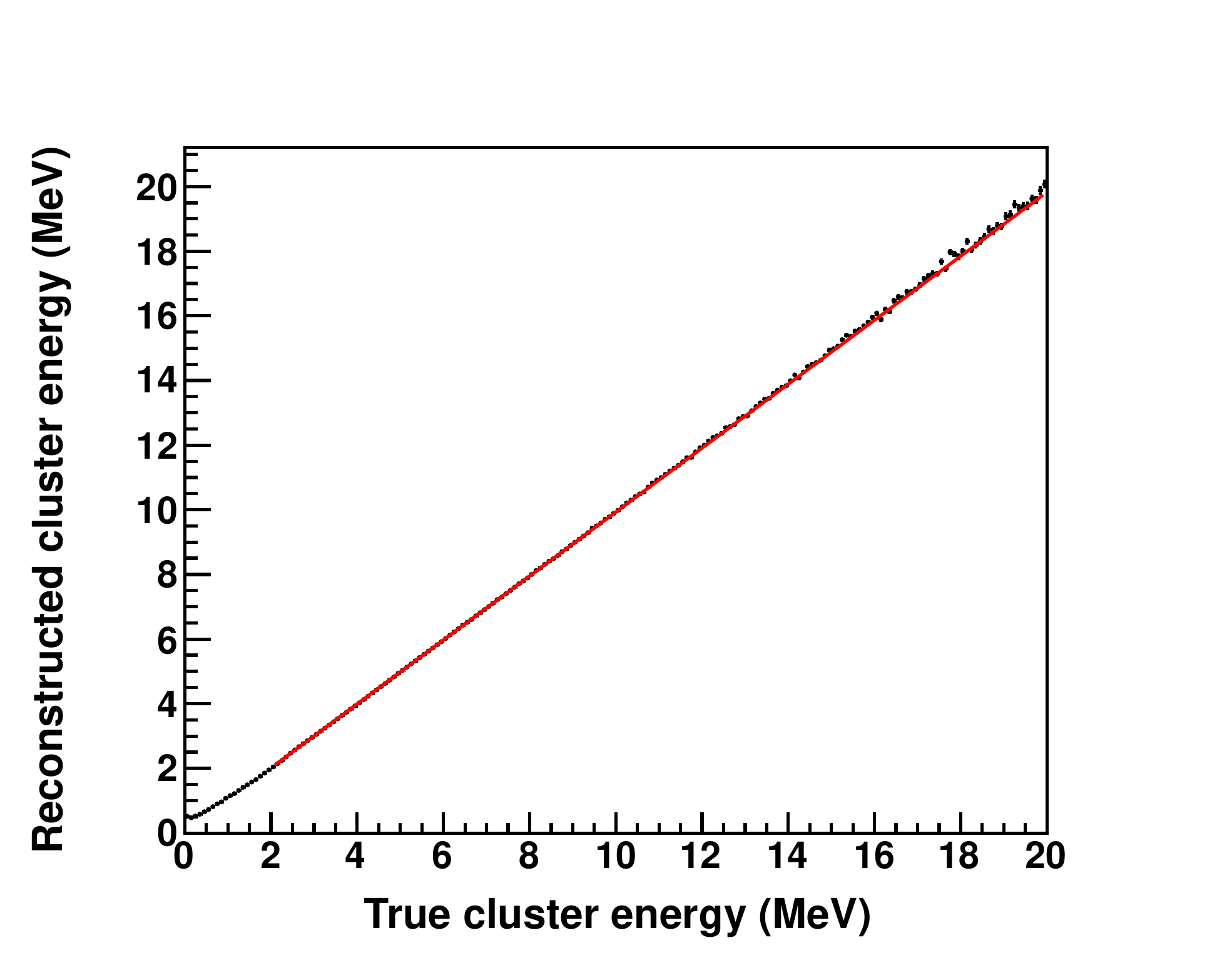}
\caption{Plot of the reconstructed cluster energy as a function of the generated cluster energy in the 
simulation, and the resulting linear fit.}
\label{fig:mc_reco_vs_true_cluster_energy_tuning}
\end{figure}

The data are first analyzed with a trial MEU factor.  The reconstructed cluster energy is the product of 
the attenuation and strip response-corrected PE and the trial MEU factor.  The peak reconstructed cluster 
energy in data and simulation (Fig.~\ref{fig:data_reco_cluster_energy_tuning}) is determined by fitting the 
region of the distribution above half-height with a fifth-order polynomial.  
Separately, the reconstructed cluster energy in the simulation is compared to the true energy.  
The true cluster energy is the total energy in the active scintillator of the strips 
composing the cluster, according to the GEANT simulation.  The reconstructed and true peaks are fit to 
a straight line as shown in Fig.~\ref{fig:mc_reco_vs_true_cluster_energy_tuning}.  The dominant uncertainty in the energy scale is due to  uncertainty in  the per plane thickness of the active scintillator.  Given the uncertainties in the material assay described in Sec. \ref{sec:tracking_modules}, the MEU uncertainty is 2\%.  

The trial MEU factor is corrected by the ratio of simulated to data reconstructed energy and by the slope of the simulated reconstructed 
to simulated true energy to give the final MEU factor.
The tuning uses the peak reconstructed cluster energy to minimize absorber effects and muon energy loss.  Absorber effects, particularly delta ray propagation and absorption, vary by subdetector due to the passive materials.  Delta rays contribute primarily to the high-side tail of the cluster energy distribution, and sampling the peak minimizes this contribution.  

Reconstructed cluster energy in the
simulation includes simulated detector effects that smear true cluster energy.  Detector smearing shifts the peak of the true cluster energy distribution while preserving the mean.  It gives the correct absolute energy scale by correcting the reconstructed cluster energy mean from the simulation 
to the true cluster energy mean using the slope of reconstructed versus true cluster energy in the simulation.

\begin{figure}[htb]
\center
\includegraphics[width=0.9\textwidth]{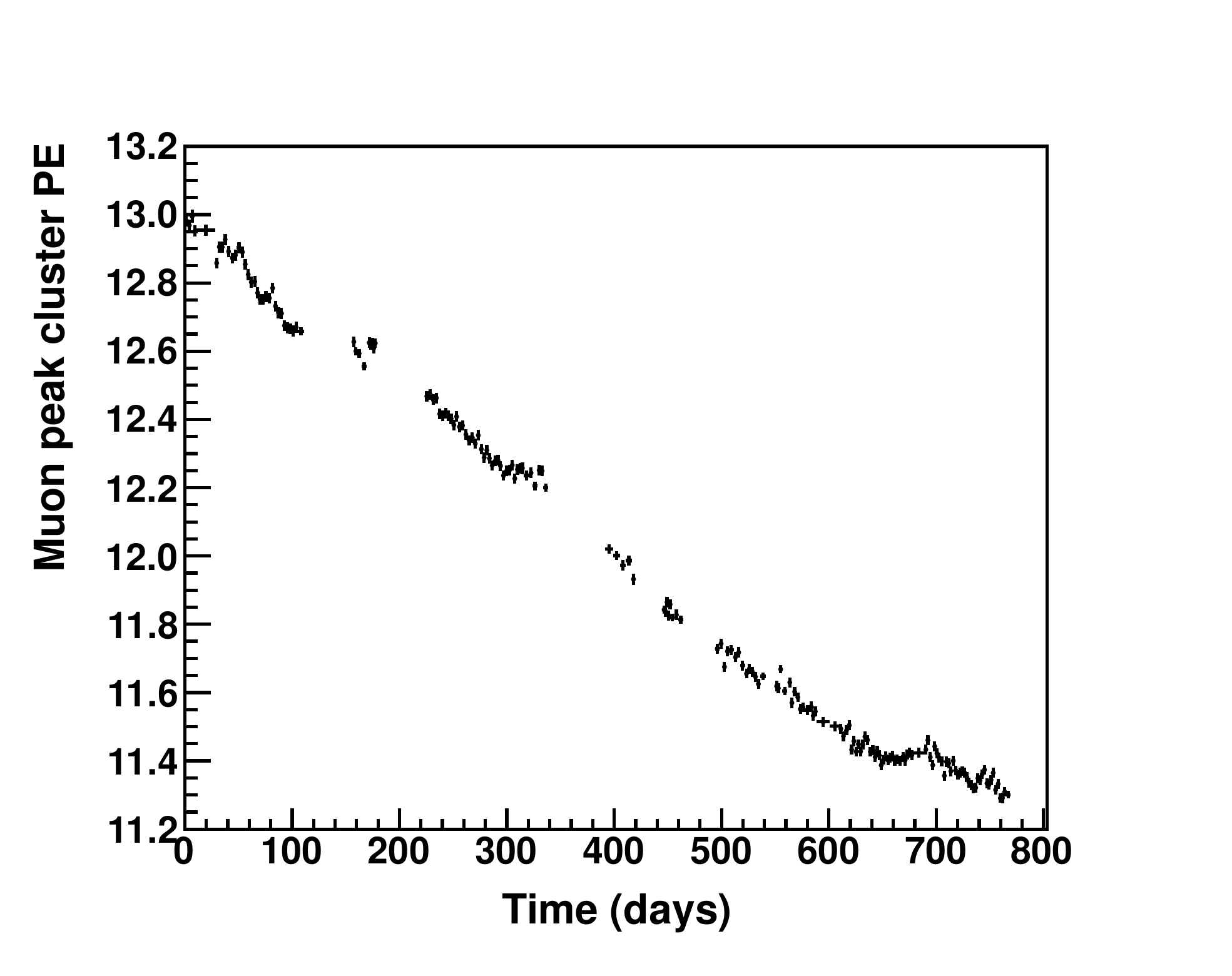}
\caption{Light level in the \minerva detector as a function of time, as demonstrated by the number of PE in the peak of the muon cluster distribution.  }
\label{fig:data_cluster_pe_vs_time}
\end{figure}

The light output of the scintillator  is found to vary with time due to scintillator aging and
 environmental conditions of the detector hall.  Figure~\ref{fig:data_cluster_pe_vs_time} 
shows the peak cluster PE as a function of time in the rock muon sample over the full detector LE data set.  
 Around day 600 a new cooling system was installed in the detector hall; subsequently the ambient temperature was reduced by about $6^\circ$ C.  A modest improvement in the peak cluster PEis indicated by the light level evolution around that time, as displayed in Fig.~\ref{fig:data_cluster_pe_vs_time}.
Time variation of the light output is accounted for in the absolute energy scale by dividing the data set 
into two-day intervals and tuning a MEU factor for each interval.  The resulting peak-calibrated reconstructed 
cluster energy as a function of time for rock muons in data over the full detector LE data set is constant.   

\makeatletter{}
\subsection{Timing Calibration}
\label{sec:timing_calibration}

A timing calibration is performed to correct for transport time in the optical fiber, time slewing, and channel-to-channel time offsets.  Time slewing is due primarily to the scintillator decay times and is a function of the PE yield within a scintillator stripo.  Channel-to-channel time offsets include cable delays between FEBs along a chain and time offsets between chains.  The time offsets between channels on the same FEB are negligible.  

\begin{figure}[htbp]
\center
\includegraphics[width=0.9\textwidth]{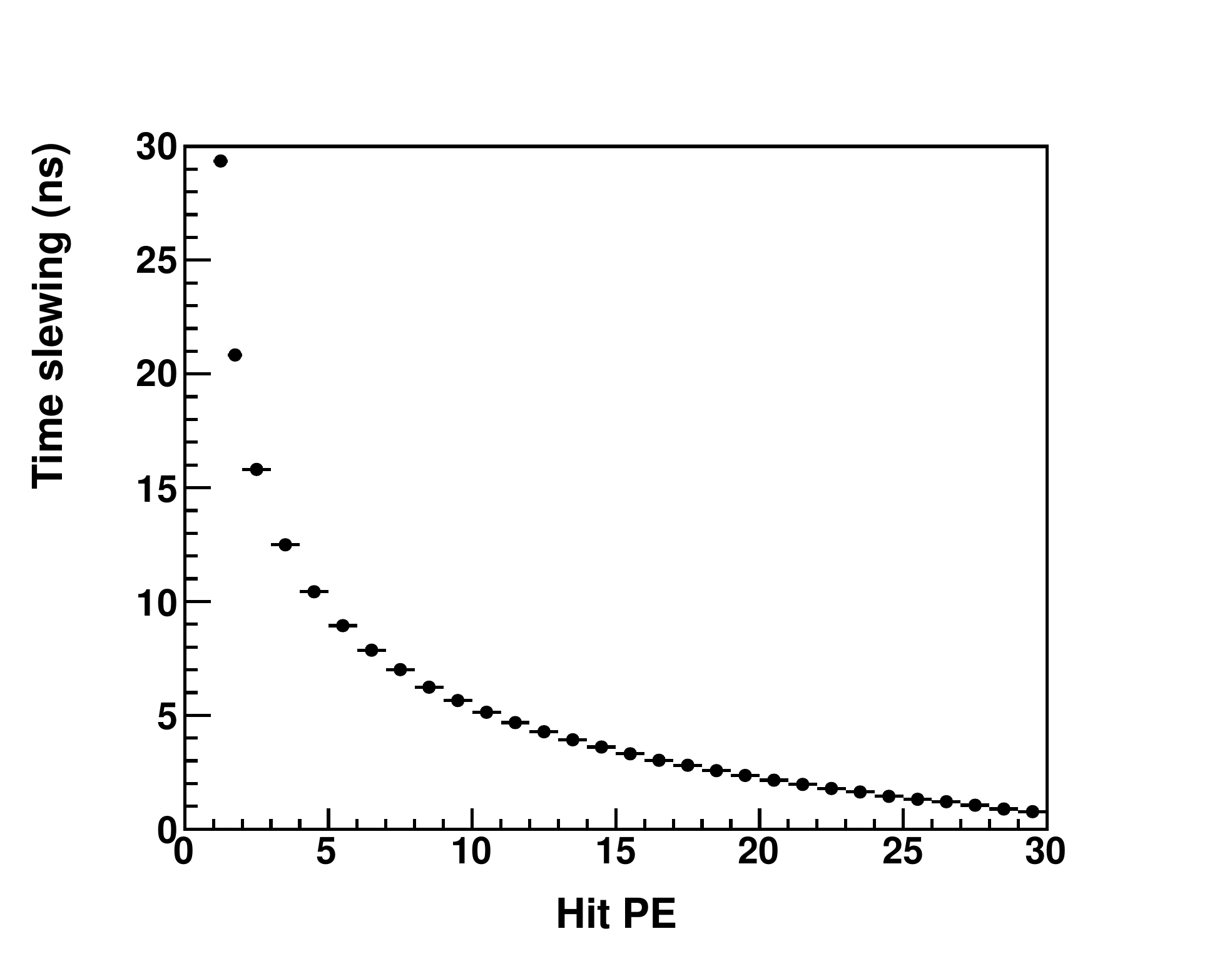}
\caption{Measured time slewing vs. hit PE along rock muon tracks.}
\label{fig:time_slewing_vs_pe}
\end{figure}

Time slewing and the channel-to-channel offsets are measured in an iterative procedure 
using hit time and PE from identified rock muon tracks.  They are measured relative to the truncated 
mean hit time along the rock muon track, which is corrected for muon time-of-flight, 
transport time in the optical fiber, and time-slewing and channel-to-channel offsets measured in the previous iteration.  
Transport time in the optical fiber is corrected using the fiber length and the speed of light in the fiber.  
The measured time slewing as a function number of PE in a single strip (''hit PE'') (Fig.~\ref{fig:time_slewing_vs_pe}) is parameterized by a third order polynomial in $1/ \sqrt{\hbox{hit PE}}$, which is used to correct for time slewing in data.  One time offset is measured for each group of channels read out by the same high-gain TriP-t chip.  This accounts for channels that have low statistics and takes advantage of the small time offsets between channels on the same FEB.  The time offset between channels on different FEBs is observed to be as large as 30 ns.  Since hardware swaps change the channel-to-channel time offsets, the timing calibration is performed after each hardware swap.

\begin{figure}[htbp]
\center
\includegraphics[width=0.9\textwidth]{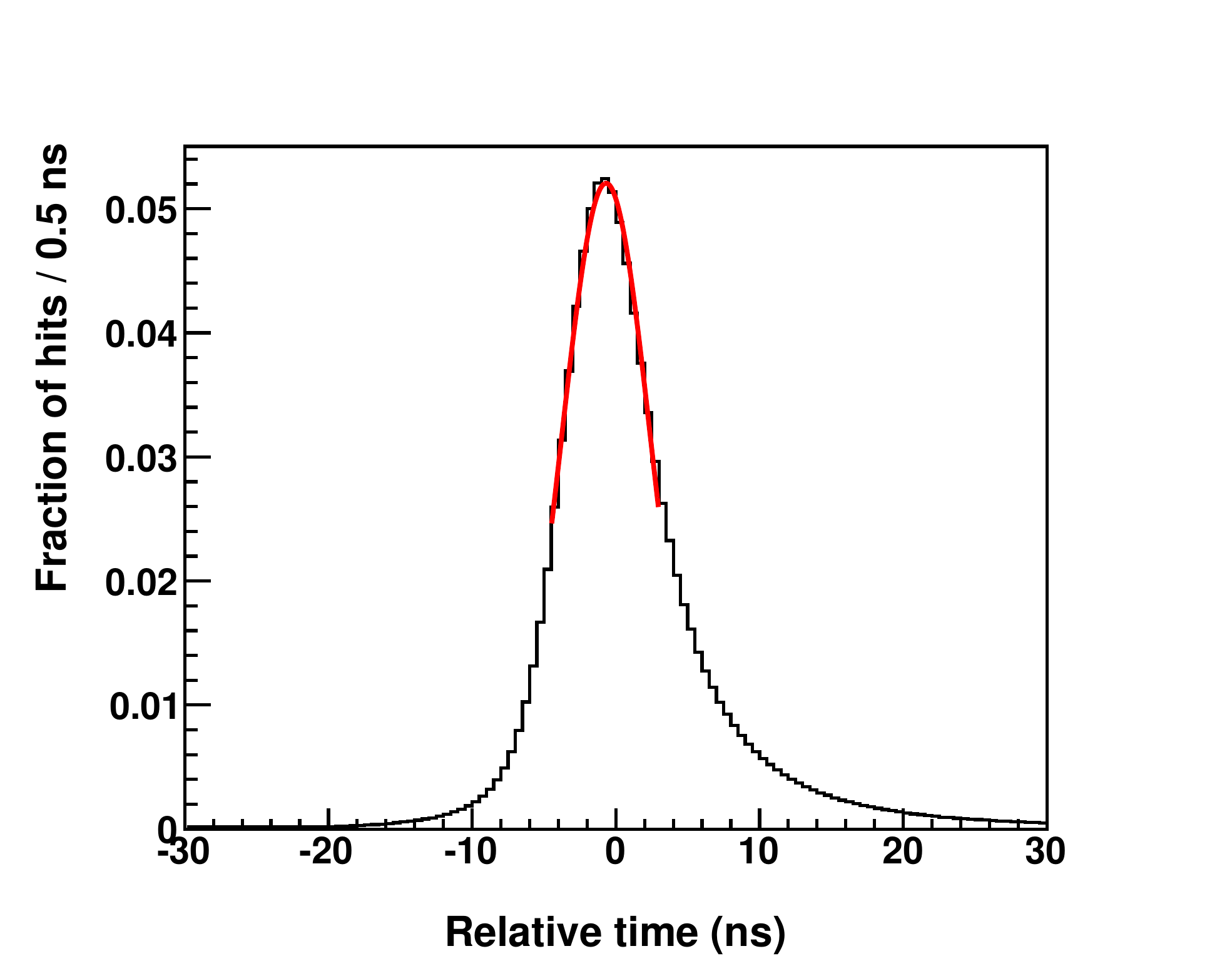}
\caption{Calibrated hit time along rock muon tracks relative to the truncated-mean calibrated hit time along the track.}
\label{fig:post_cal_exact_cor_rel_time}
\end{figure}

The calibrated time resolution is determined from calibrated hit times along rock muon tracks.  Figure~\ref{fig:post_cal_exact_cor_rel_time} shows the calibrated hit time along rock muon tracks relative to the truncated-mean calibrated hit time of the track.  A Gaussian fit to the region above half-height gives a width of 3.0 ns.  The tails are asymmetric, especially for low pulse-height hits.  This is because hit times are set by the first photon to liberate
 a photoelectron. Low pulse-height hits with a small number of scintillation photons can fluctuate at late times due to scintillator decay times.  This effect becomes insignificant at higher pulse heights where there are many scintillation photons.

\makeatletter{}
\subsection{Cross-talk Measurements}
\label{sec:cross_talk}

A number of processes can cause light incident on one PMT channel to produce a signal in another channel.
These are collectively known as  ``cross-talk.''   While these can, in principle,
 be differentiated by tests on the bench, once the detector components are assembled and installed 
it is virtually impossible to separate them from one another with any significant confidence, 
particularly at large pulse-heights.  As discussed in Sects.
\ref{sec:pmt_testing} and \ref{sec:feb_response} the 
dominant types of cross-talk in \minerva are optical (fiber-to-PMT coupling) and PMT internal 
(dynode chain).  Effects other than the readout electronics cross-talk discussed in 
Sect. \ref{sec:feb_response} will be discussed here.

The ideal probe for a measurement of either type of cross-talk in the detector would be one in 
which individual pixels were illuminated with a well-defined light pulse.  
Unfortunately, once a PMT is mounted on the detector
there is no system available to \minerva\  that can accomplish this goal.
The LI system discussed in Section \ref{sec:light_injection} cannot be used because 
it illuminates all pixels at once.  The next best option is to use
data generated by neutrino interactions.  For this measurement, we use the data-set from
rock muons described in Section \ref{sec:meu}.

\begin{figure}[htb]
\center
\includegraphics[width=0.9\textwidth]{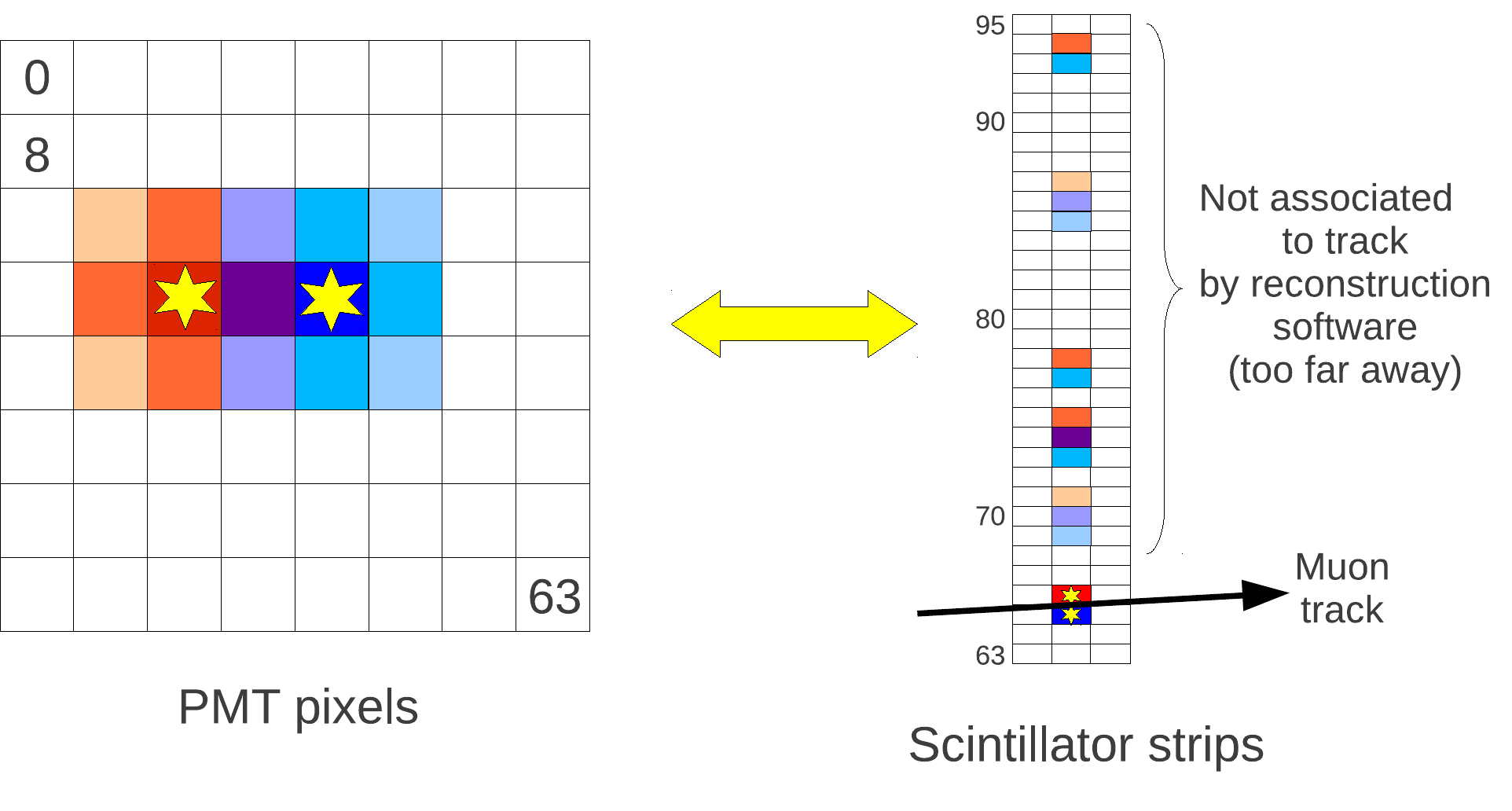}
\caption{Schematic depiction of cross-talk on the PMT face maps to scintillator strips.  The darkest blue and red (stars on the PMT diagram; strips 65 and 66 in the scintillator sketch) are the original signal from a muon track; the cross-talk energy is colored according to which original signal hit it will be associated with by the algorithm described in the text (darker means stronger cross-talk).  Purple represents cross-talk that will be associated with either hit at random since it is ambiguous.}
\label{fig:crosstalk-id-pmtface}
\end{figure}

Hits within a rock muon's time slice are classified as signal or noise based on 
whether or not they have been associated with the track by the track reconstruction software.  
The fiber weave 
described in Section \ref{sec:pmts} moves cross-talk hits far enough 
away from the track that they are not associated with it.
Cross-talk hits can be further distinguished from other noise by assuming that the hit
occurs in the same PMT as on-track activity, and associating each cross-talk hit 
to the on-track hit that is nearest to it on the PMT pixel grid.  A sketch of how this 
process would work for a typical muon event is shown in Fig. \ref{fig:crosstalk-id-pmtface}.
				
Once hits have been identified as signal or cross-talk and the rest discarded, 
an average cross-talk fraction $f_{xt}$ for the PMT is defined as the ratio
of the energy of cross-talk hits to the energy of the on-track hits.
Various permutations of this metric are also
used, most notably the ``nearest neighbor'' pixel cross-talk  $f_{\mathrm{xt}, \mathrm{NN}}$
average for each PMT,
because the strongest cross-talk is generally generally between nearest-neighbor pixels.
				
\begin{figure}[htb]
\center
\includegraphics[width=0.8\textwidth]{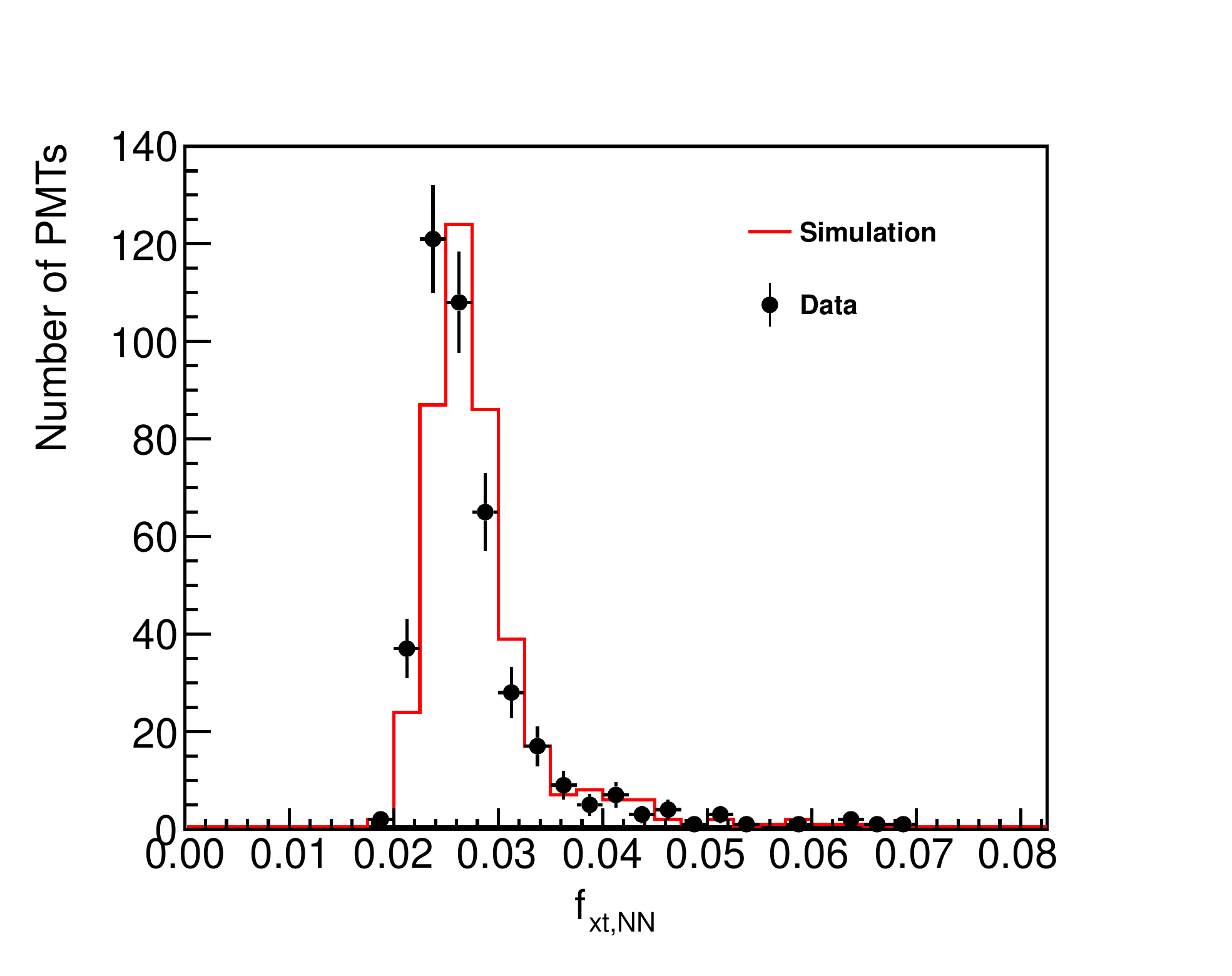}
\caption{Distribution of nearest-neighbor pixel cross-talk, $f_{\mathrm{xt}, \mathrm{NN}}$ in data (solid points) compared with simulation
(red histogram).  A detailed description of  $f_{\mathrm{xt}, \mathrm{NN}}$ is given in the text.}
\label{fig:crosstalk-frac-sim-vs-data}
\end{figure}

The measured values of $f_{\mathrm{xt}, \mathrm{NN}}$ for PMTs currently in use in \minerva are shown 
as the black points in Fig. \ref{fig:crosstalk-frac-sim-vs-data}.  The vast majority of  PMTs 
have total nearest-neighbor cross-talk fractions below 4\%.
Comparisons with  the detector simulation can also be made using this method.  
The red curve in Fig. \ref{fig:crosstalk-frac-sim-vs-data} depicts the simulation's prediction for the 
optical cross-talk component only (the dominant mechanism and the sample used
to tune the simulation).  Though the simulation implements a technique to individually scale the cross-talk for each simulated channel to that measured, agreement is modest at best.  The disagreement is likely
driven by individual channels in PMTs whose response deviates significantly from the underlying model 
used in the simulation (based on studies by MINOS~\cite{minosPMT}).  
Comparisons are made only in channels in the inner detector region, where the tracking software performs 
robustly.  Channels in the outer detector are scaled using the default parameters obtained from MINOS's study.
				
\begin{figure}[htb]
\center
\includegraphics[width=0.8\textwidth]{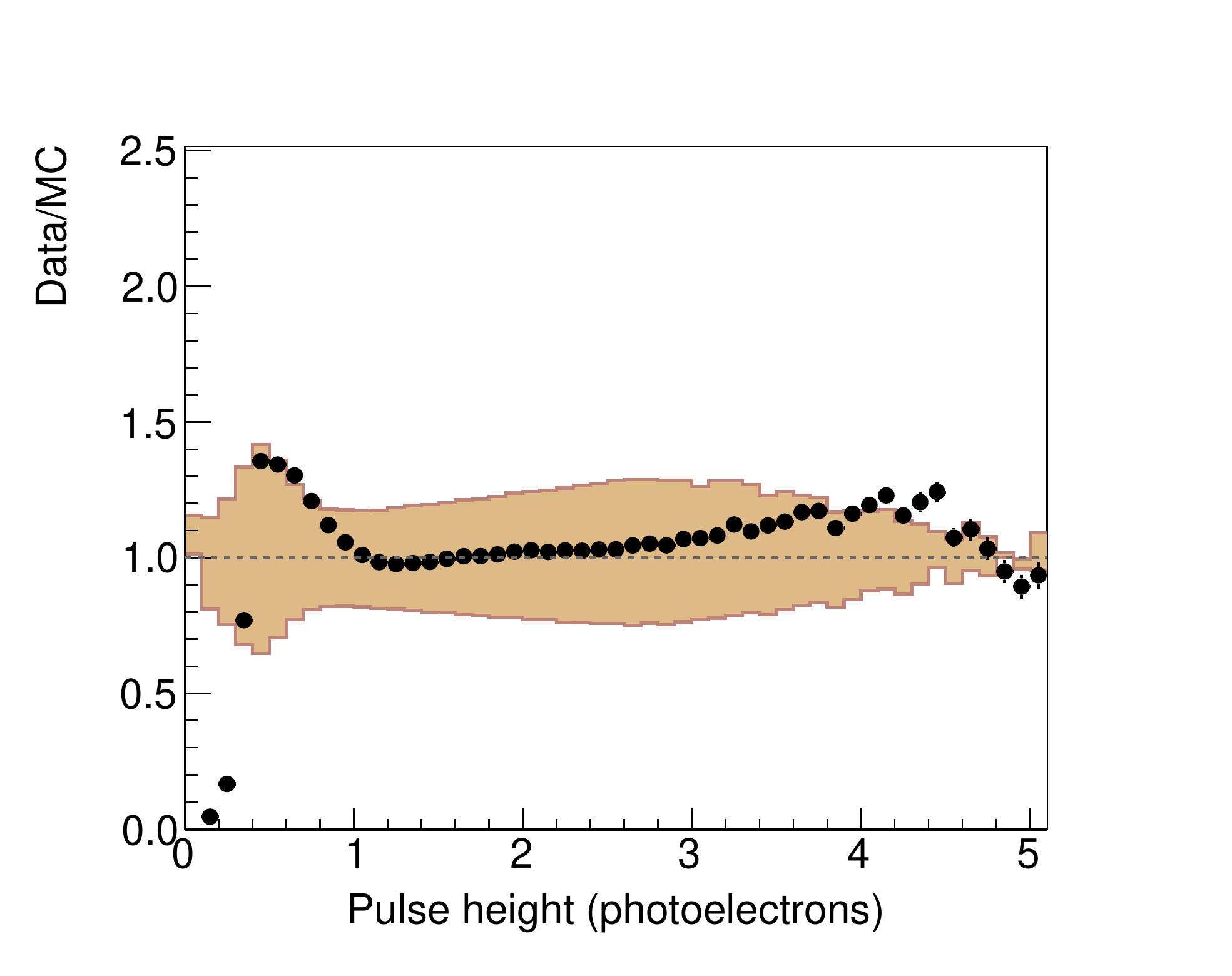}
\caption{Ratio of data to simulation for the photoelectron pulse-height spectrum of measured cross-talk hits.  The shaded region corresponds to the size of the systematic uncertainty assigned to cross-talk in neutrino analyses.}
\label{fig:crosstalk-pe-sim-vs-data}
\end{figure}

Though the simulation's per-PMT averages do not identically match the data, the individual pulse height spectra for simulated cross-talk hits do agree.  Figure \ref{fig:crosstalk-pe-sim-vs-data} illustrates this: within the 1-3.5 photoelectron range (where cross-talk is most important), the difference between simulation and data is typically less than 10\%.  Below this range, agreement is less important in practice because the electronic discriminator threshold for nearly all FEBs is roughly 0.8 PE.  Above this range, agreement is again less essential because the Poisson probability for a resultant cross-talk hit of this pulse height is extremely low. For example, the probability of a 4 PE cross-talk hit resulting from a 100 PE hit from neutrino-induced activity, assuming the nominal 1\% nearest-neighbor cross-talk leakage quoted above, is roughly 1.5\%; for a 25 PE hit, the probability drops to about 0.01\%.  Moreover, $\geq 100$ PE hits (corresponding to roughly 25 MeV of energy deposited in a single strip) are rare; they comprise less than 1\% of usable (i.e., discriminated) hits in typical LE data.
			
\makeatletter{}
\section{Muon Reconstruction}
\label{sec:MuonReconstruction}

This section describes the reconstruction of tracks which start in the 
\minerva detector and continue into MINOS. The tracks must first be found and 
associated (matched) in the two detectors. 
The momentum is measured using range or curvature in MINOS and
the sign of the track charge is determined using the direction of track curvature as described below.

\makeatletter{}
\subsection{Cluster Formation}
\label{sec:cluster_formation}

Due to the overlapping triangular profiles of the scintillating strips in the central detector, particles traversing a plane will normally pass through at least two strips in
the plane, and can induce activity in more than two strips, giving
a ``cluster'' of activity.
The first step in reconstruction of events is identification of clusters.  
Clusters
are formed from groups of hits directly adjacent to each other in a single plane,
within the same time slice.  Any strip which does not register hits between strips with valid hits leads to a new cluster
being formed. An isolated strip
without neighbors that has registered a hit is also identified as a cluster. 

A position is calculated for each cluster based on the energy deposition in
the strips.  The energy-weighted position is calculated using all hits contained within
a cluster. A time is also found for a cluster, where the time from the hit with the
most energy within the cluster is assigned as the cluster time.
The resulting clusters are classified by their composition. The types of clusters
are: {\em low activity}, {\em trackable}, {\em heavily ionizing}, {\em superclusters},
or {\em cross-talk}. 
\begin{itemize}
\item [(i) ] {\em Low activity} clusters are those with a total energy
deposit of less than 1 MeV.

\item [(ii) ] {\em Trackable} clusters meet the following requirements: 
total cluster energy between 1 and 12 MeV, four hits or fewer, and at least
one hit with more than 0.5 MeV. If more than two hits have a hit energy greater
than 0.5 MeV, they must be adjacent to each other.

\item [(iii) ] {\em Heavily ionizing} clusters must meet similar criteria
to trackable clusters: total cluster energy greater than 1 MeV and one to three hits
with energy greater than 0.5 MeV.  If two or three hits are greater than 0.5 MeV,
they must be adjacent to each other.  In addition, it must not be a trackable cluster.
Heavily ionizing clusters are important in forming high angle tracks.\\

\item [(iv) ]  {\em Superclusters} are those with 
 more than 1 MeV in energy that do not meet the criteria for either trackable
or heavily ionizing clusters. Any cluster with five or
more hits is automatically considered a supercluster.\\

\item [(v) ]  {\em Cross-talk} clusters are identified by 
inspecting the PMT pixels associated
with hits within that cluster. The PMT pixels associated with that cluster
are compared to PMT pixels associated with a particle interaction. If these cluster
PMT pixels are found to be directly adjacent to the pixels related to the particle
interaction, the cluster is considered to be a cross-talk cluster.\\

\end{itemize}

\makeatletter{}
\subsection{Track Reconstruction in MINERvA}
\label{sec:tracking}

A ``track'' is a reconstructed object that approximates a charged particle's
trajectory through the detector.  Typically, only one track is needed to reconstruct a particle trajectory, but 
multiple tracks may be necessary if the 
particle undergoes a large-angle scatter or else decays.  
A single track pattern recognition scheme is used multiple times to find all tracks in an event.

The track pattern recognition uses all ``seeds'' within a single time-slice, where a seed is 
a collection of three trackable or heavy-ionizing clusters
that meet the following requirements: no two clusters in the same plane; each cluster's plane is
in the same orientation (X, U, or V); clusters must be in consecutive planes, when the planes are
sorted by orientation and longitudinal position; and clusters must satisfy a fit to a two-dimensional line.
A single cluster may be used to make multiple seeds.  These requirements limit seeds to an angle of less 
than 70$^{\circ}$ with respect to the longitudinal axis, which in turn imposes a similar limit upon 
reconstructed tracks.

Seeds within the same plane orientation are merged to form 
track ``candidates''.  
The merger of two seeds requires that: the slope of the seeds' linear fits are consistent; the seeds share at 
least one cluster; and the seeds do not contain different clusters in the same scintillator plane.  
If the algorithm merges two seeds into a candidate, it will attempt to merge additional seeds 
to the candidate using the same criteria.  A seed may only be used by one candidate, therefore this stage of the algorithm is 
sensitive to the order of merging attempts.
After all candidates are built, they may be merged 
using criteria similar to those for seeds; they must have consistent 
fitted slopes and intercepts and may not contain different clusters in the same scintillator plane.  
Track candidates are not required to share clusters and therefore may contain gaps, i.e. tracks that
cross a scintillator plane without containing a cluster in that plane.  This allows a track candidate to accurately 
follow particle trajectories that intersect dead regions in the detector.

Two algorithms are used in succession to attempt to form 3D-track objects from track candidates.
The first algorithm examines all possible combinations of three 
candidates in which no two candidates share the same plane orientation.  
Such a combination of candidates are formed into a 3D-track if they overlap longitudinally 
and are mutually consistent with the same three-dimensional line.  By requiring a track candidate 
in each view, this algorithm is unable to form tracks that intersect fewer than 11 planes. 
The algorithm also searches for a particular topology in which a particle trajectory 
bends in only two views.
In this instance, the combination of candidates will fail the overlap requirement because 
the candidate in the view without a visible bend will be longer.  
If this topology is detected, the longer candidate is broken into two shorter candidates 
and kinked tracks are found.   

The second algorithm considers all remaining candidates that have not been used to make a track.  It forms all 
possible combinations of two candidates which do not share the same plane orientation.  
If the candidates have a similar longitudinal overlap, they are used to construct a three-dimensional line.  A search 
for unused clusters that have a position consistent with the 
candidate pair is then performed in the remaining view and a
3D-track is formed if a sufficient number of clusters is found.
This algorithm is more successful in tracking particle trajectories that are 
obscured by detector activity in one of the three 
orientations and is able to make tracks that intersect as few as 9 planes.
Both algorithms reject candidates that contain only one seed with an angle greater than 60$^{\circ}$ with respect to the neutrino beam direction; such candidates typically 
correspond to random energy deposits that happen to fit a straight line.

\begin{figure}[htbp]
\begin{center}
\includegraphics[width=0.6\textwidth]{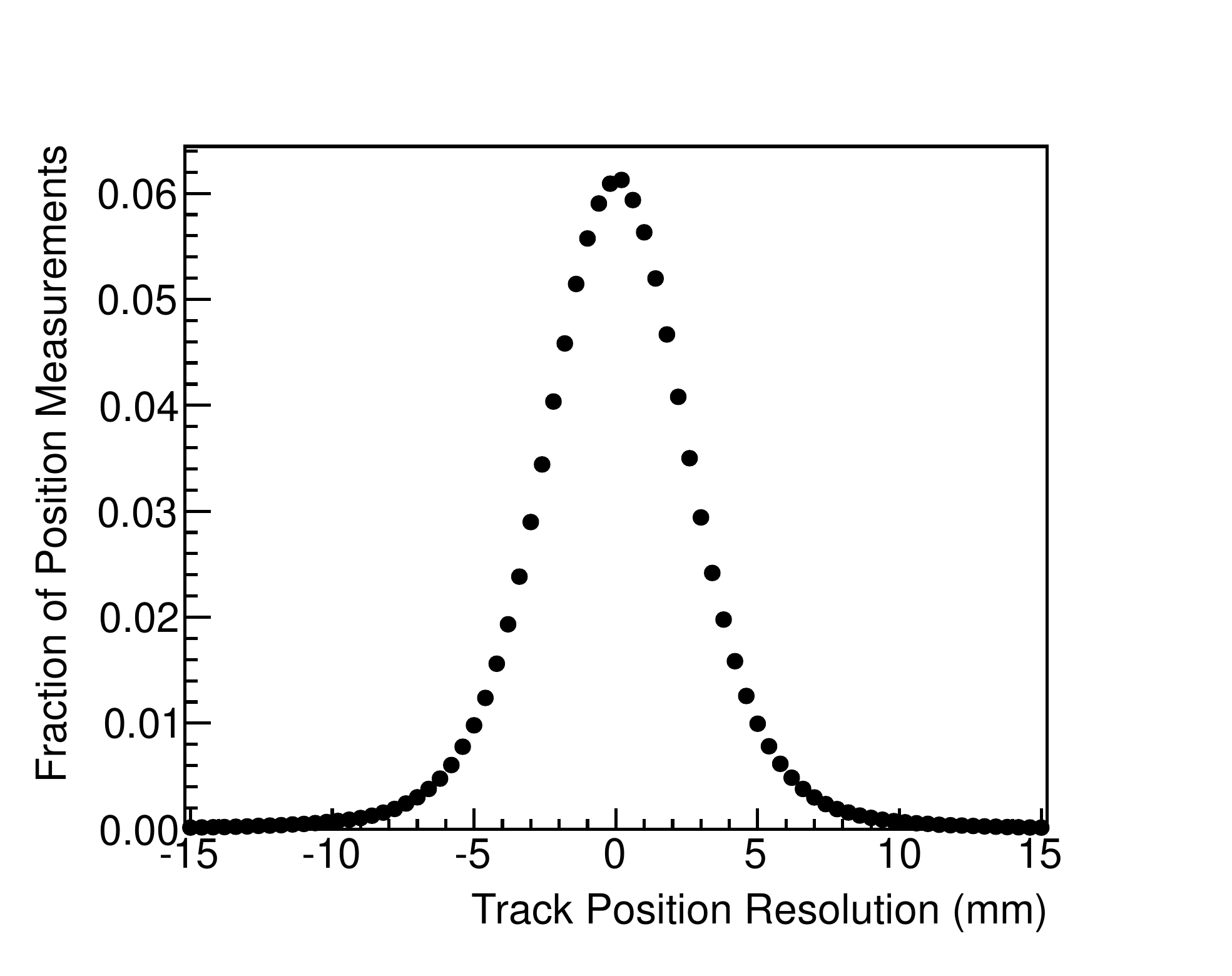}
\caption{ Resolution of the fitted positions along a track relative to the measured cluster positions for a sample of data rock muons.  
The RMS of the distribution is 3.1 mm.}
\label{fig:trkNodeResid}
\end{center}
\end{figure}

All 3D-tracks that are found are fit by a Kalman filter fit 
routine that includes multiple scattering\cite{fruhwirth}.  The fit is required to converge, but there is no 
requirement placed on the fit $\chi^2$.  Figure \ref{fig:trkNodeResid} shows the tracking position resolution after the Kalman filter fit.  
The fit is then used to add additional clusters to the track by searching nearby planes for which the track does not contain a cluster.  Superclusters are made available 
here, allowing the track 
to project into a region of shower-like activity (as in the case of a muon track that is partially obscured by a hadronic shower in all three views).  
Multiple tracks are allowed to claim the same cluster, each taking a fraction of its energy, to handle situations where two tracks intersect the same supercluster.

\begin{figure}[htbp]
\begin{center}
\includegraphics[width=0.475\textwidth]{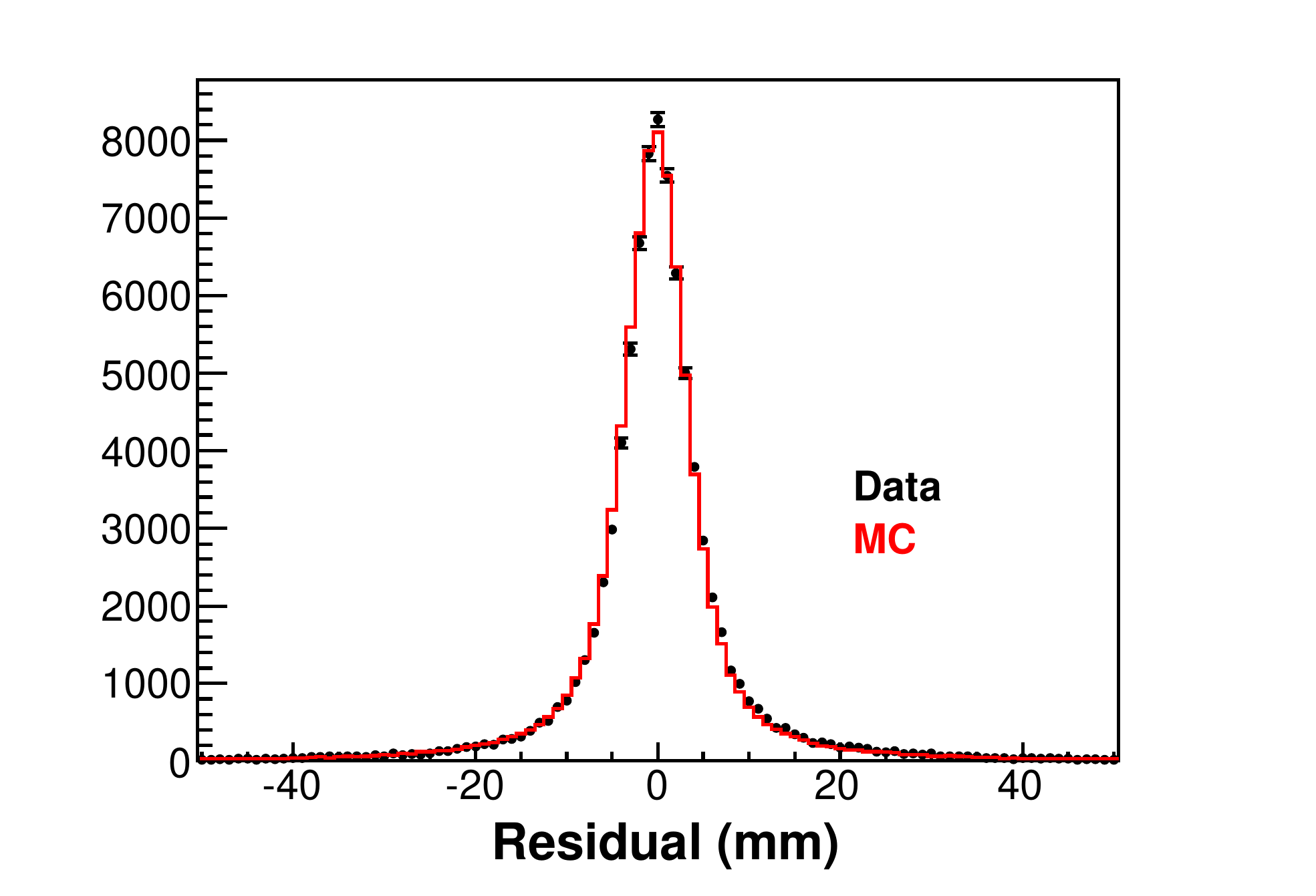}
\includegraphics[width=0.475\textwidth]{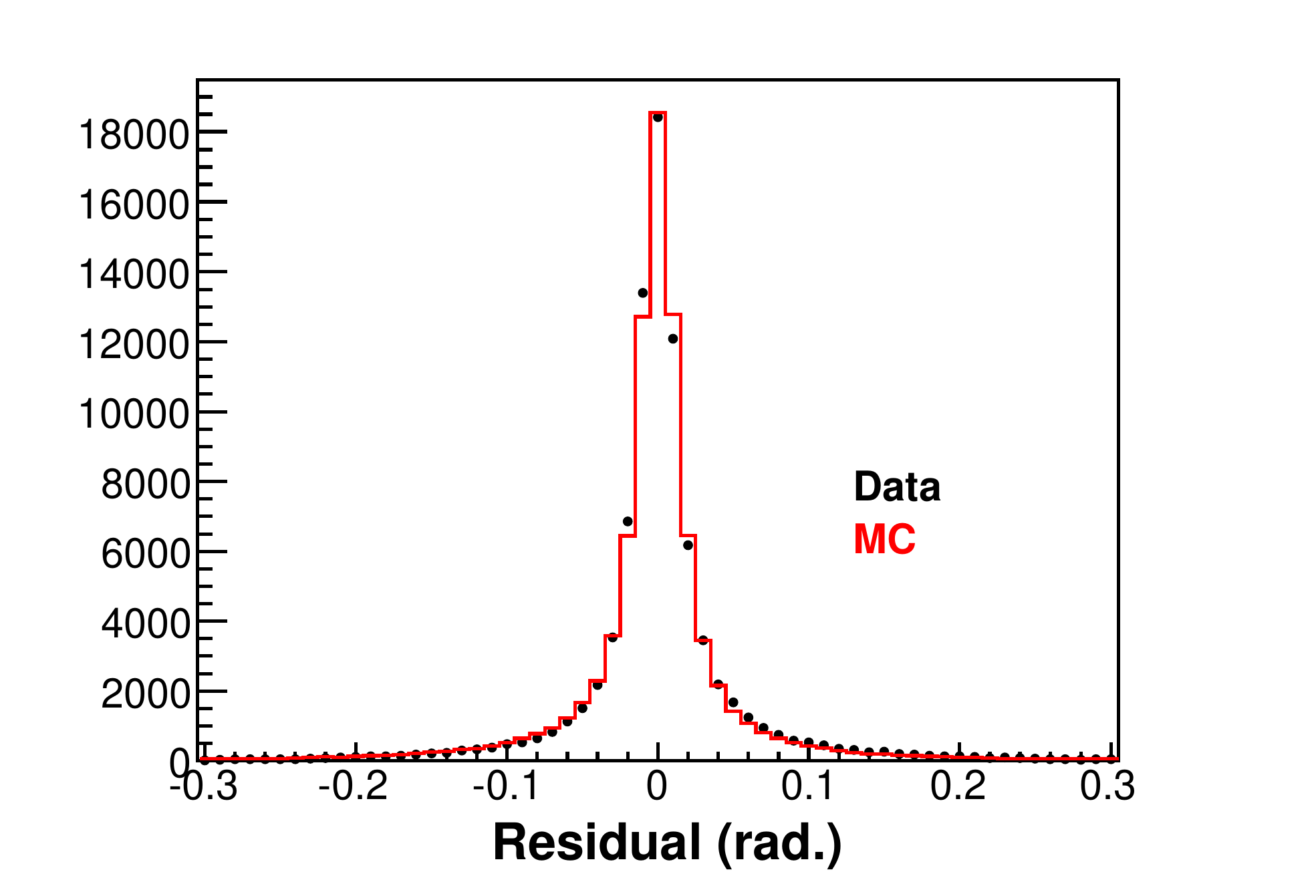}
\caption{ Residuals of fitted positions and angles between the upstream and downstream regions of tracks made in a rock muon sample. }
\label{fig:broken_track_study}
\end{center}
\end{figure}

Tracks produced by rock muons entering the front face of \minerva with 
momentum less than 20~GeV were used to study tracking performance. 
Each track is divided into two cluster groups and 
upstream and downstream clusters are then fit separately into two independent tracks.
The cluster at the break point is
included in both tracks.   The differences between the upstream and
downstream track positions and angles is then a measure of the
position and angular resolution, as shown in 
Fig. \ref{fig:broken_track_study}. 
The distributions in the data and
simulation are consistent with each other.  The position resolution is
simulated to within 1~mm and the angular resolution is simulated to within 1~mrad.  The position offsets are less than 60$\mu$m in both transverse dimensions and the angular offsets are less than 1~mrad in both directions.

\begin{figure}[htbp]
\begin{center}
\includegraphics[width=0.9\textwidth]{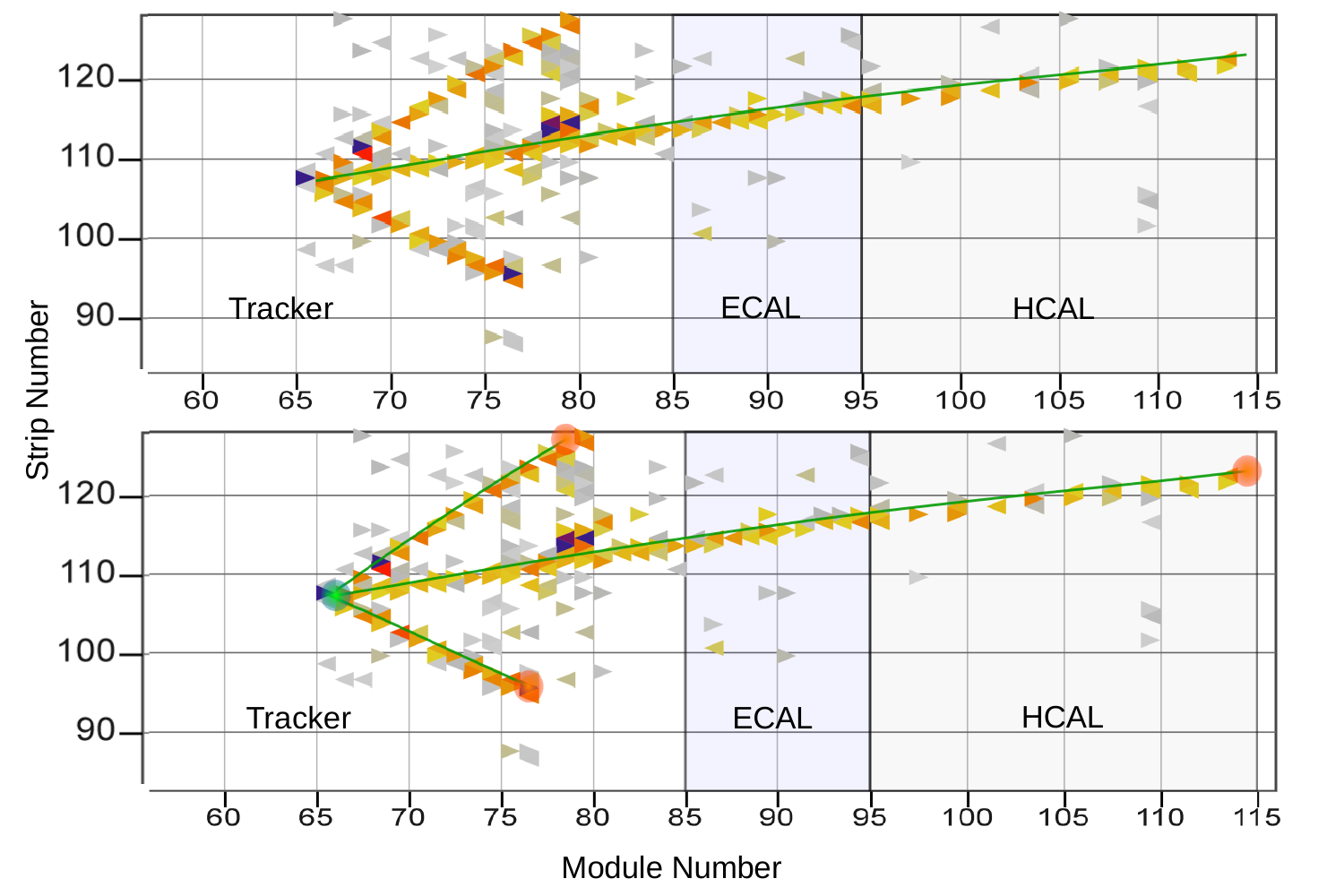}
\caption{ Display in X-view of a neutrino interaction recorded in data.  
Each triangular ``hit'' depicts a single scintillator strip with deposited energy.   
Reconstructed tracks (vertices) are indicated by superposed lines (dots).  Initially 
 the reconstruction algorithm finds a long muon-like anchor track (upper display) and the 
primary vertex location is inferred.   Then additional tracks emerging from the vertex 
are searched for.    In this event, two hadronic tracks are found and 
further constrain the vertex position (shorter lines and solid dot in the lower display).}
\label{fig:trkEvDisplay}
\end{center}
\end{figure}

The track pattern recognition scheme described above is used by a master reconstruction algorithm designed to reconstruct high-multiplicity 
final states while imposing restrictions to reduce the number of tracks formed from unrelated detector activity.  
If at least one track of 25 clusters or more is found, the longest track is  designated the
``anchor track'' and its clusters are marked as used. 
The starting point of the anchor track is used to define the neutrino interaction location, called the primary vertex.
The track pattern recognition is then re-run on the remaining un-used clusters.  
Tracks consistent with emerging from the primary vertex are identified and the primary vertex location is
re-estimated with the Kalman filter technique\cite{luchsinger} using all tracks. 
Tracks that are inconsistent with emerging from the primary vertex are deleted.  The procedure is repeated, searching for
tracks consistent with the endpoint of each track emerging from the primary vertex, in order to find particle  
trajectories that abruptly change direction due to secondary interactions.  Figure \ref{fig:trkEvDisplay} illustrates this 
procedure using event displays. 

Each found track is submitted to a procedure, called ``cleaning,'' that removes extra 
energy, which varies according to the type of track found.  
The purpose of cleaning is twofold: to remove the energy that is likely to be unrelated 
to the tracked particle so that it may be used by future iterations of the track pattern recognition, 
and to improve the vertex energy measurement.  Anchor tracks 
are typically found to be muons and are thus expected to deposit energy as a MIP.  
If the track contains energy near the vertex that is inconsistent with a MIP, or contains a supercluster, 
the extra energy is removed from the track.  Non-anchor tracks 
are assumed to be hadronic and only their superclusters are cleaned thereby minimizing bias in downstream particle 
identification algorithms.  The non-anchored track will only 
use energy from the supercluster in strips that intersect the track fit.  If the supercluster is near the 
track's endpoint, the track will incorporate as much energy as 
possible from the intersected strips in order not to disturb the energy loss profile of the Bragg peak. 
Otherwise, the track will use an energy equal to its mean cluster energy.

\makeatletter{}
\subsection{Charge Determination and Energy Reconstruction}
\label{sec:muon_reco}

The charges and momenta of muons exiting  \minerva\  are reconstructed using the MINOS near detector, which lies directly downstream of the \minerva\ detector. 
The MINOS calorimeter region is designed to measure both shower energies and muon 
momenta, while the downstream spectrometer region is only used for tracking.
The MINOS coordinate system is defined as follows: the positive $Z$-axis is parallel to the coil and points downstream; the positive $Y$-axis is 90$^\circ$ with respect to the $Z$-axis and points upward.  The $X$-axis is orthogonal to the other two axes. The center of the MINOS coordinate system is located at the intersection point between the coil and the first steel plane. 
Typically, muons produced with momenta between 0.5 and 6 GeV/$c$ within \minerva\   and matched to MINOS are contained 
in the MINOS calorimeter region. Higher momentum muons stop in the MINOS spectrometer region  
or escape MINOS completely.  Given the transverse extent of MINOS relative to \minerva, the MINOS detector provides coverage only for muons created that are at most 20$^\circ$ relative to the $Z$-axis.  

Charged particles traversing MINOS are deflected due to the magnetic field. From this deflection, information about the charge and momentum of the particle can be extracted using the same procedures used by the MINOS collaboration.~\cite{Rustem}.
In neutrino mode, if the deflection is towards the coil, the muon is negatively charged; and if the muon is deflected away from the coil, the charge is positive. The coil polarity is normally reversed in anti-neutrino mode to ensure that the dominant neutrino type is focused by the spectrometer. 

\makeatletter{}

For tracks in \minerva and MINOS to be merged into a muon candidate, the two tracks 
under consideration must be within 200 ns of each other in time, 
\minerva tracks must have activity present in at least one of the last five modules of the 
detector, and  the MINOS track must start within one of the first four planes 
of MINOS.  

Two separate methods are used to match tracks: a track projection method and a closest 
approach method.  For the track projection method, the MINOS track 
is extrapolated to the plane 
that contains the last activity on a \minerva track and 
the \minerva track is extrapolated to the plane that contains the start of the MINOS track.  
The position of the most downstream activity on a \minerva 
track is computed with the projection of the MINOS track, and likewise the 
projection from \minerva is compared to the start of the MINOS track. 
The distance between these 
points is called the match residual. If both match residuals are smaller 
than 40 cm, the tracks are considered matched tracks.  If multiple candidate matches exist,
 the match with the smallest residual is taken.

If no matches are found using the track projection method, 
a closest approach method is used.  The MINOS track is projected toward \minerva\  
and the \minerva\  track is projected toward MINOS and the point of closest approach
of the  two tracks is found.  This method can be useful if the muon undergoes a hard scatter 
in the passive material between the two detectors (for example, the support structure of the MINERvA detector or the first MINOS steel plane).  The \minerva tracks 
matched to a MINOS track are almost exclusively muons giving a high level of confidence 
to the particle identification in such cases.

The $\mu^{\pm}$ momentum in MINOS is determined by two different methods: range and curvature. The range method is based on total energy loss through interactions in the MINOS detector and is applied only to muons that are contained inside the calorimeter region. The curvature method reconstructs the momentum by means a track fitting algorithm developed by MINOS~\cite{MINOS2}. The algorithm  relates the curvature of the track ($K$), the magnetic field ($B$) and the momentum component perpendicular to the field ($P$), according to
\begin{equation}
K \equiv \frac{1}{R(cm)} = \frac{0.3\;B(kGauss)}{P(MeV)},
\end{equation}
\noindent where $R$ is the radius of curvature. 

\begin{figure}
\centering
\includegraphics[width=1.0\textwidth]{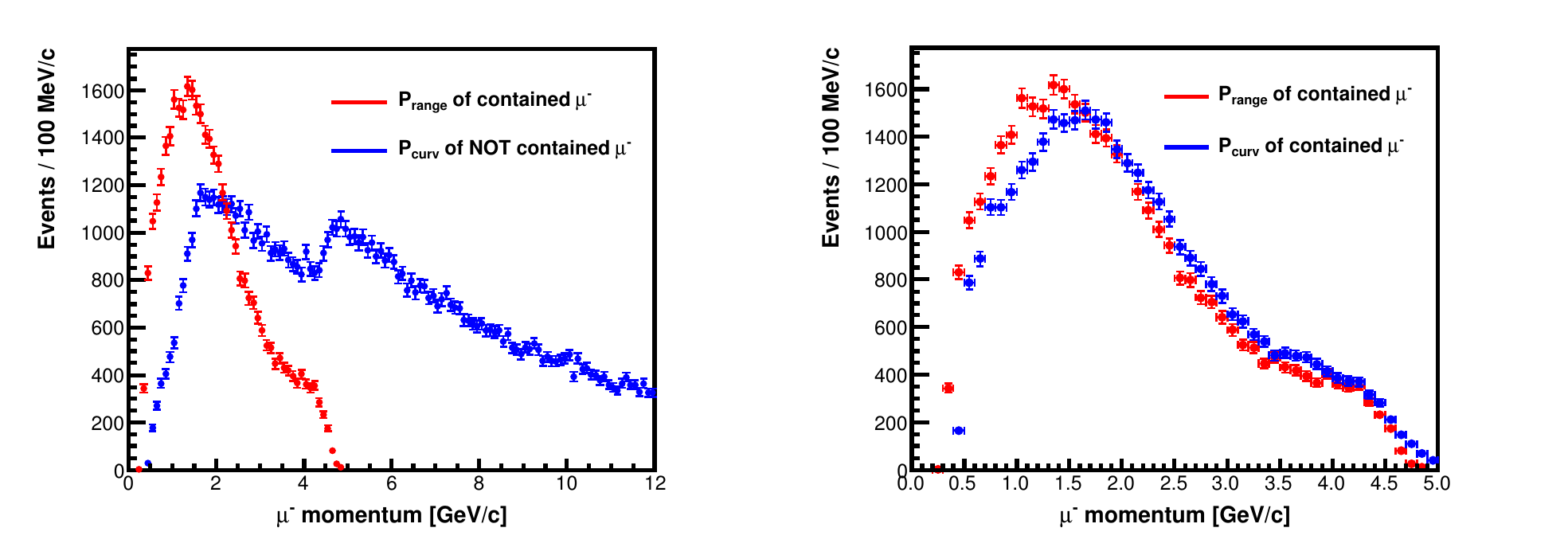}
\caption{Left:  The distribution in $P_{range}$ of muons contained in the calorimeter region versus the distribution of $P_{curv}$ 
for non-contained muons.  The contained muon sample peaked at lower momentum.  
Right:  Comparison of  $P_{range}$ and $P_{curv}$  distributions of contained muons. The plot shows that the two methods for reconstructing the momentum 
have different resolutions and reconstruction biases, especially at lower momenta.}
\label{fig:Pdistributions}
\end{figure}

\begin{figure}
\centering
\includegraphics[width=1.0\textwidth]{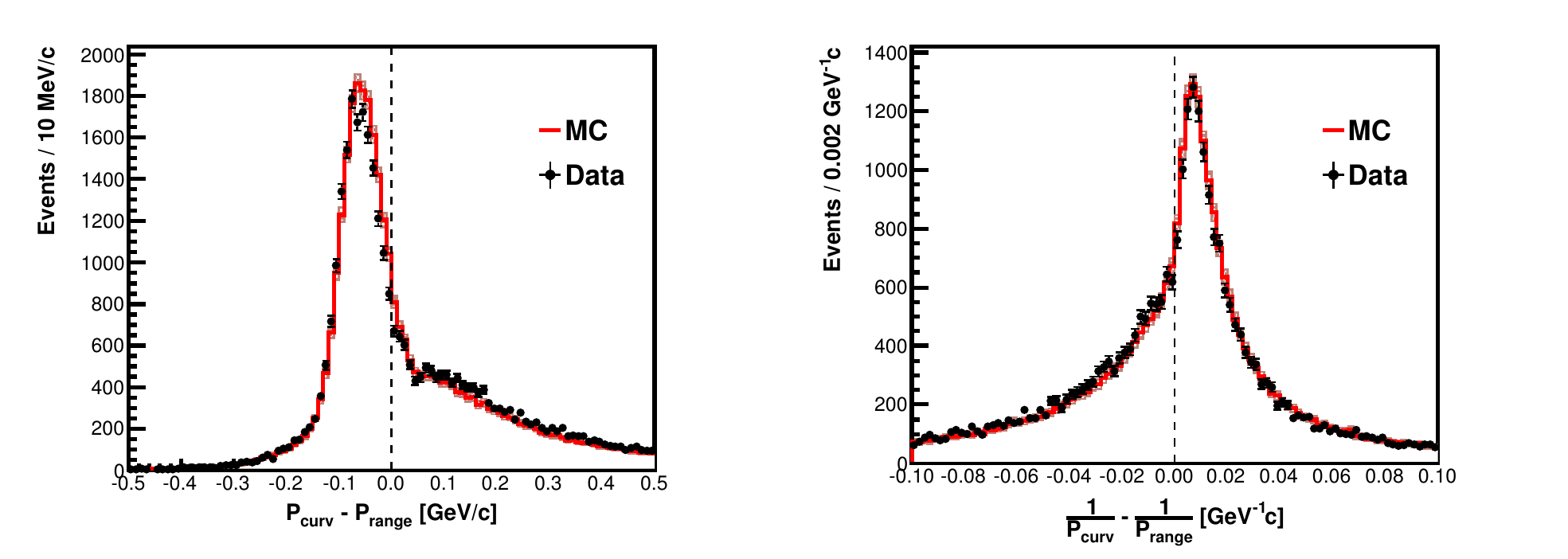}
\caption{Left: Residual $P_{curv}-P_{range}$ distribution of all CC inclusive events where the muon stops in the calorimeter region of MINOS.  Right: Residual $1/P_{curv}-1/P_{range}$ distribution for the same events.  This distribution is used to determine the systematic uncertainty for the curvature-based measurement relative to the range-based measurement.}
\label{fig:Residuals}
\end{figure}

The two methods differ in their ability to reconstruct momentum, as illustrated in Figs.~\ref{fig:Pdistributions} and~\ref{fig:Residuals}. The $P_{range}$ method is more precise; its estimated systematic uncertainty is 2\% \cite{Rustem}
derived from uncertainties on the simulation of the MINOS geometry, detector mass, and dE/dx parameterization, and track vertex reconstruction.  The momentum resolution for muons in MINOS is 10\% (5\%) for muons measured by curvature (range)\cite{minos-xsec}.  

For muons initiated in \minerva\ which are sufficiently energetic to escape from  the calorimeter region of MINOS, the momentum is reconstructed via the $P_{curv}$ method.  This is done because the coarser sampling in the MINOS spectrometer region and the fact that the signals in that region are summed together.  Both facts together result in a potential bias in momentum determinations based on range due to the high accidental activity.   \minerva\  has developed an approach to calculate the systematic uncertainty of the $P_{curv}$ method that is similar to that used by the MINOS collaboration~\cite{Rustem}.  

This study does differ from that of MINOS because it uses a high-statistics rock muon sample illuminating the entire face of the MINOS detector and it uses a well-defined track vertex at the front face of the MINOS detector. For this study, only muon tracks that are contained in the fully instrumented part of MINOS are used, so that information is available from both the $P_{range}$ and $P_{curv}$ methods.  

In order to determine the systematic error on $P_{curv}$, the ($1/P_{curv} - 1/P_{range}$) distributions of the data and the simulation were divided into six $P_{range}$ bins to determine the error as a function of the momentum. The arithmetic means for the data, $\bar{\mu}_{data}$, and simulation, $\bar{\mu}_{MC}$,  distributions are found.  Then 
the \emph{curvature difference}, defined as $\Delta K=|\bar{\mu}_{data}-\bar{\mu}_{MC}|$, or the  deviation of the $1/P_{curv}$ measurement from the $1/P_{range}$ measurement is obtained.
For small curvature uncertainty, $\Delta K$, $\Delta P_{curv} = - P^2_{curv} \Delta K$.
The additional systematic uncertainties on muon momenta measured by curvature 
are  0.6\% for muons with momenta greater than 1~GeV/$c$ and 2.5\% for those less than 1~GeV$/c$. 
The range and curvature uncertainties for muons measured by curvature are added in quadrature to obtain the total muon momentum uncertainty for muons measured by curvature of 2.1\% (3.1\%) for muons with momenta above (below) 1~GeV.

\makeatletter{}
\subsection{Muon Reconstruction Efficiency and Acceptance}
\label{sec:muon_eff_acc}

The muon reconstruction efficiencies are evaluated using simulated inclusive charged 
current muon neutrino interactions inside the \minerva detector tracker region.  All muons are 
counted in the efficiency denominator, and the efficiency numerator is determined 
by matching reconstructed tracks to the true particle trajectory that deposited 
the most energy into the clusters on the track.  A true muon is "tracked" if a 
track is matched to it in this manner.  

\begin{figure}[t]
\begin{center}
\includegraphics[width=0.45\textwidth]{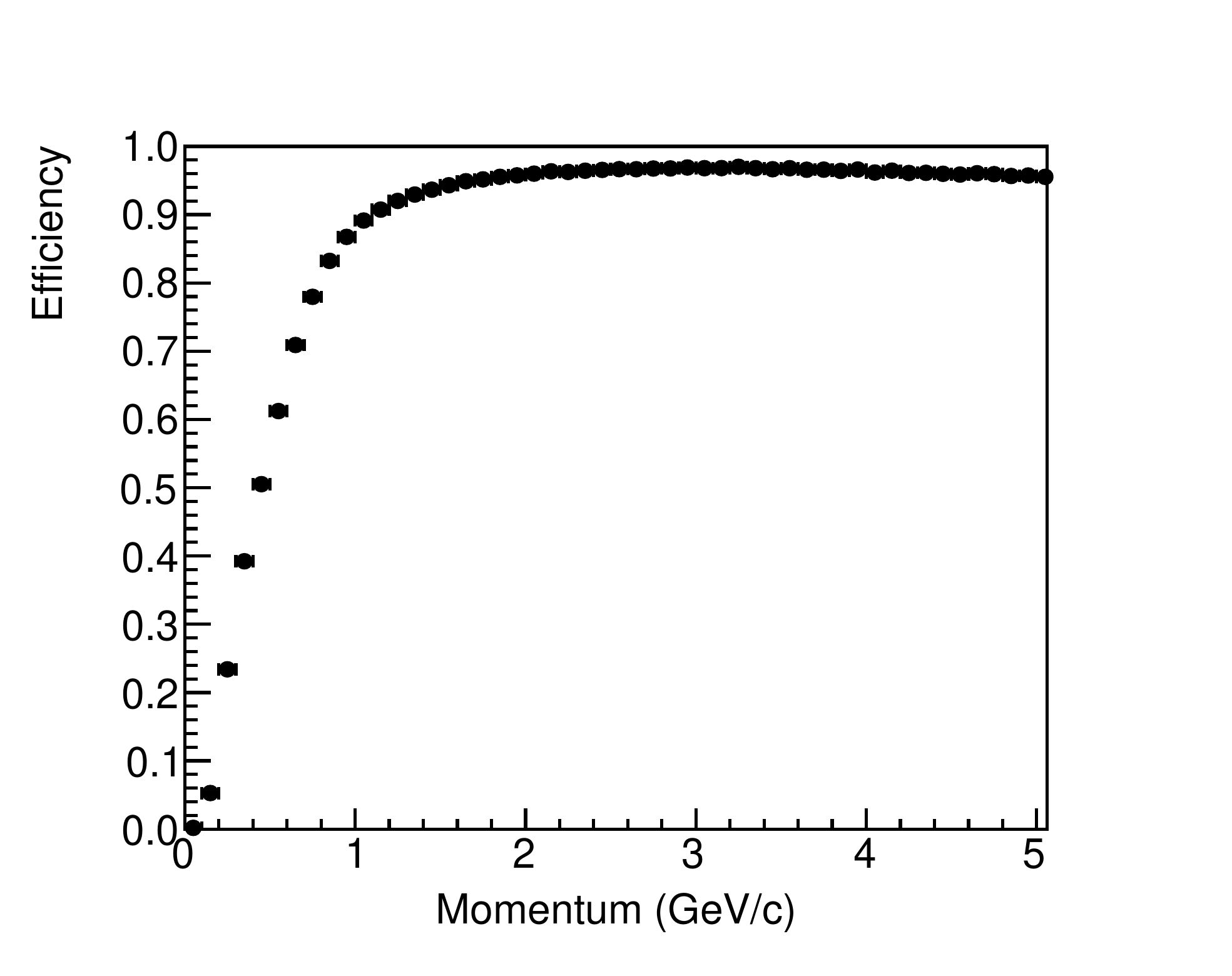}
\includegraphics[width=0.45\textwidth]{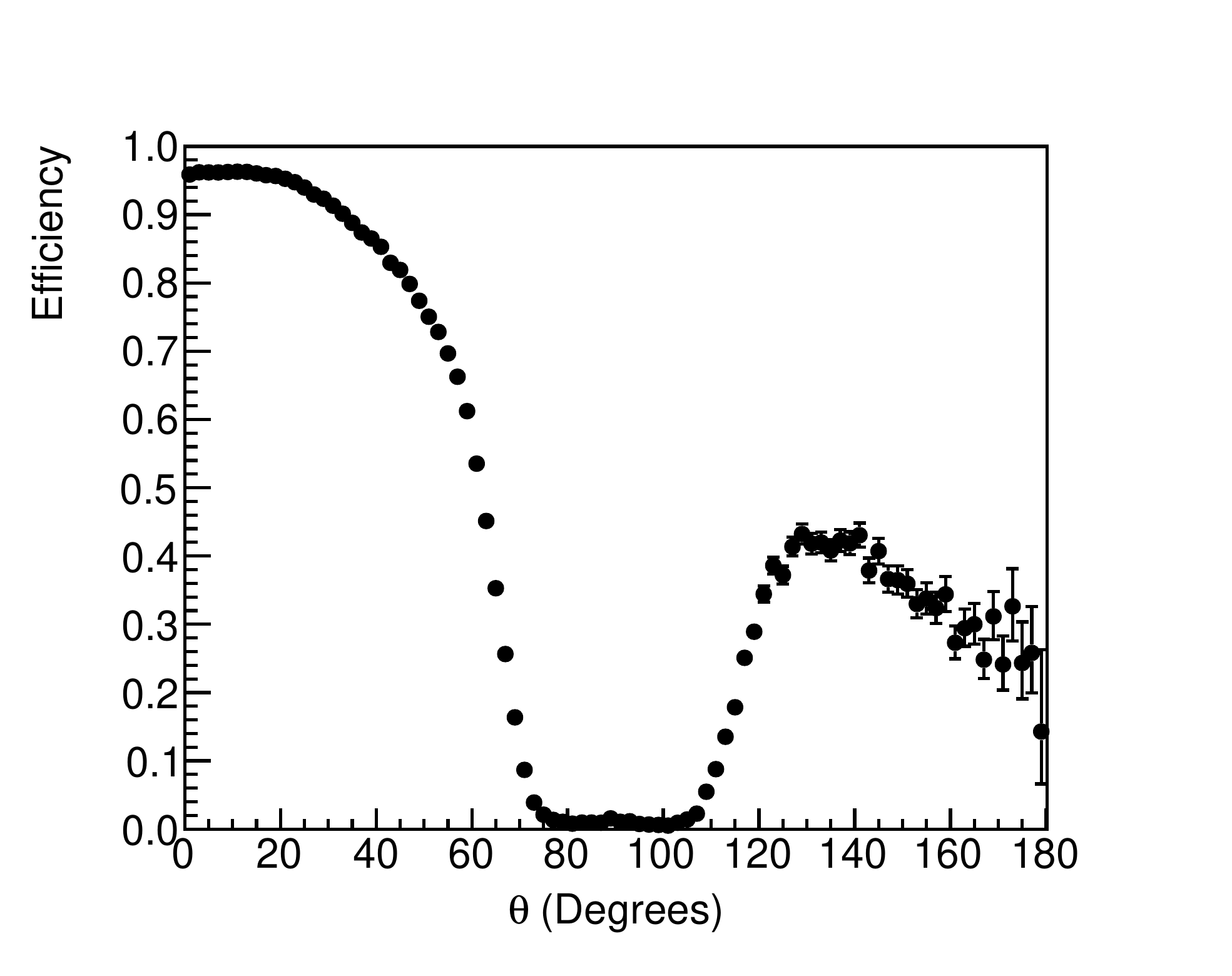}
\caption{ Muon tracking efficiency as a function of muon momentum (left) and angle with respect to Z-axis (right). 
The shape of the efficiencies is primarily a function of the number of planes intersected by the muon trajectory. 
Backwards muons are tracked less often because they typically have momenta below 1 GeV/$c$.}
\label{fig:muTrkEff}
\end{center}
\end{figure}

Figure \ref{fig:muTrkEff} shows the muon tracking efficiency as a function of momentum 
and angle with respect to the longitudinal axis, respectively.  The tracking efficiency 
decreases below 2.0 GeV/$c$ primarily due to the requirement that the anchor track contain 
at least 25 clusters.  The \minos\ muon acceptance turns on at approximately 2.0 GeV/$c$ because of the requirement that the muon pass through 25.4~cm of steel to be tracked in MINOS.  As a result, the anchoring requirement in \minerva does not inhibit charged current (CC) analyses that require 
MINOS-matched muons.  The angular efficiency decreases sharply around $60^\circ$ because only trackable and 
heavy-ionizing clusters are initially used in the track pattern recognition. The more gradual decline in efficiency between 
$20^\circ$ and $60^\circ$ is attributed to muons that exit the inner detector before crossing the minimum number of planes needed to 
form a track.  The MINOS angular acceptance is zero above $20^\circ$, so again the limitations of the track pattern recognition do not 
inhibit MINOS-matched muon analyses.

\makeatletter{}
\section{Recoil Energy Reconstruction}
\label{sec:recoil}

\newcommand{\nuEnergyRez}{$\sigma/E = 0.134 \oplus 0.290/\sqrt{E}$ }

\newcommand{\antiNuEnergyRez}{$\sigma/E = 0.163 \oplus 0.251/\sqrt{E}$ }

\newcommand{\deltaEoEdef}{$\Delta E/E^{\mathrm{true}}_{\mathrm{recoil}} = (E^{\mathrm{cal}}_{\mathrm{recoil}} - E^{\mathrm{true}}_{\mathrm{recoil}})/E^{\mathrm{true}}_{\mathrm{recoil}}$ }

\minerva is a finely grained detector, capable of identifying individual
particles in the recoil system (event energy not associated with the primary lepton)
for moderate multiplicity events.
As of early 2013, the energy of the recoil system is reconstructed by calorimetrically summing
energy depositions within the detector, without regard to the topology
of the event. In the future, a more complete algorithm will be developed
which identifies individual final state particles and compensates for
electromagnetic and hadronic depositions to improve energy resolution.

\minerva currently employs a simple calorimetric sum in which energy in the
sub-detectors not associated with the muon track is weighted to account for the active fraction of the scintillator
planes and additional passive absorber. The calorimetric constants are determined
by the $dE/dx$ of a minimum ionizing particle at normal incidence. For a given
sub-detector, the calorimetric constant is given by:
\begin{equation}
	C^{sd} = \frac{E_{\mathrm{abs}} + E_{\mathrm{sc}}}{f \times E_{\mathrm{sc}}},
\end{equation}
\noindent where $E_{\mathrm{abs}}$ is the energy loss in one absorber plane,
$E_{\mathrm{sc}}$ is the energy loss in one scintillator plane, and $f$ is the
active fraction of the scintillator plane in that sub-detector. For the central tracking region,
$E_{\mathrm{abs}} = 0$, yielding $C^{sd} = 1/f =$ 1.222 from the 81.85\% active
fraction. 
The corresponding fractions for the ECAL and HCAL are 2.013 and 10.314.
 The constant for the OD is likewise calculated assuming
normal incidence into the OD (orthogonal to the beam axis).

An overall calorimetric scale is derived by fitting calorimetric reconstructed
recoil energy to true recoil energy for simulated events. True recoil energy is
defined as the energy of the neutrino minus the energy of the outgoing lepton
\begin{equation}
	E_{\mathrm{recoil}}^{\mathrm{true}} \equiv E_\nu - E_{\mathrm{lepton}}
\end{equation}

\noindent Calorimetric reconstructed recoil energy is defined as:
\begin{equation}
	E_{\mathrm{recoil}}^{\mathrm{cal}} \equiv \alpha \times \sum_i C^{sd}_i E_i
\end{equation}
\noindent where $\alpha$ is the overall scale, $i = \{\mathrm{central~tracking~region, ECAL, HCAL, OD}\}$,
$C^{sd}_i$ is the calorimetric constant for sub-detector $i$ and $E_i$ is the total
visible recoil energy in sub-detector $i$, calculated from all clusters of hits
within a $-$20 to 35\,ns window around the event time (defined by the muon)
and not identified as cross-talk. This time window is narrower
than a typical time slice to remove pile-up from neutrino  and
background interactions that are adjacent in time. The cross-talk rejection prevents energy from
the muon track being included in the calorimetric sum.

The parameter $\alpha$ is determined by minimizing the quality factor $Q$
\begin{equation}
	Q = \sum \frac{\big[\arctan(E_{\mathrm{recoil}}^{\mathrm{cal}}/E_{\mathrm{recoil}}^{\mathrm{true}})-\pi/4\big]^2}{N},
\end{equation}
\noindent where the summation is over events of true recoil energy between 1.0 and
10.0\,GeV, and $N$ is the total number of such events. This metric 
is less susceptible to the asymmetric tails of the
$E_{\mathrm{recoil}}^{\mathrm{cal}}/E_{\mathrm{recoil}}^{\mathrm{true}}$
distribution, which is bounded below by zero, and bounded from above
by energetic hits in the active portion of the calorimeters (which are weighted up by the calorimetric
constants) and overlapping events.

After fitting $\alpha$, \deltaEoEdef is plotted in bins of true recoil
energy. A per-bin energy correction is derived in the form of a polyline mapping
$E_{\mathrm{recoil}}^{\mathrm{cal}}$ to $E_{\mathrm{recoil}}^{\prime\,\mathrm{cal}}$. Each node on the polyline
corresponds to one true recoil energy bin, with
\begin{align}
	x = E_{\mathrm{recoil}}^{\mathrm{cal}}         & = \langle E_{\mathrm{recoil}}^{\mathrm{true}} \rangle \times (1+\eta) \\
	y = E_{\mathrm{recoil}}^{\prime\,\mathrm{cal}} & = \langle E_{\mathrm{recoil}}^{\mathrm{true}} \rangle,
\end{align}
\noindent where $\langle E_{\mathrm{recoil}}^{\mathrm{true}} \rangle$ is the average true recoil energy in the bin, and $\eta$
is the mean of a Gaussian fit to the distribution. For example, if a bin with
$\langle E_{\mathrm{recoil}}^{\mathrm{true}} \rangle =$ 1.0\,GeV is 3\% low ($\eta =$ -0.03), the polyline maps 0.97\,GeV
to 1.0\,GeV. The lower limit of the polyline is fixed at (0.0, 0.0)\,GeV; the upper
limit is fixed at (50.0, 50.0)\,GeV.

\begin{figure}[htbp]
	\begin{center}
	\includegraphics[width=0.7\textwidth]{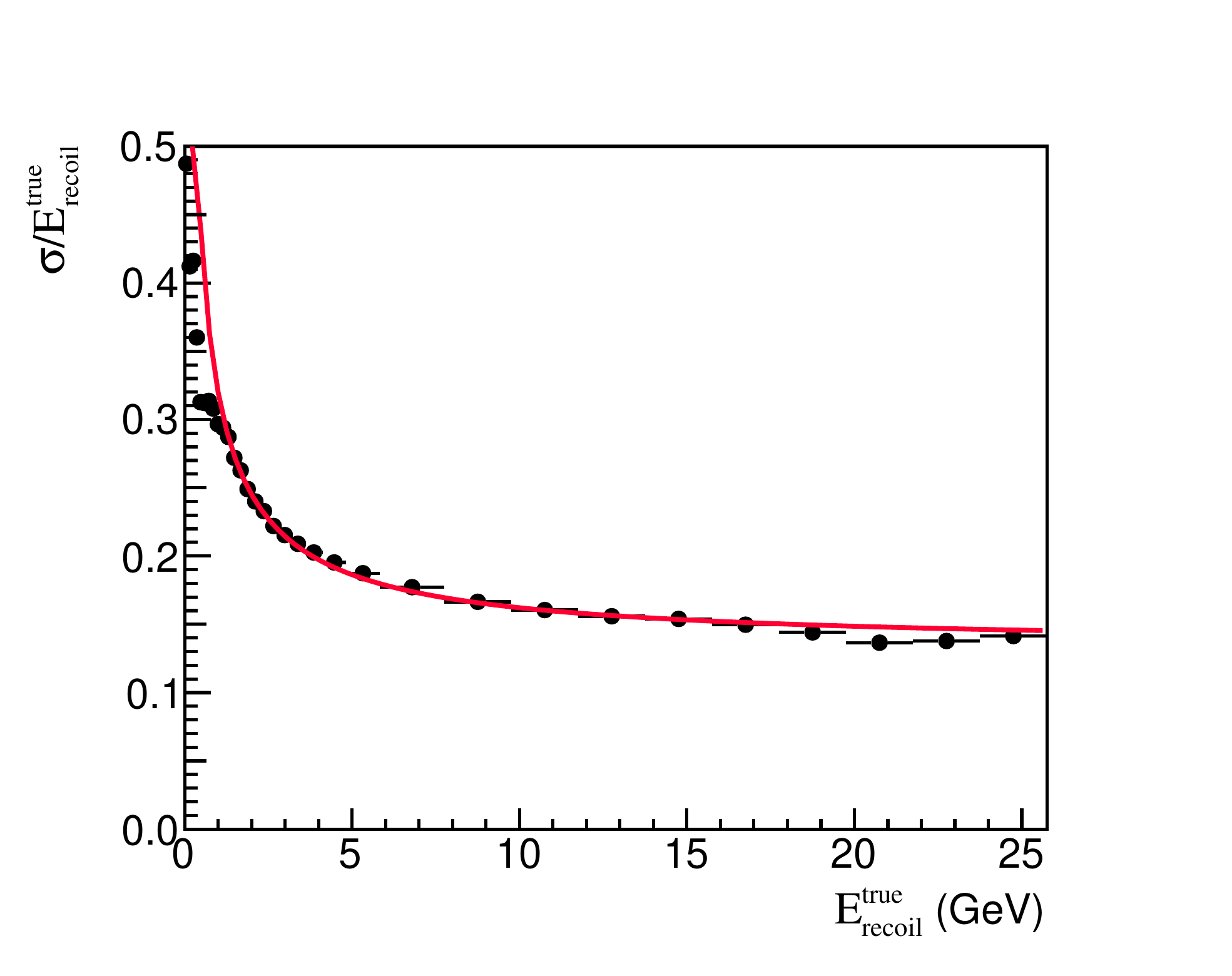}
	\caption{Calorimetric energy resolution for simulated charged-current inclusive
	neutrino events in the \minerva detector. The points show the width of a Gaussian
	fit to the difference between the measured and true recoil energy divided by the true recoil energy, binned by true recoil energy.  The line represents
	a functional fit, \nuEnergyRez.}
	\label{fig:calorimetric-energy-rez}
	\end{center}
\end{figure}

For simulated charged-current events with
MINOS-matched muons and with vertices within the fiducial tracking region, the value
$\alpha$ = 1.568 is obtained, with a calorimetric
energy resolution of \nuEnergyRez (see Fig.~\ref{fig:calorimetric-energy-rez}).
The observed calorimetric
energy resolution is a convolution of many effects: final-state interactions,
shower fluctuations to electromagnetic, hadronic and neutral components,
passive absorber deposition, scintillator, PMT and electronics response,
attenuation along scintillator strips, containment in the detector, and
overlapping events.  For the lowest recoil energies (below 1~GeV) the systematic uncertinty is considerably lower than suggested by the parameterization.  This is due to the fact that for those events the recoil system is dominated by a single proton that is not subject to many of the sources of shower fluctuations, and the particle is usually contained in the fully active region.  In these cases the resolution is better if the ionization profile is fit to a proton hypothesis, as described in Sec.~\ref{sec:dedx}.  

\section{Test Beam Detector Response Calibration}
\label{sec:testbeam}

The simulation of the calorimetric response of single final state particles
in the \minerva detector has been validated in a test beam program at the
Fermilab Test Beam Facility in 2010.  A dedicated tertiary test beam with hadron momenta between 
0.4 and 2.0 GeV/$c$ was used to study the response of a small test detector of 
similar design to the full \minerva\ detector.  The beam line included a spectrometer for
 momentum analysis and a time-of-flight system for particle identification.
 A stack of 40 scintillator planes of
1.15\,m$^2$ active area, identical to \minerva planes other than
the smaller transverse dimension 
 was exposed using two different configurations. The first configuration, 
with no absorber in front of the first 20 planes and Pb absorber before each of the downstream 20 planes, corresponds to the downstream
tracker region and ECAL. The second configuration, with lead upstream and steel downstream,
is very similar to the ECAL and HCAL region of the \minerva
detector.  

Other than the smaller transverse size, the detector design, electronics, and software are intentionally as
similar to the full \minerva\ detector as possible.  The most
important differences are as follows:  the air gap between planes and
passive absorber is larger for the tracker and ECAL configurations,
the typical light yield is doubled due to the shorter average fiber length, sets of four planes alternate readout on
opposite sides instead of all from the same side, and the muon energy
response calibrations are primarily done with cosmic-ray muons.  These 
produce differences in response that are accounted for
in the simulation and through the identical in-situ calibration procedures. 
Calibrations have been performed for the beamline momentum, the detector response to cosmic ray muons, and unrelated particle activity from the beamline.  The uncertainties in the material assay are comparable to the uncertainties in the calibrations.

These data are compared to a Monte Carlo simulation of the testbeam detector
geometry and its response using the same software and calibration
infrastructure as the full \minerva detector.  In order to assure that simulation and data share the same beam phase space, the particles trajectories
 input to the detector simulation are derived from data events with added resolution smearing based on multiple
scattering in the beamline.  The hadronic interaction model  used in the
simulation is the Bertini Cascade model packaged with the 9.4.p02
version of GEANT4\cite{geant}.

\subsection{Pions}

\begin{figure}[htbp]
	\begin{center}
	\includegraphics[width=0.7\textwidth]{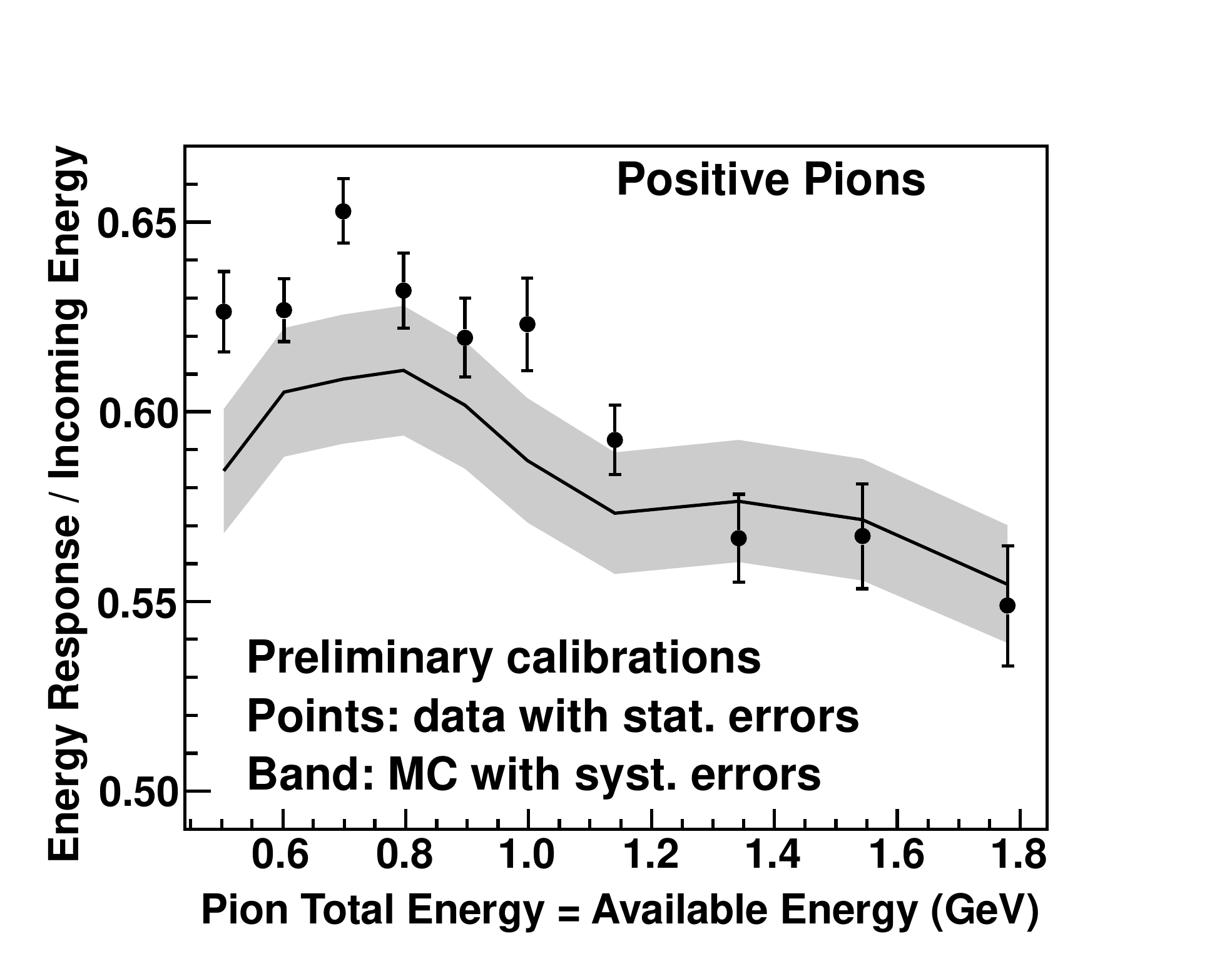}
	\caption{Results from calorimetry analysis for positive pions.
          The error band represents relative uncertainties between the
          data and simulation; there is
          an additional systematic uncertainty in the relative  scale between simulation and data of 3\%.  
        }
	\label{fig:testbeam-pion-response}
	\end{center}
\end{figure}

The  GEANT4 simulation of pions is approximately consistent with the pion data in the ECAL +
HCAL configuration as illustrated in Fig. \ref{fig:testbeam-pion-response}. At the low end of the momentum range, differences
between the data and simulation are 5\% on average, with the simulated response 
somewhat lower. Consequently a 5\% uncertainty is assigned to the calorimetric energy response for pions in that energy regime.
The energy resolution is well reproduced in the simulation.

\subsection{Protons}
 
A proton sample is used to set constraints on the
proton calorimetry response.  An analysis
similar to the pion case for the ECAL + HCAL configuration gives an
uncertainty in the calorimetric response of 5\% for protons
of momenta between 1 and 2~GeV/$c$.

\subsubsection{Stopping protons and Birks' parameter}

The proton sample is also used to study the Birks' law behavior of stopping protons in the central tracking region.
The central tracking + ECAL configuration has too little material to
contain interacting protons of momenta exceeding 1 GeV/$c$,  
but does provide constraints on calorimetry for lower momenta.  
A subset of the latter sample is used to study saturation behavior of the scintillator, which is commonly parameterized using Birks' Law, where the ionization is scaled by a factor of $(1 + k_B\times dE/dx)^{-1}$.  Events are selected which terminate in a tracking region plane between 11 and 19 and have a gap of no more than one plane upstream of that point. 

\begin{figure}[htbp]
	\begin{center}
	\includegraphics[width=0.7\textwidth]{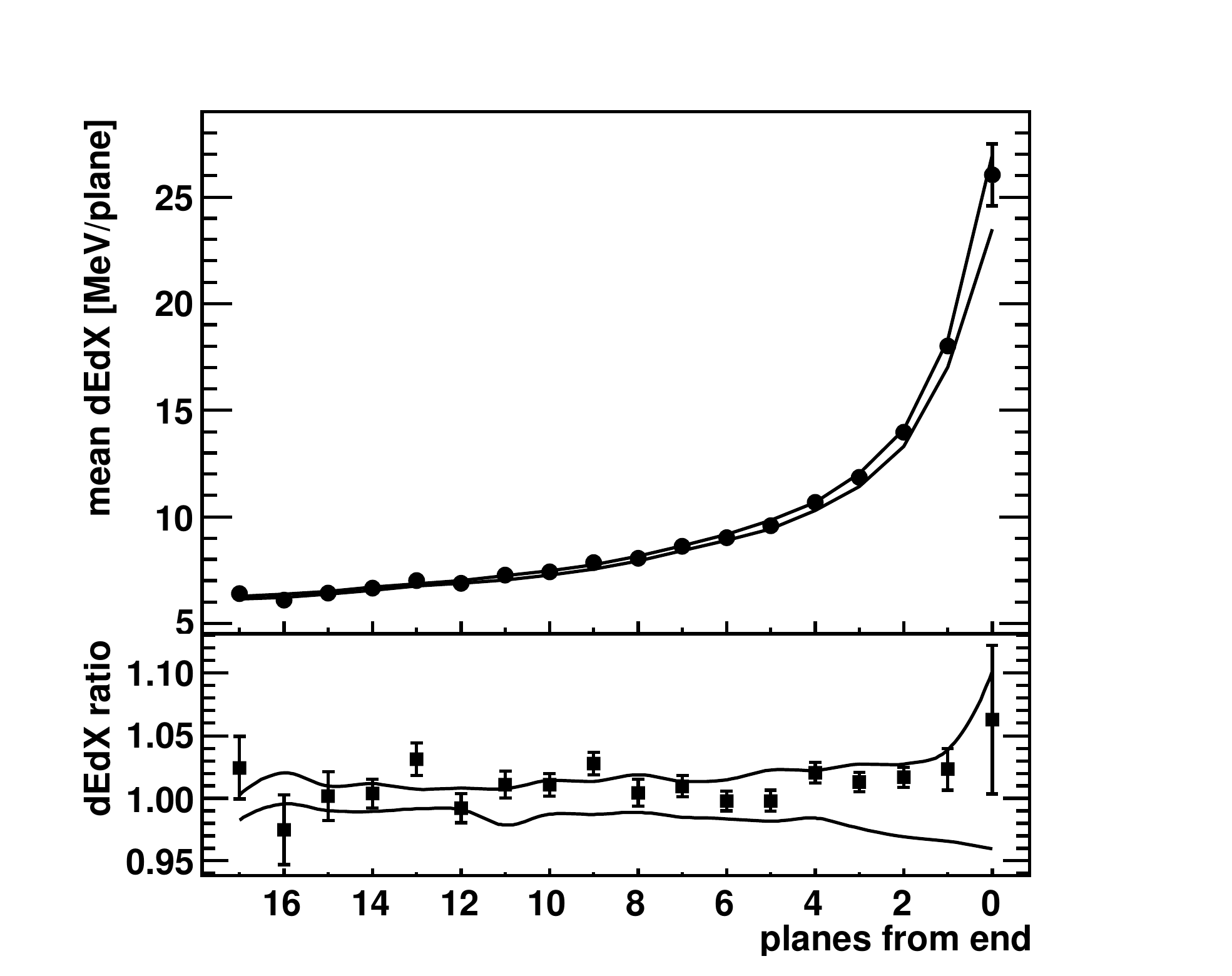}
	\caption{Energy loss profile for protons that stop at the end
          of their range.  The two smoothed lines in each part of the figure show the high-statistics 
          simulated profile with a Birks
          parameter that is 30\% lower (top) and higher (bottom),
          corresponding to higher and lower response, respectively.  The ratio is
          taken relative to the default simulation (not shown).        }
	\label{fig:testbeam-birks}
	\end{center}
\end{figure}
For all planes of all selected proton events, the distribution of
energy loss per plane is formed as a function of the number of planes from the end of the
event and shown in Fig. \ref{fig:testbeam-birks}. The data in the figure are constructed from a Gaussian fit to the
energy-per-plane distribution. The  central value from the Gaussian fit is shown for the data and
simulation, while the fit uncertainty is also shown for the data.
The uncertainty at the end of the track is larger because that final
distribution is not Gaussian.  There is a range of observed
energy loss from protons that travel only a short distance into the
plane to protons that stop just before they would have exited the
plane.  This leads to a larger spread in the peak energy loss.
The data lie between two alternative Birks' parameters at $\pm 30$\% of the nominal value ($k_B=0.133$mm/MeV).  
The $\pm$30\% is taken to be the estimate of the uncertainty in the parameter.
 For physics in this region that is sensitive to both the
total proton response and the Birks' parameter, such as activity near the neutrino
interaction vertex, the uncertainty in the response (~3.5\%) and the Birks' parameter
uncertainty are added in quadrature.

A full description of the test beam program and associated measurements is the subject of a separate manuscript which currently is in preparation.

\makeatletter{}
\section{Detector Performance}
\label{sec:DetectorPerformance}
Once all calibration procedures are complete, several checks are done to ensure that the detector's energy scales are accurate, well-modeled in the simulation, and constant over time.  Charged current events in the detector are used as checks, since their analysis requires all the calibration steps described in this article, and occur at high enough rates to provide precise measurements of the energy scales as a function of time.  Another check of the reconstruction is the energy deposition at the end of a track for particles that stop in the active region of the detector.  Other calibration cross-checks use the electrons that come from muon decays, for those muons that stopped in the detector (Michel electrons).  This chapter demonstrates the performance of the calibrated detector using these three data samples.
  
\makeatletter{}
\subsection{Charged Current Interactions}
\label{sec:cc_events}

Muon neutrino and muon antineutrino charged current (CC) interactions occurring within the ID volume provide high statistics checks of both the muon and recoil energy reconstruction, since the CC reactions $\nu_{\mu} (\bar\nu_\mu) N \rightarrow \mu^{-} (\mu^+) + X$, include both a muon and a hadronic recoil system. 

Events are selected as follows: first, the event must contain a reconstructed track 
which matches a muon reconstructed in the \minos\ detector where its momentum and charge are determined; 
second, the reconstructed muon vertex must lie within a fiducial volume in the scintillator-only 
section of the \minerva ID (see Sect.~\ref{sec:detectordesign}). The event must not contain any dead channels 
induced by previous interactions. Channels whose discriminators trip due to detector activity 
experience a dead-time, where they are insensitive to new energy, during an approximately 100~ns 
push-and-reset period which follows the $\sim$150 ns charge integration window.  We require that there  
be no more than one such dead discriminator in a path projected upstream of the muon track.  
This selection is essential to prevent rock muons with tracks partially lost to dead time from 
being confused with fiducial events. 

\begin{figure}[htb]
	\center
	\includegraphics[width=0.48\textwidth]{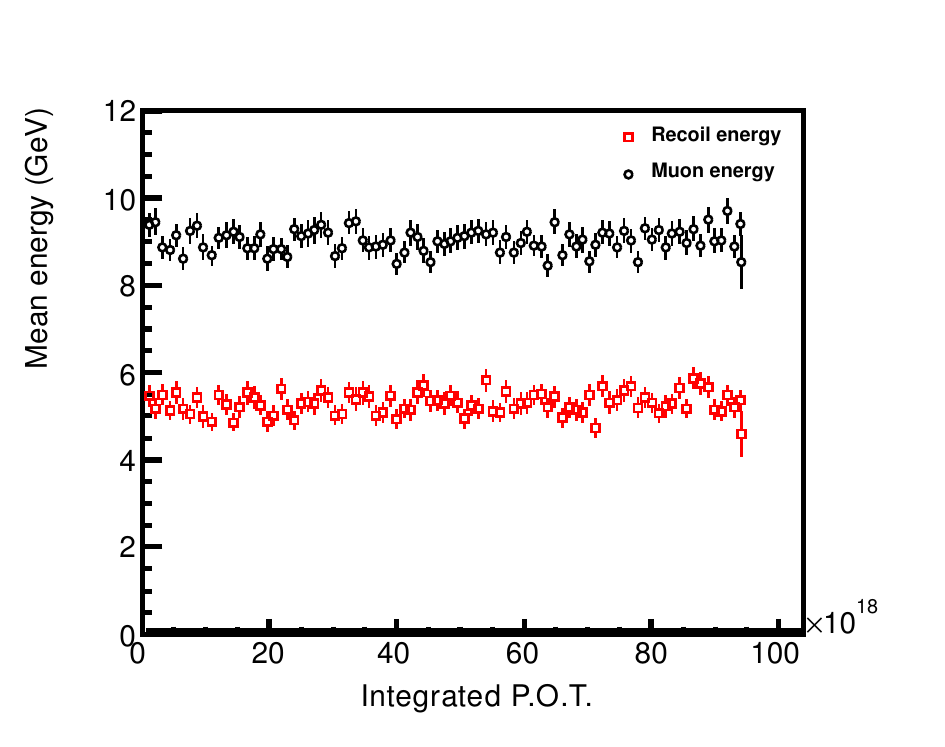}
	\includegraphics[width=0.48\textwidth]{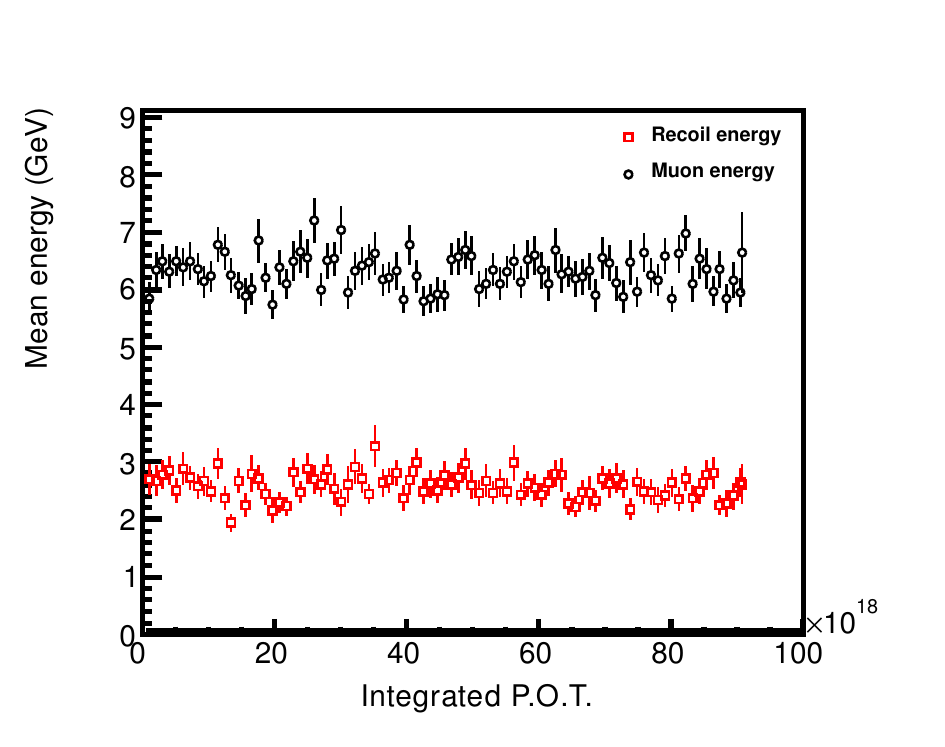}
	\caption{Mean muon and recoil energies in CC neutrino (left) and antineutrino (right) events as a function of integrated protons on target (POT).  Each point corresponds to $10^{18}$POT, which is approximately one day of running.}
	\label{fig:cc_energy_stability}
\end{figure}

Figure~\ref{fig:cc_energy_stability} shows the mean muon energy and the mean recoil energy, as functions of integrated protons on target over the 
first four months of neutrino data taking (left) and the first four months of antineutrino data taking (right), corresponding to $10^{20}$ protons integrated on the NuMI target.  The mean muon energies are constant to within 1.5\% (1.2\%) over the time shown for the neutrino (antineutrino) running.  The mean recoil energies are constant to within 3.0\% (3.5\%) over the time shown for the neutrino (antineutrino) running.  

\makeatletter{}
\subsection{Energy Loss for Stopping Particles}
\label{sec:dedx}

The granularity and light yield of the detector make it possible to use $dE/dx$ profiles near the ends of the tracks
to identify some of the particles that stop in the detector.
In cases where the hadron 
loses energy via electromagnetic processes, decays in flight, elastically 
scatters, or undergoes minimum inelastic hadron scattering, the $dE/dx$ 
distinguishes between minimum and heavily
ionizing particles. However,
because hadrons traversing the detector can undergo various other
processes, such as inelastic scattering, pion charge exchange, and 
absorption in flight, the 
particle's $dE/dx$ profile cannot always be used in this fashion.

In practice, for every track that is found, a  $\chi^2$ is determined by comparing the energy deposited per scintillator plane to templates derived from the $dE/dx$ profile expected in the detector for different momenta and for two different particle types: pion and proton.

\begin{figure}[htbp]
	\center
	\includegraphics[width=0.69\textwidth]{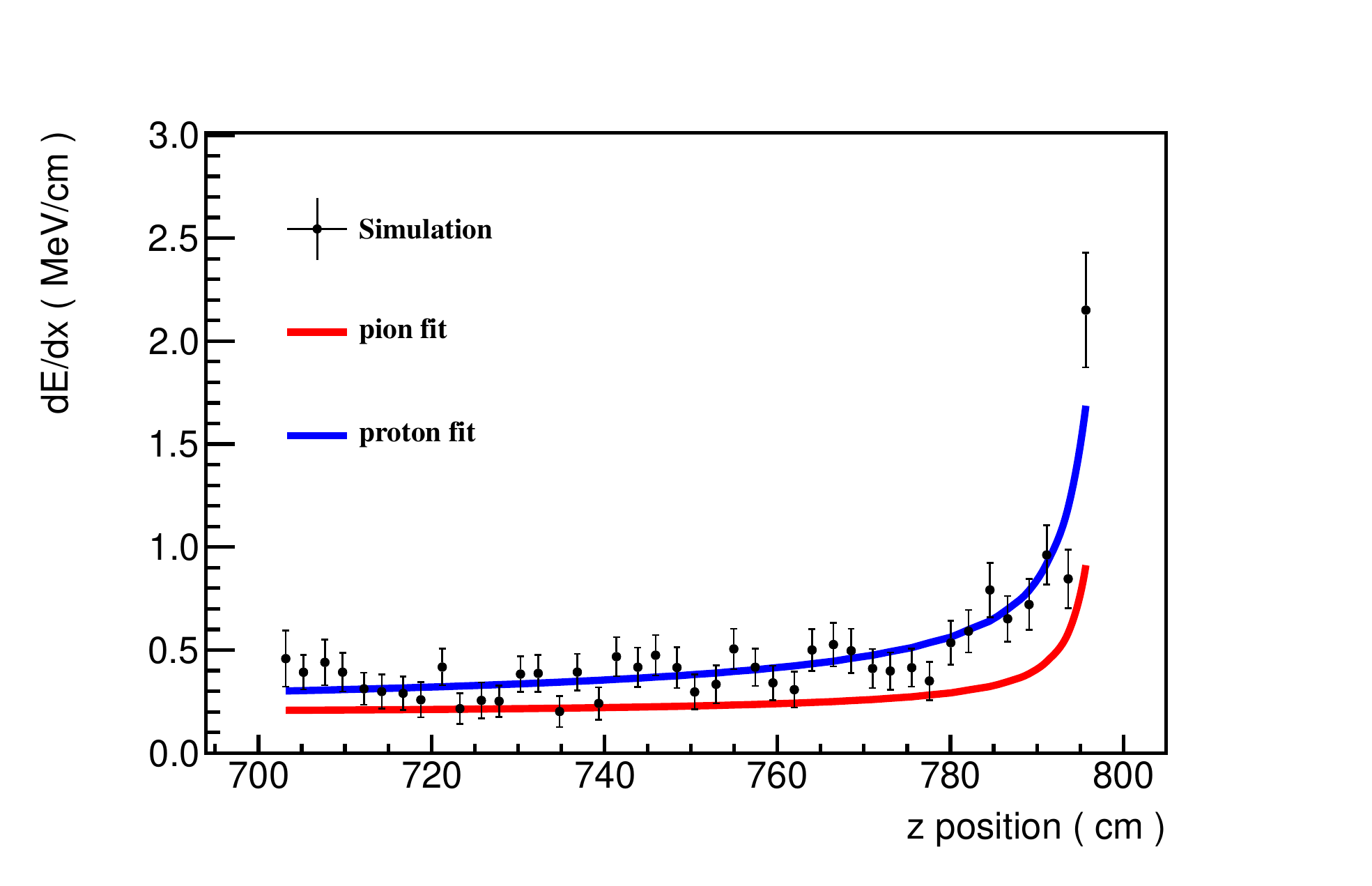}
	\includegraphics[width=0.69\textwidth]{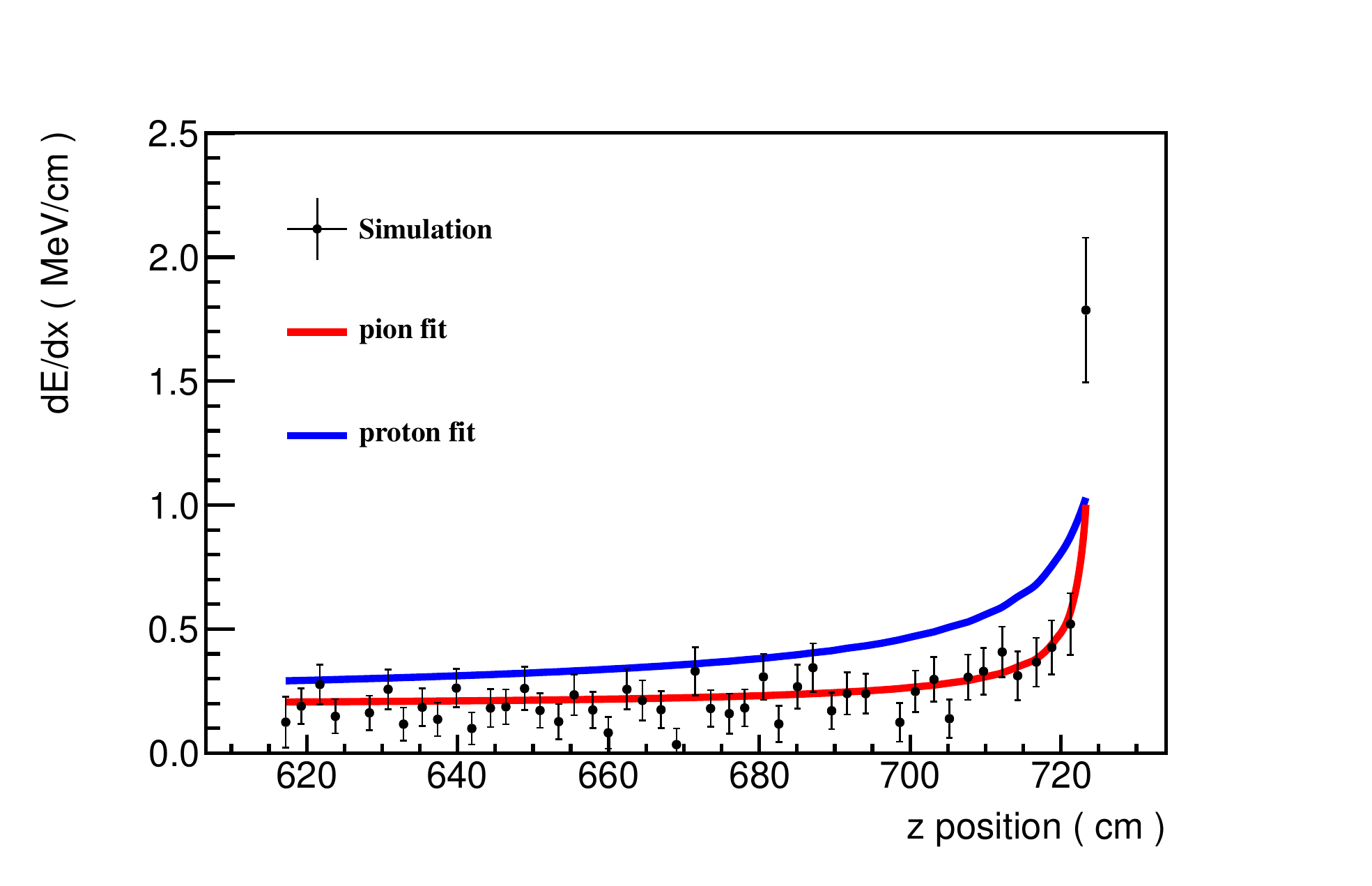}
	\includegraphics[width=0.69\textwidth]{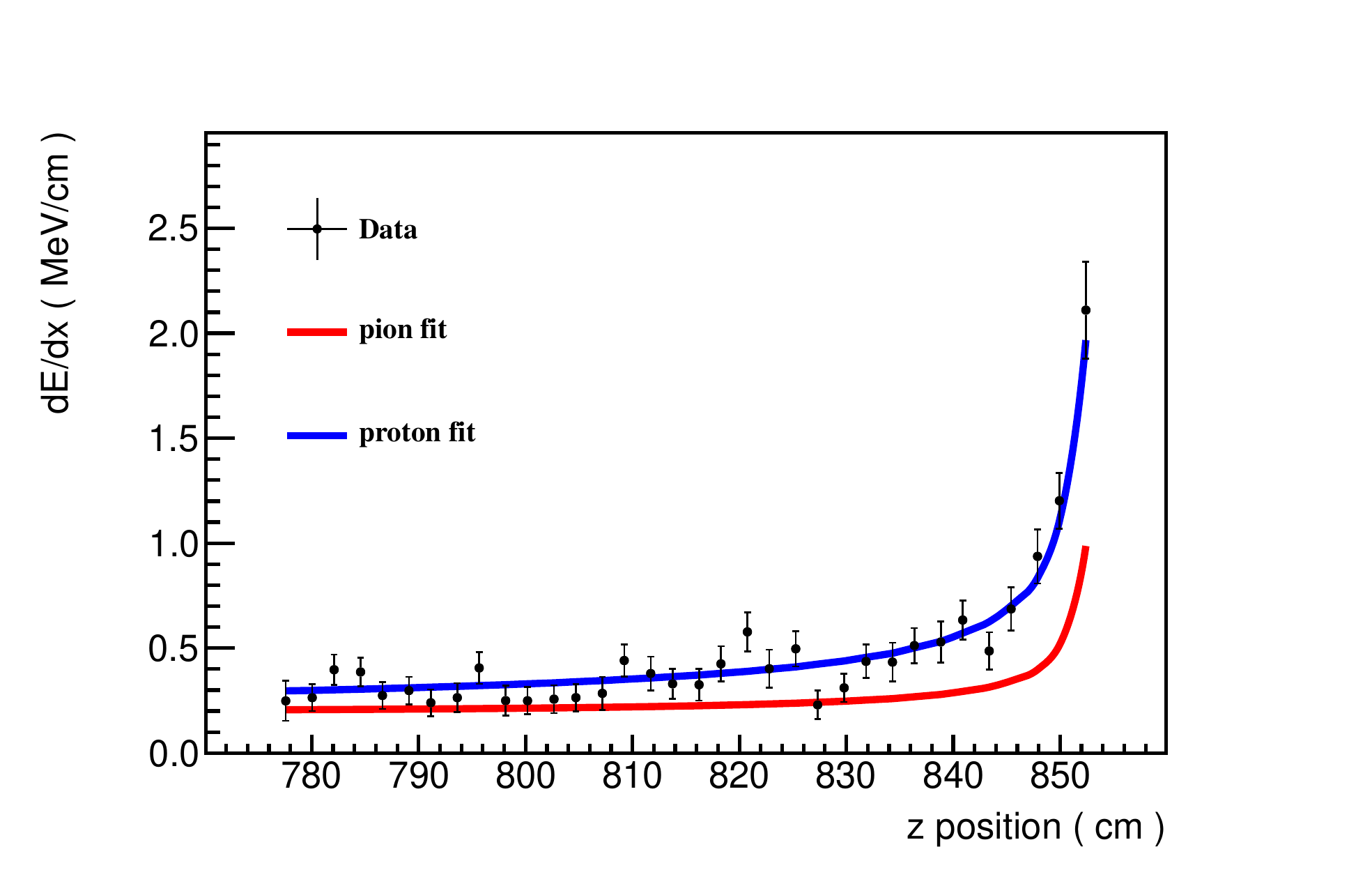}
	\caption{$dE/dx$ profiles for a typical simulated proton (top) and pion (middle) interaction, and for a clearly identified proton candidate in the data (bottom). }
	\label{fig:dedx_profiles}
\end{figure}

Figure \ref{fig:dedx_profiles} shows the $dE/dx$ profiles for two events from the detector simulation and an event from the data compared to pion and proton templates. The topmost  profile shows a proton generated with a momentum of 1.0~GeV/$c$; the $\chi^2$ for the proton (pion) hypothesis is 34 (177) for the 41 degrees of freedom (planes traversed).  The reconstructed momentum for the proton (pion) hypothesis is 1.2 (0.475)~GeV/$c$.  The center profile shows a simulated pion of 0.385~GeV/$c$ momentum;  the $\chi^2$ for the pion (proton) hypothesis is 31 (185) for the 39 planes traversed.  The reconstructed momentum for the pion (proton)  hypothesis is 0.36 (1.00)~GeV/$c$.  The lower profile is taken from a reconstructed track in data, where the measured proton (pion) momentum is 1.00 (0.36)~GeV/$c$ and the $\chi^2$ for the proton (pion) hypothesis is 29 (197) for 33 degrees of freedom.  

\begin{figure}[htbp]
	\center
	\includegraphics[width=0.49\textwidth]{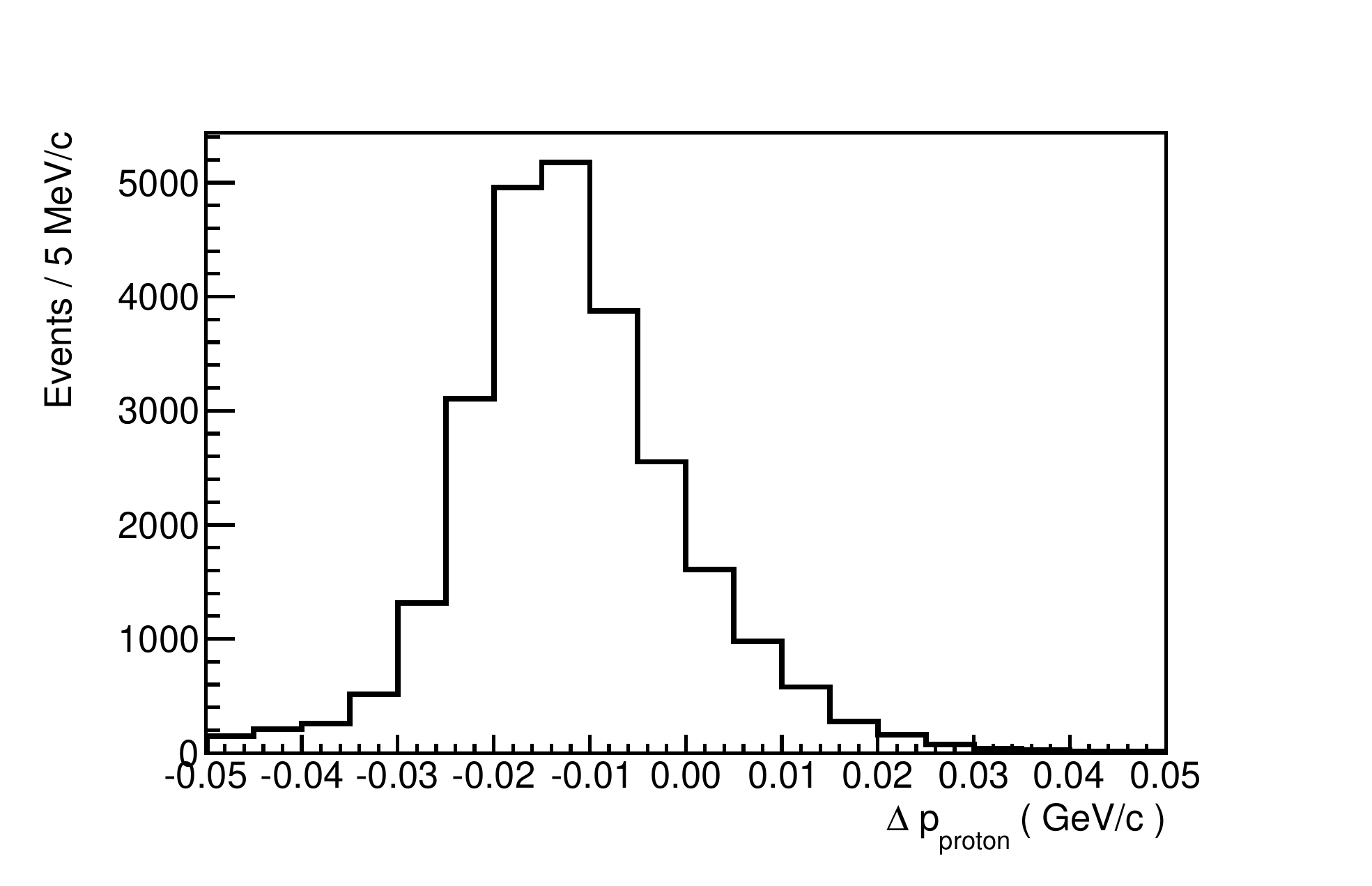}
	\includegraphics[width=0.49\textwidth]{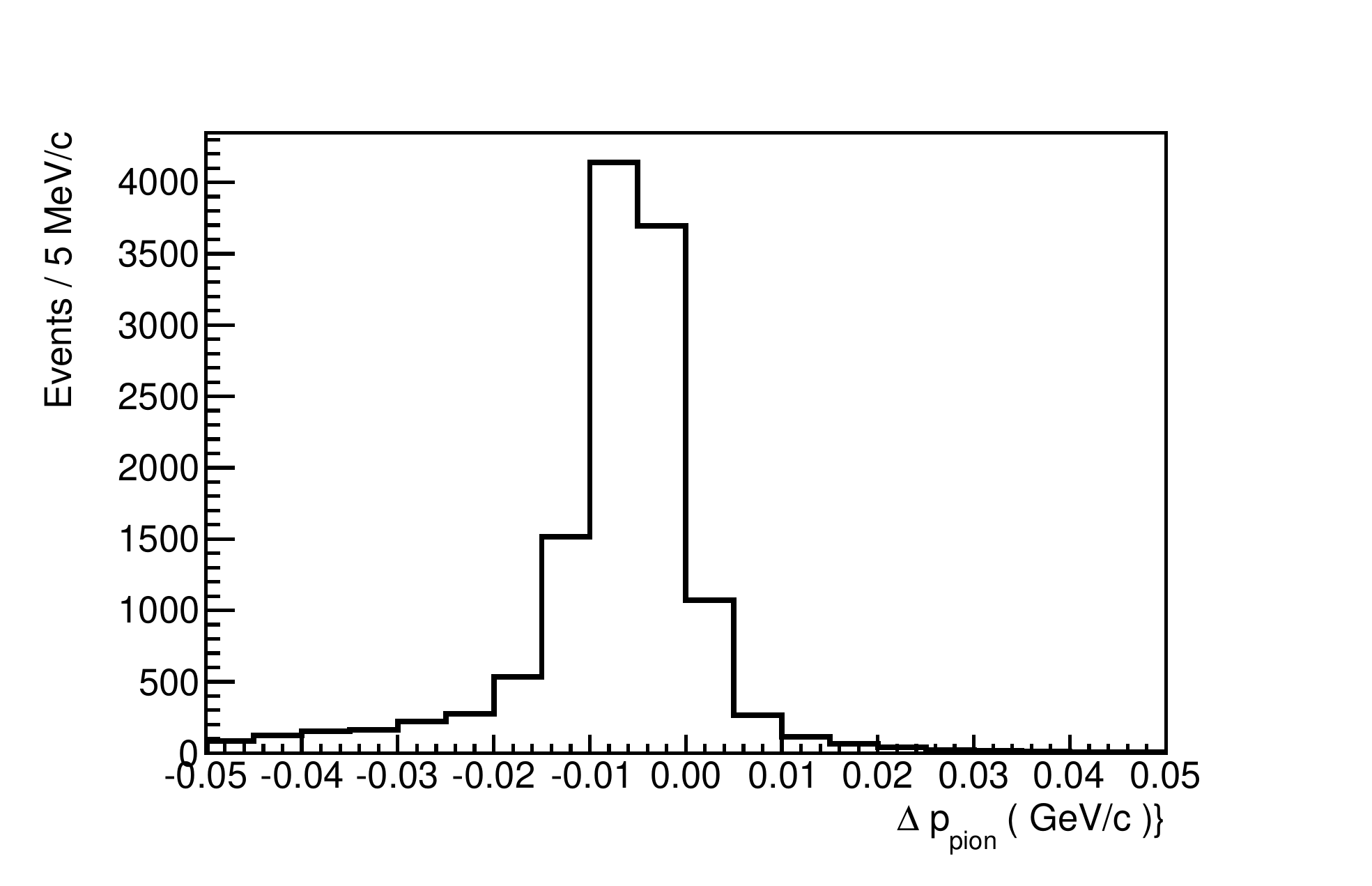}
	\caption{Momentum resolution for protons (left) and pions (right) obtained for simulated reconstructed tracks that stop in the \minerva\  inner tracking detector. $\Delta$ p is the difference between the reconstructed and true momentum of the single particle.}
	\label{fig:dedx_residuals}
\end{figure}

When available, the $dE/dx$ information is also used to measure energy for stopping particles
more precisely than a calorimetric energy sum. 
Figure \ref{fig:dedx_residuals} shows the predicted momentum resolution derived from $dE/dx$  information for protons and pions that stop in the inner tracking region of the \minerva\ detector. 

\makeatletter{}
\subsection{Michel Electrons}
\label{sec:michel_electrons}

Michel electrons are produced by a stopped (anti-) muon from a neutrino interaction or  the decay chain of $\pi^\pm$.
The response of the detector to Michel electrons at different locations provides a cross-check of the relative calibration.
The overall electromagnetic energy scale can also be checked by comparing the Michel electron spectrum in the data to that predicted by a simulation tuned to muon energy depositions.  

In principle, a Michel electron is most cleanly identified by searching for a delayed signal near the endpoint of a stopped muon track. 
However, isolated energy depositions in time slices with no other detector  activity are found to be predominantly due to delayed Michel electrons.  The full sample of such energy depositions can therefore be used without requiring an identified precursor muon.

\begin{figure}[ht!]
\begin{center}
\includegraphics[width=0.7\textwidth]{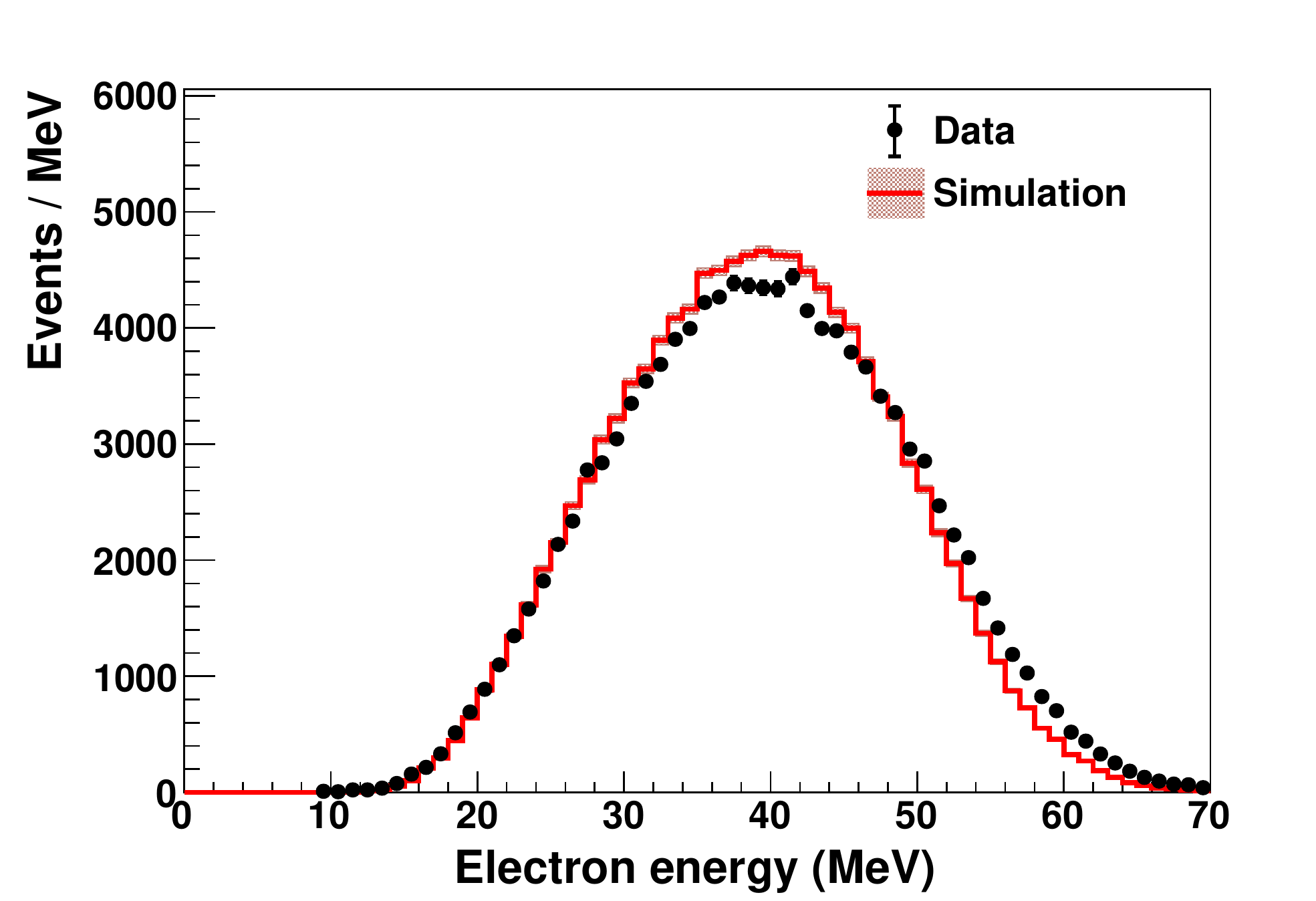}
  \caption{\small Energy distribution of Michel electron events observed in the detector
(solid points) compared to the full detector simulation (histogram).}
  \label{fig:Michel_energy}
  \end{center}
\end{figure}

Figure \ref{fig:Michel_energy} shows the energy spectrum comparison 
between data and simulation, where the simulation only includes neutrino interactions in the detector and the data includes all low energy candidates. The lack of a high-side tail in the energy distribution shows that the non-Michel background is low and can be ignored in this comparison.  The means of the distributions agree to within 3\%.  

\makeatletter{}
\subsection{Energy loss by  Electrons and Photons}
\label{sec:electron_photon}

\begin{figure}[ht!]
\begin{center}
\includegraphics[width=0.65\textwidth]{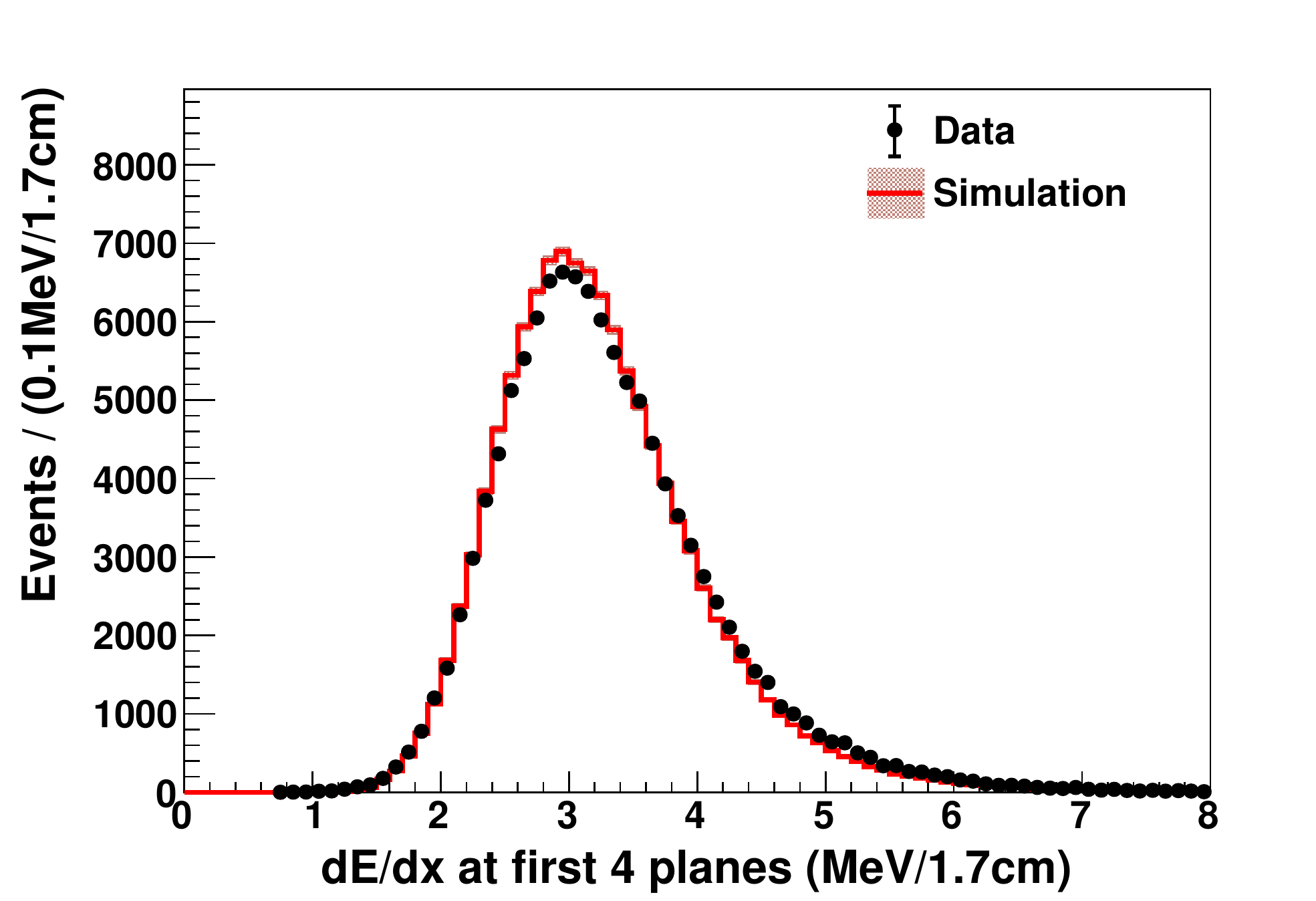}
  \caption{Average $dE/dx$ over the first 4 planes of Michel Electron events.} 
  \label{fig:Michel_dEdx_4planes}
  \end{center}
\end{figure}

The separation of electrons from photons is important for studying electromagnetic 
final states in \minerva.
The high granularity and low $Z$ nuclei of the central tracking region allows the 
$dE/dx$ near the beginning of electromagnetic showers to be used to distinguish electrons from photons.
Electrons lose energy as a single highly-ionizing particle near the start of a track. 
Because a photon produces an electron-positron pair when it converts to form a track, the
energy loss is then twice that of an electron.
This difference in $dE/dx$ is only valid near the start of a shower because an
electromagnetic (EM) shower develops stochastically as it propagates.
Consequently, the average $dE/dx$ over the first four planes at the start of an EM shower is a good discriminant.
A distribution of the average $dE/dx$ over the first 4 planes for Michel electron events is shown in Fig.~\ref{fig:Michel_dEdx_4planes}.

\makeatletter{}
\section{Conclusions}
\label{sec:Conclusions}

The \minerva\ detector has been operating since March 2010, and has achieved a total integrated live time exceeding 97\% for the "Low Energy" run that extended through the end of April 2012.  During that time over 99.5\%  of the channels were live and calibrated, using 
{\em in situ} neutrino beam-induced calibration samples and {\em ex situ} tests.  

The detector energy scale for the signal initiated by minimum ionizing particles is understood at the 
2\% level and the response has been  calibrated to be constant to better than 1\% across the 200 scintillator planes in the detector.  Although the light level decreased by over 15\% due to scintillator aging during the course of the initial two-year run, a calibrated energy stability of better than 2.5\%
and a timing resolution of better than 4 ns has been achieved over the length of the run.
The detector alignment, light 
collection, and calibration procedures result in a detector whose position resolution per plane 
for muons impinging perpendicular to the scintillator plane is 3.1~mm.  
The calorimetric energy resolution of the detector is estimated to be 
$\sigma/E = 0.134 \oplus 0.290/\sqrt{E(GeV)}$ for hadron showers generated by neutrino interactions. 
 Using a test beam and a smaller version of the detector, the hadron energy scale has been determined to 5\% for pions, and the Birks' law constant to characterize saturation in the scintillator has been constrained to $0.133\pm 0.040$mm/MeV.  Knowlege of this constant is particularly important for measuring energy depositions of low energy protons which may be in the final state of neutrino quasi-elastic interactions.

The detector composition is an important part of any cross section measurement.
The fiducial masses and chemical makeup of this detector and its various nuclear 
targets have been measured at the 1.5\% level for the scintillator planes
and at the 1\% level for the solid nuclear targets.  

The \minos\ near detector plays an integral role in any \minerva\ CC cross section measurement.
The \minerva\ implementation of the \minos\ near detector geometry and reconstruction are 
shown to give an absolute muon energy scale to 2.6\% (3.1\%) for muons above (below)  1.0~GeV for muon momenta measured by curvature.  The momentum uncertainty for muons measured with range is 2\%, as determined by MINOS material assay, dE/dx parameterization, and track reconstruction uncertainties.   

In summary, detailed neutrino cross section measurements for both exclusive and
inclusive channels for a wide range of target nuclei is now underway as the result of implementation by \minerva\ of the detector design and calibration approaches
reported in this article.

\section{Acknowledgments} 

This work was supported by the Fermi National Accelerator Laboratory, which
is operated by the Fermi Research Alliance, LLC, under contract No.~DE-AC02-07CH11359, including the \minerva\ construction project, with the United States
Department of Energy. Construction support also was granted by the United States
National Science Foundation under NSF Award PHY-0619727 and by the University
of Rochester. Support for participating scientists was provided by DOE and NSF (USA) by CAPES and CNPq (Brazil), by CoNaCyT (Mexico), by CONICYT
(Chile), by CONCYTEC, DGI-PUCP and IDI/IGI-UNI (Peru), by the Latin American Center 
for Physics (CLAF) and by FASI(Russia).  The \minerva\ Collaboration wishes to express its thanks to 
the \minos\ Collaboration for the use of its near detector data, reconstruction, calibration 
and simulation. Finally, the authors are grateful to the staff of Fermilab for their 
contribution to this effort, during the design, construction, data taking and data analysis phases of the experiment.

\makeatletter{}
\bibliographystyle{elsarticle-num}

\end{document}